\documentclass[10pt, a4paper]{report}
\usepackage[english]{babel}
\usepackage{amsmath}
\usepackage{amsfonts}
\usepackage[T1]{fontenc}
\usepackage{graphicx}
\usepackage{tikz}
\usetikzlibrary{shapes, arrows, matrix}
\usepackage{float}
\usepackage[utf8]{inputenc}
\usepackage{url}
\usepackage[colorinlistoftodos]{todonotes}
\usepackage{listings}
\usepackage{subcaption}
\usepackage[multiple]{footmisc}
\usepackage{algorithm}
\usepackage{longtable}
\usepackage{algpseudocode}
\usepackage{wrapfig}
\usepackage{natbib}
\setcitestyle{authoryear,open={(},close={)}}
\usepackage{csquotes}
\usepackage[paperwidth=8.27in, paperheight=11.69in, margin=0.8in]{geometry}
\usepackage{musicography}
\usepackage{hyperref}
\hypersetup{colorlinks=true,linkcolor=black,urlcolor=black,citecolor=black,bookmarksopen=true}
\usepackage{bookmark}
\usepackage{pdfpages}
\usepackage{appendix}

\begin{document}


\tikzstyle{decision} = [diamond, draw, text width=5em, text badly centered, inner sep=0pt]
\tikzstyle{block} = [rectangle, draw, text width=6.3em, text centered, rounded corners]
\tikzstyle{rect} = [rectangle, draw, text width=6.3em, text centered]
\tikzstyle{double_block} = [rectangle, draw, text width=6em, text centered, rounded corners, double, minimum height=4em]
\tikzstyle{line} = [draw, -latex']
\tikzstyle{double_line} = [draw, style=double]

\begin{titlepage}

\newcommand{\HRule}{\rule{\linewidth}{0.5mm}} 

\center 
 

\textsc{\LARGE Technical Report}\\[1.5cm]


\HRule \\[0.4cm]
{ \Large \bfseries DECIBEL: Improving Audio Chord Estimation for Popular Music by Alignment and Integration of Crowd-Sourced Symbolic Representations}\\[0.4cm] 
\HRule \\[1.5cm]
 

\begin{minipage}{0.4\textwidth}
Daphne \textsc{Odekerken}\\ 
Hendrik Vincent \textsc{Koops}\\
Anja \textsc{Volk}\\
Utrecht University


\end{minipage}\\[2cm]


{\large \today}\\[2cm] 

\vfill 

\end{titlepage}

\chapter*{Abstract}
Automatic Chord Estimation (\textsc{ace}) is a fundamental task in Music Information Retrieval (\textsc{mir}), which has applications in both music performance and in research on other \textsc{mir} tasks. The \textsc{ace} task consists of segmenting a music recording or score and assigning a chord label to each segment. \textsc{Ace} has been a task in the \textsc{mirex} competition for 10 years, but is not yet a solved problem: current methods seem to have reached a glass ceiling. Moreover, many recent methods are trained on a limited data set and consequently suffer from overfitting. In this study, we propose \textsc{decibel}\footnote{An implementation of \textsc{decibel} is available on \url{https://github.com/DaphneO/DECIBEL}.} (DEtection of Chords Improved By Exploiting Linking symbolic formats), a novel system that utilizes multiple symbolic music representations in addition to audio in order to improve \textsc{ace} on popular music.

The input for \textsc{decibel} not only consists of the audio file, but also contains a set of \textsc{midi} and tab files that are obtained through web scraping and manually matched to the audio file. Given the audio file and matched \textsc{midi} and tab files, the system first estimates chord sequences from each file, using a representation-dependent method. For audio files, \textsc{decibel} uses existing state-of-the-art audio \textsc{ace} techniques: in our experiments, we use the output of the nine \textsc{ace} submissions from \textsc{mirex} 2017 and/or 2018, as well as a commercial state-of-the-art method. \textsc{Midi} files are first aligned to the audio file, using a Dynamic Time Warping-based method; subsequently, chord sequences are estimated from the re-aligned \textsc{midi} files using an algorithm based on template matching. Tab files are first parsed, resulting in untimed chord sequences and then aligned to the audio, using an existing algorithm based on a Hidden Markov Model. In a final step, \textsc{decibel} uses a data fusion method that integrates all estimated chord sequences into one final output sequence. 

The main contributions of this study are twofold. First, by aligning different symbolic formats to audio, \textsc{decibel} automatically creates a heterogeneous harmonic representation that enables large-scale cross-version analysis of popular music. Second, our results show that \textsc{decibel}'s data fusion method improves each of the ten evaluated state-of-the-art audio \textsc{ace} methods in terms of estimation accuracy. Exploiting the musical knowledge that is implicitly incorporated in \textsc{midi} and tab files, it breaks the observed glass ceiling,  without requiring a lot of additional training, thereby prohibiting further overfitting to the existing chord annotations. 


\chapter{Introduction}\label{ch:introduction}

Automatic Chord Estimation (\textsc{ace})\footnote{\textsc{Ace} is also referred to as \textit{Chord Recognition}, \textit{Automatic Chord Transcription} or \textit{Automatic Chord Detection}.} is a fundamental problem in Music Information Retrieval. The \textsc{ace} task is concerned with estimating chords in audio recordings or symbolic representations of music. Basically, \textsc{ace} segments a song in such a way that the segment boundaries represent chord changes and each segment has a chord label. This is typically represented by a sequence of $\langle$start time, end time, chord label$\rangle$ 3-tuples. 

Figure~\ref{fig:example-chord-annotation} shows an example of such a chord sequence. We see that the first 2.6 seconds of this song do not contain any chords, hence the no-chord symbol \texttt{N}. This no-chord segment is followed by a 8.8 second segment in which an E major chord sounds, followed by an A major chord, etcetera. 

\begin{figure}[ht]
	\centering
    \begin{verbatim}
0.000000 2.612267 N
2.612267 11.459070 E
11.459070 12.921927 A
12.921927 17.443474 E
17.443474 20.410362 B
20.410362 21.908049 E
21.908049 23.370907 E:7/3
23.370907 24.856984 A
24.856984 26.343061 A:min/b3
26.343061 27.840748 E
27.840748 29.350045 B\end{verbatim}\caption{Chord sequence annotation for the beginning of the Beatles song \textit{I Saw Her Standing There}. This is a sequence of $\langle$start time, end time, chord label$\rangle$ 3-tuples.}\label{fig:example-chord-annotation}
\end{figure}

A chord can be defined as multiple notes that sound simultaneously. As we will see in Section~\ref{sec:chord-notation}, any chord can efficiently be represented by a text string (for example the chord labels in Figure~\ref{fig:example-chord-annotation}: \texttt{E}, \texttt{A}, \texttt{B}, \texttt{E:7/3} and \texttt{A:min/b3}). The progression of chords through time defines the harmonic structure of a piece of music. Therefore, the estimation of chords in a music piece has many applications in both music performance and in Music Information Retrieval (\textsc{mir}), as described below.

For music students, it is common to play along with songs on on-line streaming services like YouTube~\citep{stowell2011mir}. \textsc{Ace} can make this easier for the less experienced musician. Chordify\footnote{\url{https://chordify.net/}} is a web-service that uses \textsc{ace} techniques to display the chords of any audio file to the user. From the fact that Chordify is used by 1.5 million users every month~\citep{bountouridis2016data}, we can conclude that there is great interest in systems using \textsc{ace}.

Chord sequences have also been used by the MIR research community in high-level tasks such as cover song identification \citep{khadkevich2013large}, key detection \citep{papadopoulos2012modeling}, genre classification \citep{ajoodha2015single}, lyric interpretation \citep{kolchinsky2017the} and audio-to-lyric alignment \citep{mauch2012integrating}. To conclude, a well-performing \textsc{ace} system, which estimates reliable chord sequences, would be of great use for both music performers and \textsc{mir} researchers. 

\section{Stagnation and subjectivity}\label{sec:stagnation-subjectivity}
\cite{fujishima1999realtime} was the first who considered \textsc{ace} as a problem on its own and since this pioneering publication, many researchers have worked on the task. Despite these efforts in the past two decades, \textsc{ace} is not yet a solved problem. 

\textsc{Ace} has been a task in the annual benchmarking evaluation Music Information Retrieval Evaluation eXchange (\textsc{mirex}) since 2008. The main evaluation measure is Weighted Chord Symbol Recall (\textsc{wcsr}), which reflects the proportion of correctly labeled chords in a data set, weighted by the total length of the data set. State-of-the-art \textsc{ace} methods yield \textsc{wcsr}s of around 80\%, given a chord vocabulary of major and minor chords\footnote{\url{http://www.music-ir.org/mirex}}. However, from recent \textsc{mirex} results, \citet{humphrey2015four} and \citet{scholz2016cross} observe a stagnation in \textsc{ace} performance. \citet{scholz2016cross} give two suggestions to recover from this stagnation: \textsc{ace} methods should either make better use of musical expert knowledge, or they should use new techniques, for example deep learning, instead of relying on the small set of commonly used techniques.

Besides, \citet{humphrey2015four} throw light on another issue of \textsc{ace}: chord annotations are inherently subjective. Even between musically trained human annotators, there can be a discrepancy of over 15\% in the chord annotation of a song. 
This subjectivity matter was earlier identified by \cite{ni2013understanding}, who asked 5 musicians to each annotate the same 20 songs and show a 10\% discrepancy between the annotations. 
More recently, \cite{koops2019annotator} introduced a 50-song data set of popular music, annotated by 4 professional musicians, and found only 73\% overlap on average for the traditional major-minor vocabulary. 
The currently common practice to evaluate \textsc{ace} by comparing the results to a single reference annotation is disputed by \cite{humphrey2015four, ni2013understanding} and \cite{koops2019annotator}. 
\cite{ni2013understanding} and \cite{koops2019annotator} even claim that modern \textsc{ace} systems have started to overfit the \textsc{mirex} data set, mimicking the subjective aspects of \textsc{mirex}'s reference annotations. 
Although none of these papers presents a practical, scalable solution to the subjectivity issue, it becomes clear that subjectivity in \textsc{ace} is an important issue that is often overlooked in existing research.

In conclusion, there is need for a new strategy that overcomes existing stagnation in \textsc{ace} without further overfitting to existing (subjective) data sets. In this study, we propose a novel method that uses \textsc{midi} and tab files, which implicitly contain crowd-sourced musical knowledge, to improve existing state-of-the-art methods, requiring a minimal amount of additional training.

\section{Research goal}
The aim of this research is to show that audio \textsc{ace} can be improved by exploiting symbolic representations of popular music. For this purpose, we designed and implemented \textsc{decibel} (DEtection of Chords Improved By Exploiting Linking symbolic formats). \textsc{Decibel} is a novel system that exploits multiple heterogeneous symbolic music representations for improving \textsc{ace}. \textsc{Midi} and tab files can be considered as crowd-sourced note and chord transcriptions respectively. By using these symbolic representations, \textsc{decibel} implicitly integrates musical knowledge into existing \textsc{ace} methods, which may be a strategy to overcome the stagnation issue identified in the previous section. As \textsc{decibel} only relies to a small extent on training on reference annotations, our method prohibits further overfitting to existing data sets. 
To evaluate \textsc{decibel}, we compare its performance to state-of-the art \textsc{ace} systems submitted in the \textsc{mirex} competitions of 2017 and 2018, as well as a commercial \textsc{ace} method.
\textsc{Decibel} improves each of the tested \textsc{ace} systems.

\section{Related work}\label{sec:introduction-related-work}
Our work builds on previous approaches to integrate symbolic music and audio in the chord estimation task, as we will see in this section.

\cite{ewert2012towards} introduce a cross-version analysis framework for comparing harmonic analysis results from different musical domains. After collecting a \textsc{midi} file for each audio file in a 112-song subset of the Isophonics \citep{mauch2009omras2, harte2010towards} data set, they use two state-of-the-art chord recognition methods for \textsc{midi} data and align each \textsc{midi} file to the corresponding audio recording. This way, they create a harmonic representation for each of the 112 songs, which contains three chord label sequences: the ground truth labels and the re-aligned outputs that were obtained by the two \textsc{midi} chord recognition systems. They show that this harmonic representation can be used for quantitative evaluation of \textsc{midi} chord recognition methods, using annotations for corresponding audio recordings. In addition, by visualizing this harmonic representation, they demonstrate how it can be used for qualitative error analysis of automatically generated chord labels, and by that contributes to the understanding of an \textsc{ace} algorithm's behavior and the properties of the underlying music material. The research by \citeauthor{ewert2012towards} lays the foundation for the work proposed in this study, in which we expand the harmonic representation with chord labels from multiple \textsc{midi} and tab files for each audio recording and show how this enriched harmonic representation can be used to improve \textsc{ace}.

The integration of tab files and audio with respect to chord estimation was earlier researched by \cite{mcvicar2010enhancing} and \cite{mcvicar2011leveraging, mcvicar2011using}. In these three papers, they show that a HMM-based system for audio \textsc{ace} can be significantly improved by incorporation of external information from guitar tabs. In a preprocessing step, they obtain a set of tab files by a web scrape and consequently parse them. In the resulting format, to which they refer as Untimed Chord Sequences (\textsc{ucs}s), only the chord labels and line information of each tab is retained. As a next step, they align each \textsc{ucs} to the corresponding audio file. For this purpose, the authors introduce four variations on the traditional Viterbi algorithm. The most promising variation is Jump Alignment, which aligns the \textsc{ucs} to the audio file, thereby allowing jumps from the end of any annotation line to the beginning of any line. We implemented Jump Alignment as part of the \textsc{decibel} system and will explain the implementation more thoroughly in Section~\ref{sec:jump-alignment}.

The integration of heterogeneous output of multiple \textsc{ace} algorithms was proposed by \cite{koops2016integration}. The authors experiment with three different techniques to combine chord sequence estimates from different sources 
(the \textsc{mirex} 2013 \textsc{ace} submissions, applied to the Billboard data set) 
into one final output sequence for each song. They show that their data fusion method yields the best results in terms of \textsc{wcsr}. Also, they show that the output sequence found by their data fusion method is an improvement to the best scoring team.
In our study, we use a similar data fusion method to combine the chord labels obtained from audio, \textsc{midi} and tab files.

\section{Outline}
The remainder of this study is structured as follows: Chapter~\ref{ch:musical-background} provides the required music theory and terminology information in order to understand the concepts that will be used in the remainder of this study. In Chapter~\ref{ch:framework}, we describe \textsc{decibel}'s framework. As we will see in this chapter, \textsc{decibel} consists of four subsystems: three of these subsystems compute chord labels in their own representation-specific way and the fourth subsystem combines these results in a data fusion step. The audio, \textsc{midi} and tab subsystems are described in Chapters \ref{ch:subsystem-audio}, \ref{ch:subsystem-midi} and \ref{ch:subsystem-tabs} respectively. \textsc{Decibel}'s data fusion strategy is described in Chapter~\ref{ch:data-fusion}. Finally, we will present our conclusions in Chapter~\ref{ch:conclusion}.

\chapter{Musical background}\label{ch:musical-background}
This chapter gives an introduction to pitch, music notation, intervals and chords. The material in Section~\ref{sec:introductionpitch} to \ref{sec:chords} is based on \cite{taylor1989ab} and \cite[Chapter~1 and 5]{muller2015fundamentals}. Section~\ref{sec:musicrepresentations} describes the three music representations we consider in this research: audio, \textsc{midi} and tabs.

\section{An introduction to pitch}\label{sec:introductionpitch}
Music consists of tones, and each tone has some properties, for example its duration, start time and \textbf{pitch}. Pitch refers to the degree of highness of sound. If you hit a key on the left side of the piano and subsequently strike a key on the right, the second produced tone will be higher pitched than the first one. Similarly, men's voices are generally lower pitched than women's voices and a tuba sounds lower than a piccolo. 

\subsection{Pitch in physics: waves and frequencies}\label{sec:pitchphysics}
From a physical point of view, sound is generated by vibrating objects, for example the string and soundboard of a violin or the vocal cords of a singer.  These vibrations cause displacements and oscillations of air molecules, resulting in local regions of compression and rarefaction. This alternating pressure travels through the air as a \textbf{longitudinal wave}, from its source to a perceiver. 

The change in air pressure at a certain location can be graphically represented by a  \textbf{waveform} of the sound. A waveform plots the deviation of the air pressure from the average air pressure over time.

If the points of high and low air pressure repeat in an alternating and regular fashion, the resulting waveform is called periodic. The \textbf{period} is defined as the amount of time that is required for completing a cycle. The \textbf{frequency} is the reciprocal of period, and is measured in Hertz (Hz). For example, a period of 2.5 ms corresponds with a frequency of $\frac{1}{0.0025\text{ s}} = 400\text{ Hz}$. The higher the frequency of a sinusoidal wave, the higher the corresponding tone sounds.

However, real-world sounds, such as the tones that are produced by a musical instrument, are much more complex than a pure tone resulting from a single sinusoid. A musical tone can be described as a superposition of sinusoids, each with their own frequency. A \textbf{partial} is any of the sinusoids by which a musical tone is described. The frequency of the lowest partial present is called the \textbf{fundamental frequency} of the sound. The pitch of a musical tone is usually determined by the fundamental frequency.

Two tones with fundamental frequencies in a ratio equal to any power of two, are perceived as similar. All tones with this kind of relation can be grouped in the same \textbf{pitch class}. The distance between one musical tone and another tone with half or double its fundamental frequency is called an \textbf{octave}. For example, a tone with a frequency of 440 Hz sounds similar to a tone with a frequency of 220 Hz. These tones are an octave apart and belong to the same pitch class.

\subsection{Twelve-tone equal-temperament}\label{sec:temperament}
We have seen that pitch and frequency are closely related, but musicians typically do not specify the height of a tone in Hertz. Instead, the space of all different pitches is discretized using a \textbf{tuning system}. Although many different tuning systems have been suggested and used in history, the standard tuning in modern Western music, which is the tuning used in this research, is \textbf{twelve-tone equal-tempered tuning}. This tuning system is the standard system used as basis for tuning the piano.

In the twelve-tone equal-tempered tuning system, an octave is divided into twelve scale steps - often simply referred to as tones. The fundamental frequencies of these steps are equally spaced on a logarithmic frequency axis, as the human perception of pitch is logarithmic in nature. This means that the frequency ratio of two subsequent scale steps is constant and equals $2^{\frac{1}{12}} \approx 1.059463$. The distance, or \textbf{interval}, between two subsequent scale steps is called a \textbf{semitone}. In other words, if we multiply the frequency of an arbitrary pitch by $2^{\frac{1}{12}}$, this pitch is raised by a semitone.

These semitones can be further divided into \textbf{cents}. By definition, each octave is divided into 1200 cents, so a semitone consists of 100 cents. The frequency ratio of two subsequent cents equals $2^{\frac{1}{1200}} \approx 1.0005777895$. Note that both the cent and the semitone are logarithmic measures of distance between pitches.

\subsection{Note names and octave numbers}
\begin{wrapfigure}{R}{0.5\textwidth}
 \centering
 \includegraphics[width=0.5\textwidth]{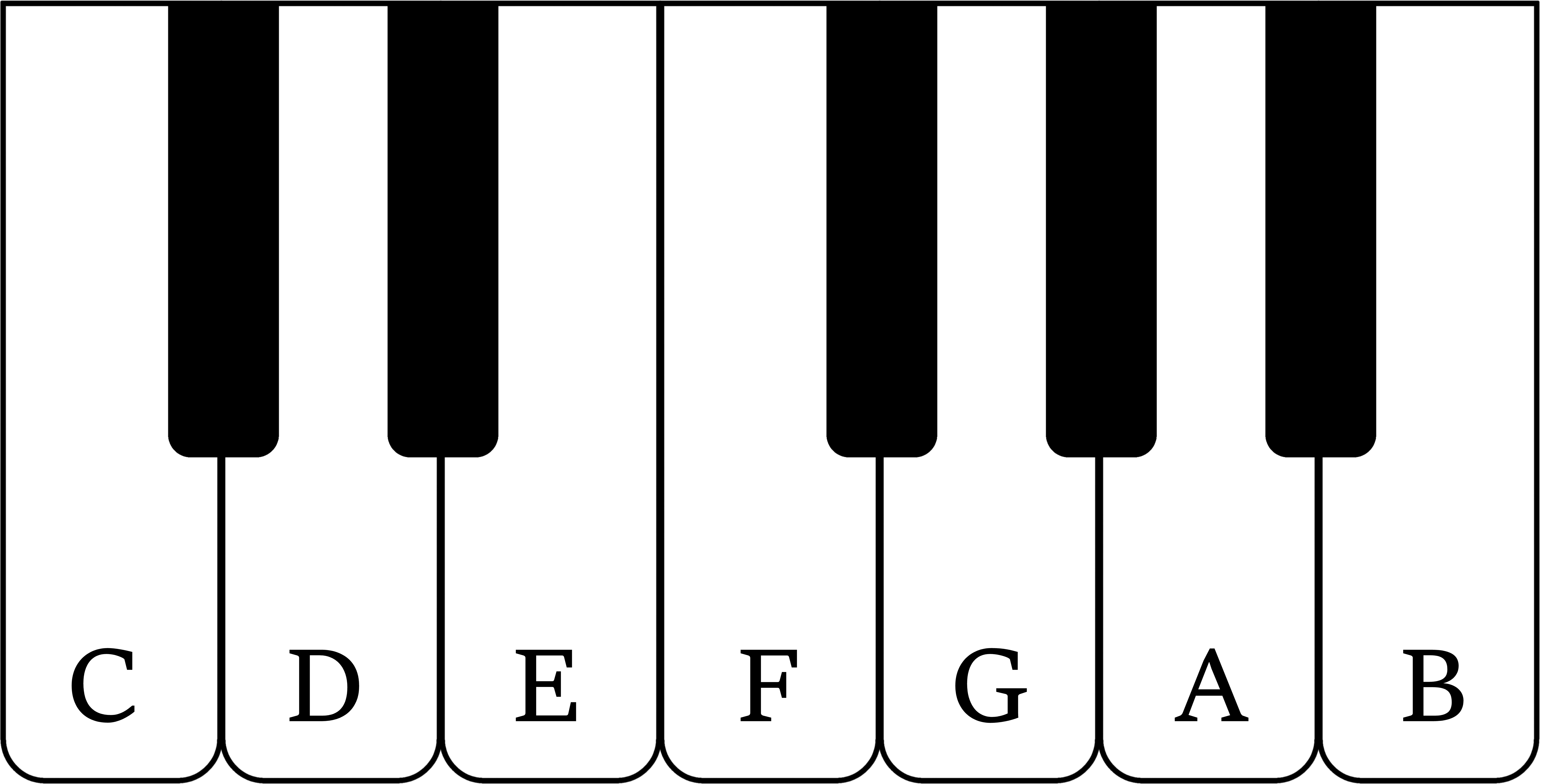}
 \caption{An octave on the piano. Pitch classes corresponding to the white keys of the piano can be named by a single letter. The remaining five pitch classes are named by a combination of a letter and an accidental.}
 \label{fig:piano}
\end{wrapfigure}
In Section~\ref{sec:pitchphysics}, we have seen that two tones that differ in pitch by one or more octaves can be grouped in the same pitch class and in Section~\ref{sec:temperament} we learned that the twelve-tone equal-tempered tuning system divides the octave into twelve steps of one semitone. We can therefore deduce that there are twelve pitch classes. 

In modern Western music notation, each pitch class has a \textbf{pitch class name} or note name, which consists of a letter and possibly an accidental. There are seven pitch classes which can be named by just one letter from \{A, B, C, D, E, F, G\}. These pitch classes correspond to the white keys of the piano, as illustrated in Figure~\ref{fig:piano}. The remaining five pitch classes are named by a combination of a letter and an \textbf{accidental}. An accidental raises or lowers the corresponding note. The most common accidentals are the sharp (\musSharp) and flat (\musFlat), which respectively raise and lower the note with a semitone. So the black key on the piano positioned between the C- and D-keys produces a note that can be denoted by either C\musSharp~or D\musFlat. Similarly, the other ``black-key'' pitch classes can be referred to by either \{D\musSharp, F\musSharp, G\musSharp~and A\musSharp\} or \{E\musFlat, G\musFlat, A\musFlat~and B\musFlat\}. We see that two different names can refer to the same pitch class. This phenomenon is called \textbf{enharmonic equivalence}.

Following \textbf{Scientific Pitch Notation}, a pitch is not only specified by the pitch class name, but also by an \textbf{octave number}. The higher the octave number, the higher the pitch. For instance, an A1 sounds an octave higher than an A0, which is the lowest tone that can be produced by most pianos. The note A4 has a frequency of 440 Hz in modern twelve-tone equal-tempered tuning and is used as a reference note for tuning.

Note that the distance between notes denoted by two subsequent letters is not always the same. For example, D4 and E4 differ by two semitones: in Figure~\ref{fig:piano} we see that there is a distance of two piano keys, so two semitones, between D and E. On the other hand, E4 and F4 are just one semitone apart: there is no black key on the piano between these notes.

\section{Music notation}

\begin{figure}
 \centering
 \begin{tikzpicture}
 \draw (0,0) -- (15,0);
 \draw (0,0.3) -- (15,0.3);
 \draw (0,0.6) -- (15,0.6);
 \draw (0,0.9) -- (15,0.9);
 \draw (0,1.2) -- (15,1.2);
 \footnotesize
 \node[] at (-1, 0) {Line 1};
 \node[] at (-1, 0.3) {Line 2};
 \node[] at (-1, 0.6) {Line 3};
 \node[] at (-1, 0.9) {Line 4};
 \node[] at (-1, 1.2) {Line 5};
 
 \node[] at (7, 0.15) {Space 1};
 \node[] at (7, 0.45) {Space 2};
 \node[] at (7, 0.75) {Space 3};
 \node[] at (7, 1.05) {Space 4};
 \end{tikzpicture}
 \caption{The staff, with its lines and spaces}
 \label{fig:staff}
\end{figure}
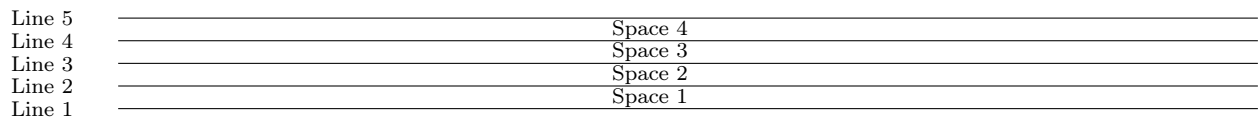

\begin{figure}
 \centering
 \includegraphics[width=\textwidth]{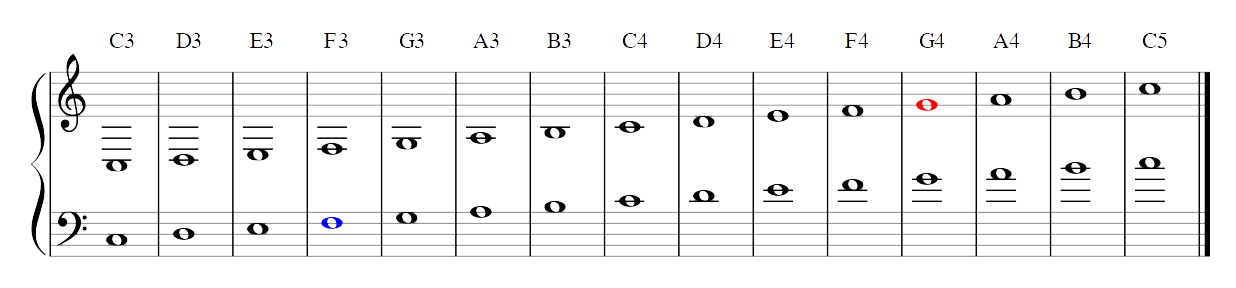}
 \caption{Notes on the staff, with their names described above. Here we see that when using different clefs, the same note is notated at a different (vertical) position on the staff.}
 \label{fig:notestaff}
\end{figure}

Music is notated on a \textbf{staff}, which consists of five parallel horizontal lines, counted from the bottom, see Figure~\ref{fig:staff}. Notes can be notated on the lines or between them. The pitch of a note is determined by its vertical position on the staff, combined with the \textbf{clef}. The two most common clefs are the treble clef or `G' clef (\includegraphics[scale=0.2]{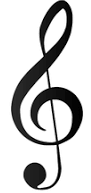}) and the bass clef or `F' clef (\includegraphics[scale=0.05]{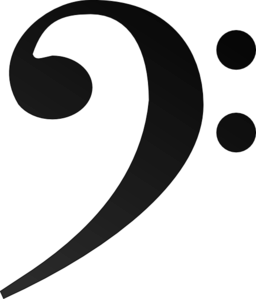}). The working of these clefs is illustrated in Figure~\ref{fig:notestaff}: if there is a treble clef at the beginning of the staff, this means that the note on the second line is G4. By way of illustration, this note is colored red. When using the bass clef, the note that is placed on the fourth line is F3 - colored blue.

If notes lie above or below the limits of the staff, short additional lines called \textbf{ledger lines} are used. We see for example that we need one ledger line for the notation of C4 in the treble clef as well as in the bass clef. As ledger lines deteriorate the readability of sheet music, it is common to use the treble clef for high notes (C4 and higher) and use the bass clef for lower notes. 

\section{Intervals}
Now that we have become acquainted with tones, pitches and their notation, it is time to introduce the notion of intervals. An \textbf{interval} is the distance between two pitches. We already know that, assuming twelve-tone equal-tempered tuning, we can divide the octave in twelve semitones. Based on the notion of semitone, we can specify other intervals that are used in Western music theory. 

Interval names consist of a \textbf{number} and a \textbf{quality}.  In counting the number, both notes are included. For example when determining the interval from C4 to F4, we count four notes (C4, D4, E4, and F4), so this interval is called a fourth. Similarly, the interval from E4 to C5 is a sixth. The interval from one note to another note with exactly the same pitch is a \textbf{unison}.

\begin{figure}
 \centering
 \includegraphics[width=\textwidth]{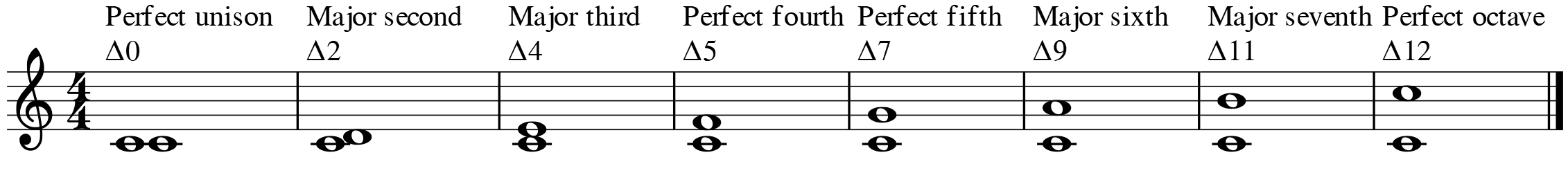}
 \caption{The eight intervals of the major scale. Note that the distance between the third (E) and fourth (F) note and between the seventh (B) and eighth (C) note is only one semitone.}
 \label{fig:intervalsbasic}
\end{figure}

Traditionally, intervals are named on the basis of the \textbf{major scale}, which consists of seven notes and an eighth note one octave apart from the first note. In the major scale, there is only one semitone difference between note 3 and 4 and between note 7 and 8, while there are two semitones difference between all other subsequent notes. An example of a major scale is C major: C - D - E - F - G - A - B - C. In Figure~\ref{fig:intervalsbasic} we see the full names of all intervals between C, which is the root note of the scale, and each of the other notes. In each major scale, the intervals unison, fourth, fifth and octave get the quality ``perfect'', while the intervals second, third, sixth and seventh get the quality ``major''.

Using only perfect and major intervals, we cannot express all different intervals. For instance, we have no name yet for an interval of three semitones. We can solve this by raising or lowering the upper note of the interval with a semitone, following these rules:
\begin{itemize}
\item Given a perfect or major interval: if either the upper note is raised a semitone or the lower note is lowered a semitone, the interval becomes \textbf{augmented};
\item Given a major interval: if either the upper note is lowered a semitone or the lower note is raised a semitone, the interval becomes \textbf{minor};
\item Given a perfect or minor interval: if either the upper note is lowered a semitone or the lower note is raised a semitone, the interval becomes \textbf{diminished}.
\end{itemize}

\begin{figure}
 \centering
 \includegraphics[width=\textwidth]{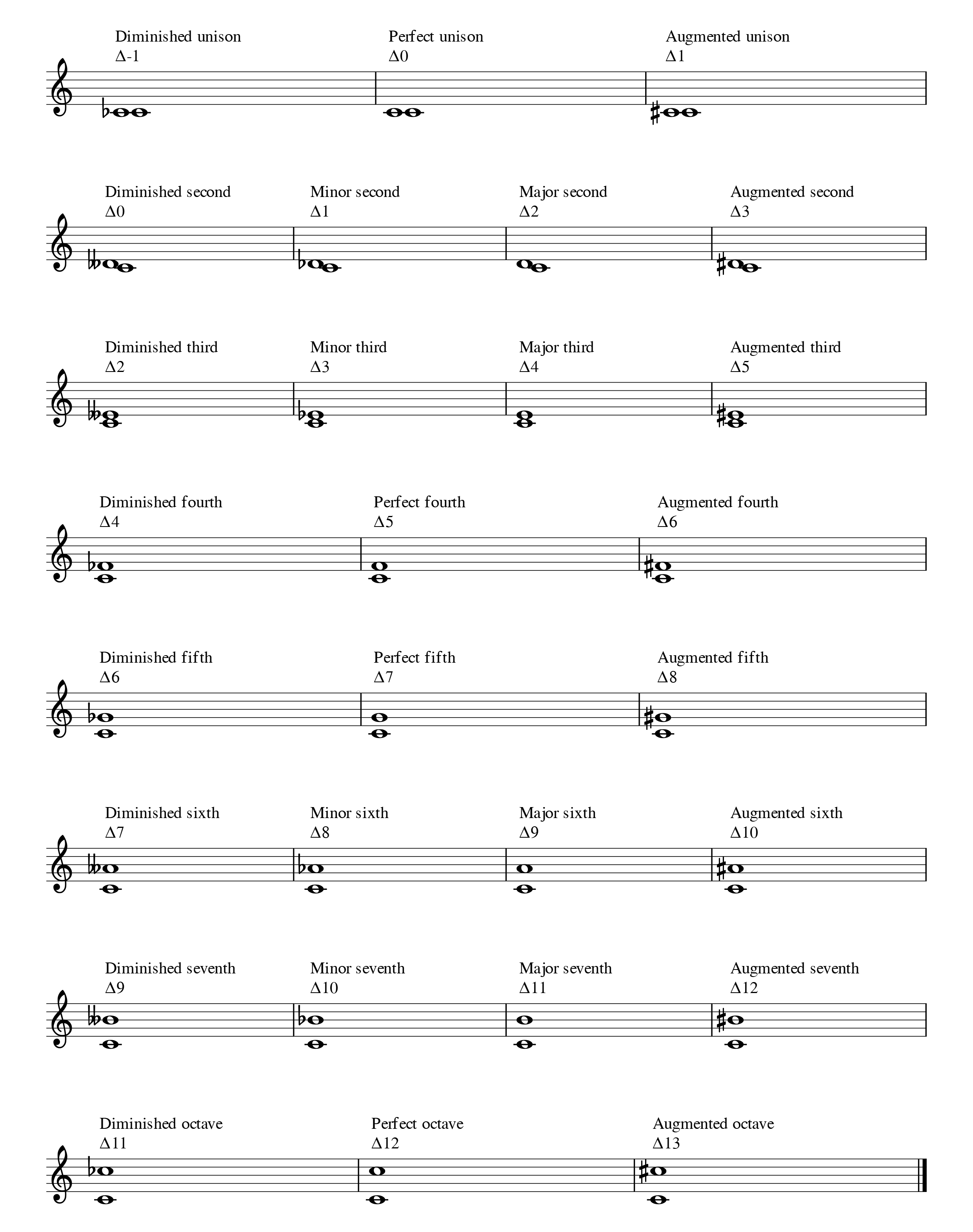}
 \caption{Major, minor, augmented and diminished intervals, with their distance in semitones}
 \label{fig:intervalsadvanced}
\end{figure}

Figure~\ref{fig:intervalsadvanced} shows all possible major, minor, augmented and diminished intervals in (or just beyond) the octave, with their distance in semitones. Note that two different interval names can refer to the same difference in semitones, e.g. both the augmented fourth and the diminished fifth consist of six semitones. These intervals are enharmonically equivalent.

\section{Chords}\label{sec:chords}
A \textbf{chord} can be loosely defined as a group of tones sounding at the same time. Most researchers agree that a chord should consist of tones from at least three distinct pitch classes.

\begin{figure}
 \centering
 \includegraphics[width=0.9\textwidth]{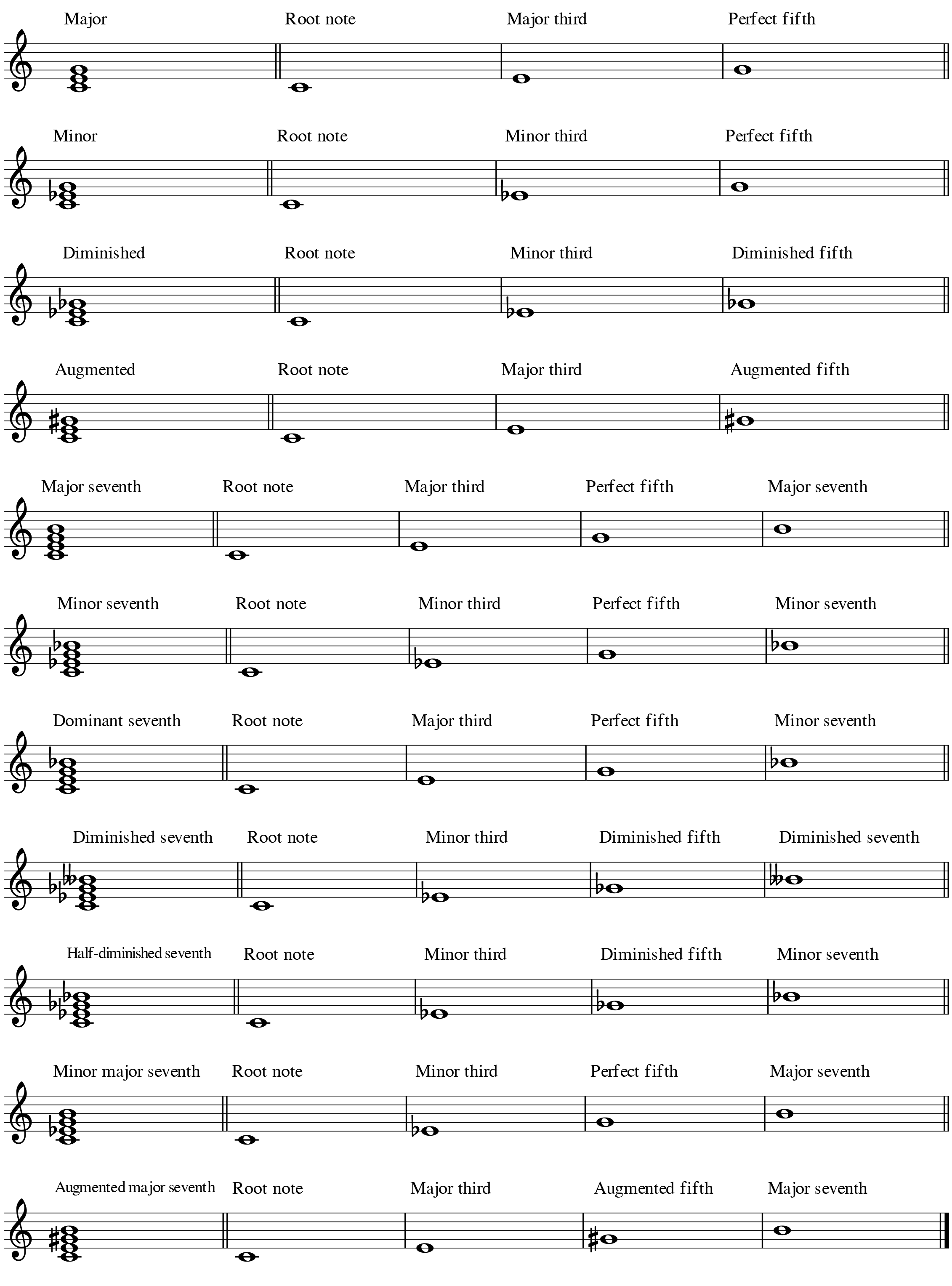}
 \caption{The most common chords based on triads and tetrads}
 \label{fig:chords}
\end{figure}

Chords that consist of tones from three pitch classes are called \textbf{triads}. In Western music, most triads can be stacked in thirds and consist of the note on which the triad is based (\textbf{root}), plus the third and the fifth above it. The root note determines the name, while the quality of the other intervals determines the type of chord. For example: a C with a major third and a perfect fifth forms a C major chord, and a D with a minor third and a perfect fifth is a D minor chord. It is possible to extend chords by stacking more thirds upon them. \textbf{Seventh} chords consist of a ``normal'' triad and an added seventh, forming a tetrad. The most common seventh chord is the dominant seventh, made out of a major tetrad and a minor seventh. Figure~\ref{fig:chords} summarizes the most used triad and tetrad chords in Western music.

For one chord, there are many variations possible. First, there exist multiple \textbf{inversions} for each interval. When the chord's lowest note is its root (like in the examples of Figure~\ref{fig:chords}), the chord is said to be in \textbf{root position}. When the lowest note is the third, for example the E in a C major chord, this chord is in \textbf{first inversion}. When the lowest note is the fifth, the chord is in \textbf{second inversion}. A seventh chord can even be in \textbf{third inversion} if the seventh is the lowest note. 
Second, notes of the same pitch class may be doubled, for example in a C major chord consisting of C3, E3, F3 and C5. This is called \textbf{octave doubling}.
Third, the notes of a chord may not be played simultaneously, but after each other. In this case, we speak of a \textbf{broken chord}.

The concept of harmony and chords is enriched by the existence of \textbf{non-harmonic tones}. These tones are not part of the chord. Instead, they are used to create a smooth melody line, to  prepare the transition to another chord or to add a dissonant element, creating musical tension. Though non-harmonic tones certainly contribute to the beauty of music, they provide a challenge for \textsc{ace} systems as it is difficult to automatically determine whether a given note is either harmonic of non-harmonic.

\subsection{Chord notation and representations}\label{sec:chord-notation}
\begin{figure}[ht]
 \centering
 \includegraphics[width=\textwidth]{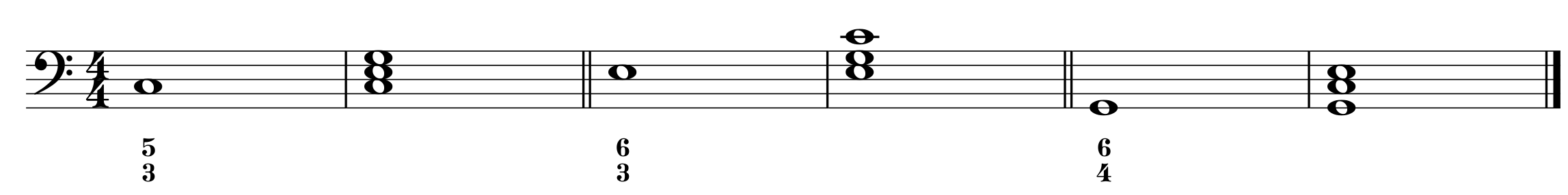}
 \caption{Three inversions of the C chord in figured bass notation}
 \label{fig:figuredbass}
\end{figure}

There exist various notations for chords, which can differ between and even within genres. \citep{taylor1989ab} In baroque music for example, the chord notation consists of a bass line with figures written underneath the notes. This notation is called \textbf{figured bass} and is illustrated in Figure~\ref{fig:figuredbass}. The figures represent the intervals that should be played above the bass note. For example: the first bar is the figured bass notation for a C major chord in root notation, and can be played like the second bar in Figure~\ref{fig:figuredbass}. Similarly, the figured-bass notation in the third bar means that we should add the third and the sixth to the bass note. The bass note is an E, so we add a G and C to the bass note, resulting in a  C major chord in first inversion - as written out in the fourth bar. The fifth and sixth bar represent a second-inversion C major chord in figured bass and full notation respectively.

\begin{figure}[ht]
 \centering
 \includegraphics[width=\textwidth]{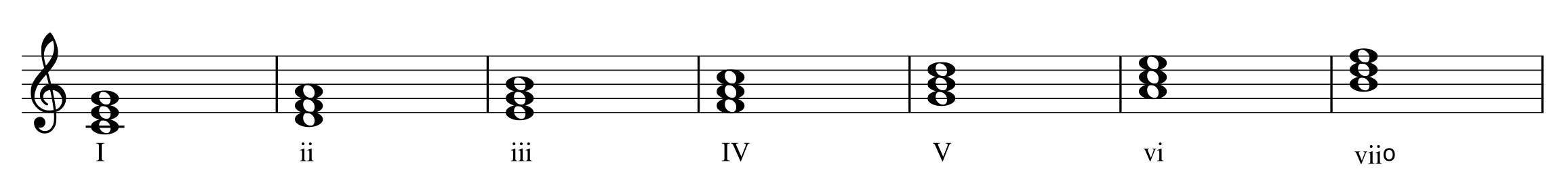}
 \caption{Seven chords in the scale of C major in Roman numeral style notation notation}
 \label{fig:romannumeralanalysis}
\end{figure}

\begin{figure}[ht]
 \centering
 \begin{subfigure}[b]{0.25\textwidth}
 	\includegraphics[height=1in]{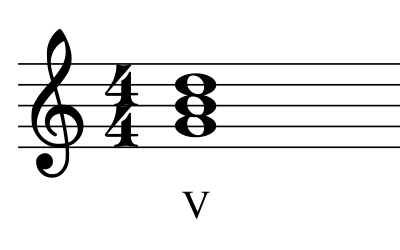}
    \caption{C major key}
 \end{subfigure}
 \begin{subfigure}[b]{0.25\textwidth}
 	\includegraphics[height=1in]{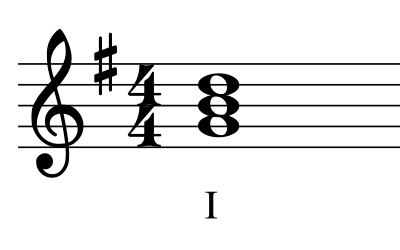}
    \caption{G major key}
 \end{subfigure}
 \caption{Roman numerals are dependent on the current key}
 \label{fig:romannumeralkey}
\end{figure}

In Classical Harmony Analysis, chords are studied and notated in relation to the current key. This notation is called \textbf{Roman numeral notation} and is, unsurprisingly, characterized by Roman numerals under the chords. These Roman numerals indicate the scale degree on which the chord is built, as illustrated in Figure~\ref{fig:romannumeralanalysis}. In this figure, we see seven chords in the key of C major. The first chord is a C major chord. The interval between the key C and the root note of the chord (C) is a unison, so we need the Roman equivalent of the digit ``1''. As a rule, major chords are capitalized, so the correct Roman numeral for this chord in this key is ``I''. The second chord is a D minor chord. The key (C) and the root note of the chord (D) are a second apart and the chord is in minor, so the corresponding numeral is ``ii''. The final bar in our example piece is a diminished B chord. Diminished chords are indicated with a small ``0'' after the Roman numeral. Similarly, augmented chords (not present in this example) are indicated with a small ``+'' after the Roman numeral. Note that the Roman numeral that a chord gets, is dependent on the key in which it occurs. In Figure~\ref{fig:romannumeralkey}, we see for example that a G major chord gets the Roman numeral ``V'' in the key of C major, while ``I'' is the right numeral in the key of G major.

\begin{figure}[ht]
 \centering
 \includegraphics[width=\textwidth]{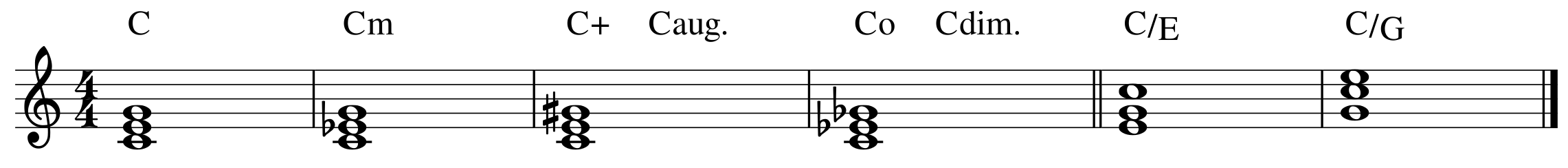}
 \caption{Major, minor, augmented and diminished C chords in root position; Major C chords in first and second inversion}
 \label{fig:popjazzchords}
\end{figure}

Yet another chord representation is used in \textbf{jazz and popular music}. These genres are characterized by improvisation in performance. The basic harmonies are generally quite simple, so the main purpose of the chord representation is to be easy to read and interpret. As we will see in Section~\ref{sec:tabs}, this chord representation is common in tabs (an alternative for regular sheet music). Figure~\ref{fig:popjazzchords} shows some variations of the C chord and their chord notation in jazz and other forms of popular music. The letter name of the root of the chord is shown as a capital letter. A chord is assumed to be major unless stated otherwise. Minor chords get a ``m'' after the chord letter; ``+'' or ``aug.'' stands for an augmented chord and a chord with a ``0'' or ``dim.'' is diminished. It is unusual to specify the bass note, so a chord without bass indication can be in any inversion. In the rare cases where the bass note is specified, it is notated with a slash after the letter name of the root of the chord, like in the last two bars of our example. It is important to notice that there exist multiple variations of chord notation in jazz and pop music, and some of them cause ambiguity issues. For example: some musicians use the symbol $\Delta$ for major chords, so C$\Delta$ is a C major chord and C$\Delta^7$ is a C major seventh chord. Others use the $\Delta$ symbol as a synonym for major seventh chords, so they would notate the C major seventh chord as C$\Delta$. To interpret chords like C$\Delta$ correctly, one should therefore be well aware which variation of notation is applied.

Each of the three chord notations is suitable for the genre or field of study in which they are used. However, none of them is well-suited for the chord annotations which we need in order to train and test \textsc{ace} systems: they are either hard to write in flat text, key-dependent or ambiguous. \textbf{\textsc{ace} chord annotations} require a unambiguous, context-independent notation that is easy to write and intuitive to interpret. \cite{harte2005symbolic} propose a chord grammar in which millions of chords can be defined unambiguously, whereas they give a succinct short-hand notation for the most common chords. Chords names are based on their root note, independent of the present key. In this notation, the C minor chord can either be represented by its components, like \texttt{C:(b3,5)}, or by a shorthand string: \texttt{C:min}. We can also compose more complex chords using the component notation: \texttt{A:(3,5,b7,9)} is a dominant ninth chord, consisting of A, C\#, E, G and B. The asterisk is used as ``omit symbol'', so \texttt{D:maj7(*3)} would be equivalent to \texttt{D:(5,7)}. The bass can be specified with a slash, followed by the interval from root to bass note. A D diminished seventh chord in third inversion would be represented as \texttt{D:(b3,b5,bb7)/bb7} or \texttt{D:dim7/bb7}. This chord grammar has become the standard notation for \textsc{ace} reference annotations, including the Isophonics annotations \citep{mauch2009omras2, harte2010towards} which we use in this project.

\section{Music representations}\label{sec:musicrepresentations}
Apart from sheet music, there exist multiple other music representations. Each of them has its own way of storing pitch or chord information. In this section, we investigate the three music representations that are used in \textsc{decibel}: audio, \textsc{midi} and tabs.

\subsection{Audio representation}\label{sec:musicrepresentations-audio}
Sound is generated by vibrations and travels through the air as a longitudinal wave, which can be graphically represented by a waveform (see Section~\ref{sec:pitchphysics}). This waveform plots the deviation of the air pressure from the average air pressure over time. A \textbf{digital} audio representation (e.g. a .wav or .mp3 file) is an approximation of this waveform, in which the sound wave of the audio signal is digitized by sampling and quantization. \cite[Chapter 2]{muller2015fundamentals}.

\textbf{Sampling} refers to the process of reducing a continuous-time (CT) signal to a discrete-time (DT) signal, which is defined only on a discrete subset of the time axis. The typical sampling rate for CD recordings is 44.1 kHz, so a CD has 44100 samples per second.

In \textbf{quantization}, the continuous range of possible amplitudes is replaced by a discrete range of possible values. For CD recordings, a 16-bit coding scheme is used, which allows representation of $2^{16} = 65536$ possible values.

Although the tones that are present at a given time cannot be read directly from the waveform, there exist methods to estimate them. Remember from Section~\ref{sec:pitchphysics} that a musical tone is basically a superposition of sinusoids, each with their own frequency. Using \textbf{Fourier analysis}, we can decompose our signal (the digitized waveform) into the sinusoids it consists of - and their frequencies. In the remainder of this section, we give a short summary of the discrete Fourier transform, short-time Fourier transform and Constant-Q transform. A detailed explanation of Fourier analysis is beyond the scope of this study. For more information about Fourier transforms, we refer the reader to \citet[Chapter 2]{muller2015fundamentals}.

The \textbf{Fourier transform} converts a signal that depends on time into a representation that depends on frequency. The formula for the discrete Fourier transform is given in Equation~\ref{eq:fourier}. $X_k$ is the $k$th Fourier coefficient, which is the amount of frequency $k$ that is present in a signal $x$. The signal $x$ consists of $N$ time samples.
\begin{equation}\label{eq:fourier}
X_k = \sum_{n=0}^{N-1} x(n) \exp\{-2\pi i k n / N\}
\end{equation}

\begin{figure}
 \centering
 \includegraphics[width=0.95\textwidth]{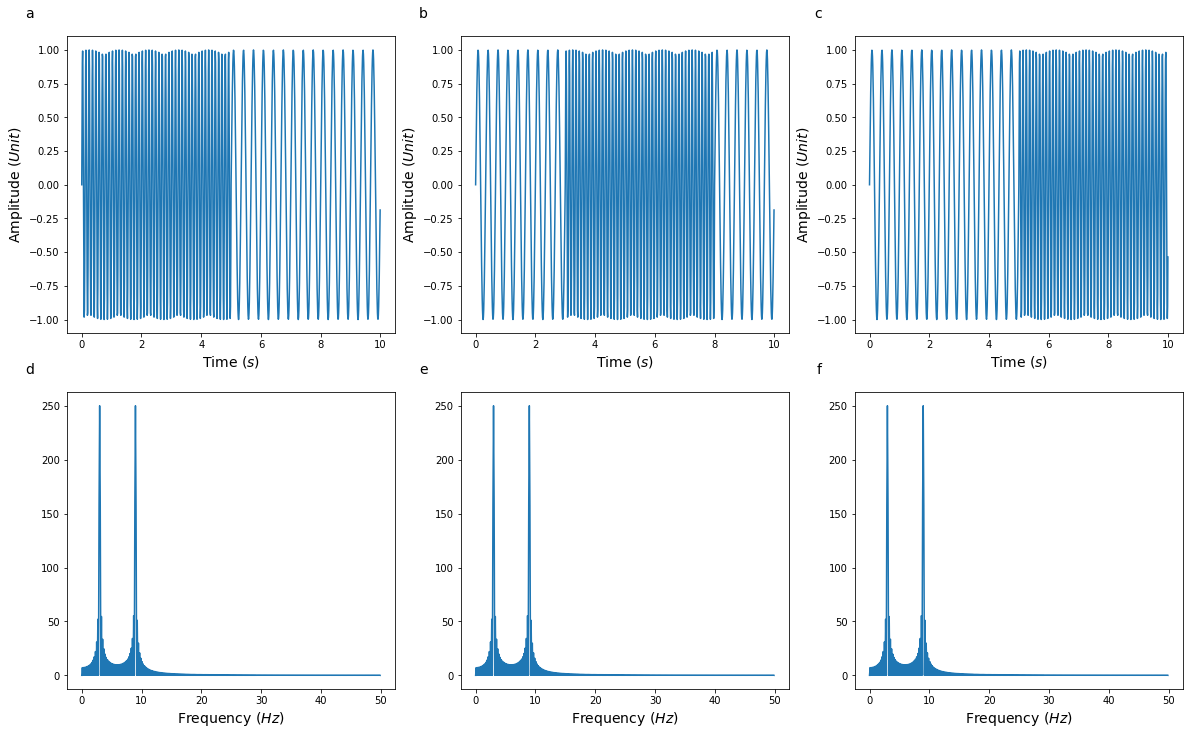}
 \caption{Three different signals (a, b, c) with the same frequency distribution yield identical Fourier transforms (d, e, f)}
 \label{fig:fourier}
\end{figure}

The magnitude of the Fourier transform tells us about the signal's frequency content: if $X_k$ is high for a frequency $k$, then this frequency $k$ is important in the signal. However, we cannot infer at which time the frequency content occurs. This is illustrated in Figure~\ref{fig:fourier}: each of the three signals at the top consist of a low note (3Hz) that sounds for 5 seconds and a higher note (9Hz) that sounds for the other 5 seconds. However, signal $a$ starts with the higher note, while in signal $b$ this note sounds in the middle and in signal $c$ the high note is placed at the end. Though the signals would sound different, the magnitude of the Fourier transform is exactly the same: we see peaks at 3 and 9 Hz.

\begin{figure}
 \centering
 \includegraphics[width=0.95\textwidth]{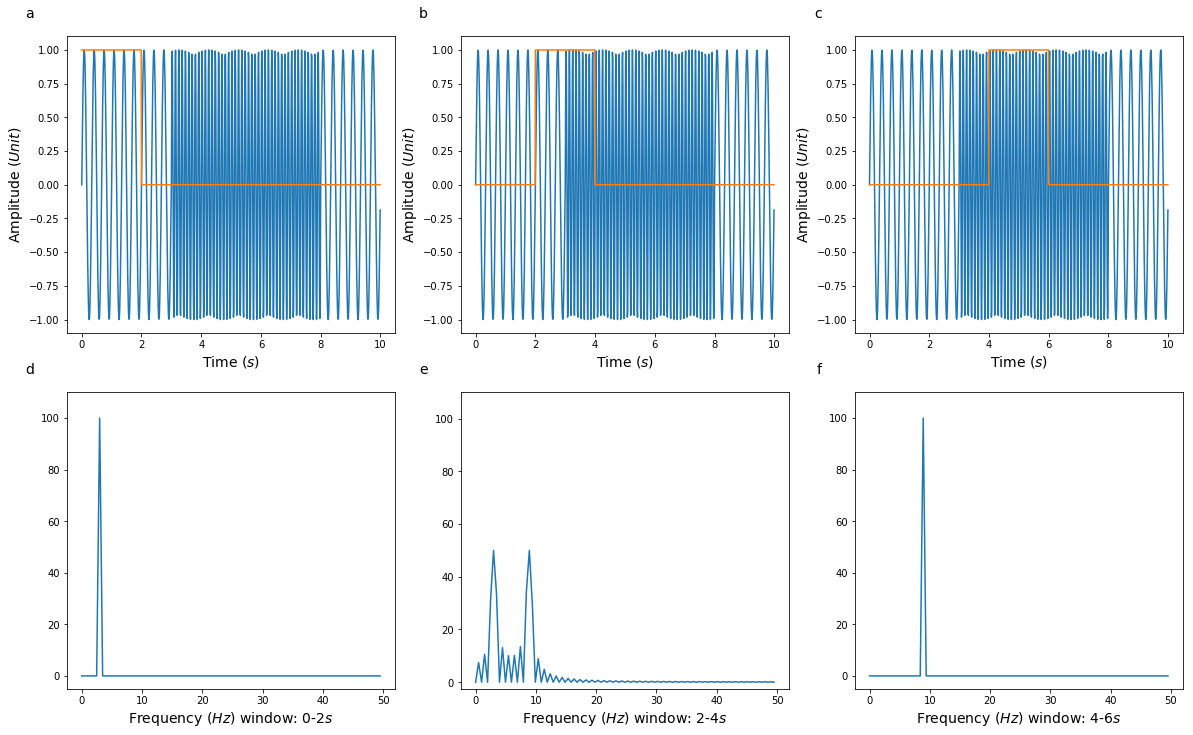}
 \caption{Short-time Fourier transform with shifting window: (\textbf{a, d}) Window centered at $t = 1s$ (\textbf{b, e}) Window centered at $t = 3s$ (\textbf{c, f}) Window centered at $t = 5s$}
 \label{fig:stft}
\end{figure}

The \textbf{short-time Fourier transform} (STFT) \citep{gabor1946theory} offers a solution for this problem. The STFT considers only a small part of the signal, which is dependent on the window function $w$. $w$ is a function with $w(n) \in \mathbb{R}$ if $n \in [\,0, N-1]\,$ and $w(n) = 0$ otherwise. $N - 1$ is the \textbf{window size}. The window is shifted every $H$ samples. $H \in \mathbb{N}$ is the \textbf{hop size}. In general, a smaller hop size gives more precise results, but is computationally more expensive than a large hop size. The choice for the perfect hop size value is therefore dependent on the application.
\begin{equation}\label{eq:stft}
X_{m,k} = \sum_{n=0}^{N-1} x(n+mH)w(n) \exp\{-2\pi i k n / N\}
\end{equation}

The formula for the STFT is given in Equation~\ref{eq:stft}. In Figure~\ref{fig:stft}, we see an example of the STFT with a rectangular window of 200 samples (2 seconds with a sampling rate of 100 Hz). The hop size is 200. These window and hop sizes are way to large for real-life applications, but are chosen in order to be visible in the figure.

\begin{figure}
 \centering
 \includegraphics[width=\textwidth]{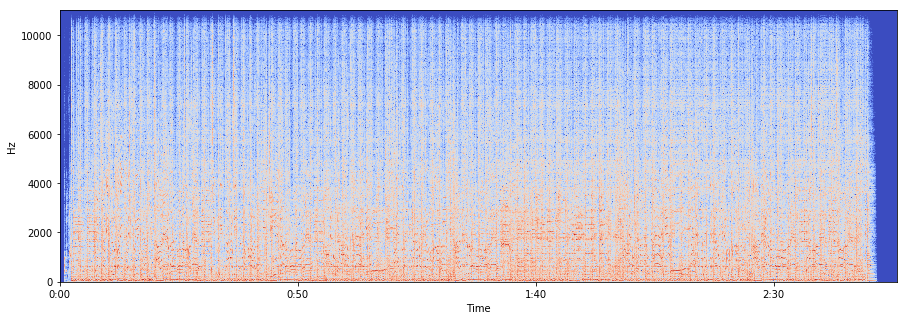}
 \caption{STFT spectrogram of \textit{I Saw Her Standing There} by The Beatles. Note that the $y$-axis represents the frequency on a linear scale.}
 \label{fig:stft-spectrogram}
\end{figure}

We can visualize the intensity of frequencies over time in 2D using a \textbf{spectrogram}. A spectrogram is the squared magnitude of the STFT:
\begin{equation}
S(m, k) = |X_{m,k}|^2
\end{equation}
Figure~\ref{fig:stft-spectrogram} shows the spectrogram of \textit{I Saw Her Standing There} by The Beatles. For this spectrogram, we chose a sampling rate of 22050Hz, window size of 2048 samples and hop size of 512 samples. Each point in the spectrogram represents the frequency intensity at a given time point. The warmer the color, the higher the frequency.

The STFT is a common method to extract feature information from a signal. Nevertheless, this method has some problematic properties: first, one needs to specify a window function. The window must be large enough to capture the lowest frequencies. On the other hand: the larger the window, the lower the time resolution. Another disadvantage of the STFT is that the frequencies calculated by the STFT are separated by a constant frequency difference, while we have seen that the frequencies of notes of a scale of Western music increase exponentially. Therefore, the frequencies from the STFT do not map directly to the frequencies of music notes.

In the \textbf{Constant-Q transform} \citep{brown1991calculation}, the window size is not constant, but dependent on the coefficient $k$: the window $N_k$ grows with higher frequencies. Also, the frequency filters $f_k$ are not spaced linearly (like in the STFT), but logarithmically: the $k$'th filter is $f_k = (2^{1/b})^k f_{min}$. 

\begin{equation}\label{eq:constq}
X_{m,k} = \frac{\sum_{n=0}^{N_k-1} x(n+mH)w(k,n) \exp\{-2\pi i k n / N_k\}}{N_k}
\end{equation}

The formula of the Constant-Q transform is given in Equation~\ref{eq:constq}. Note the differences with Equation~\ref{eq:stft}: the window function now has two parameters ($k$ and $n$); the window size $N_k$ is dependent on $k$ and we normalize by $N_k$ to compensate for high values at high frequencies. When choosing the right values for $f_{min}$ (the lowest frequency we can detect) and $b$ (the number of bins per octave), the Constant-Q transform maps directly to frequencies of musical notes.

\begin{figure}
 \centering
 \includegraphics[width=0.8\textwidth]{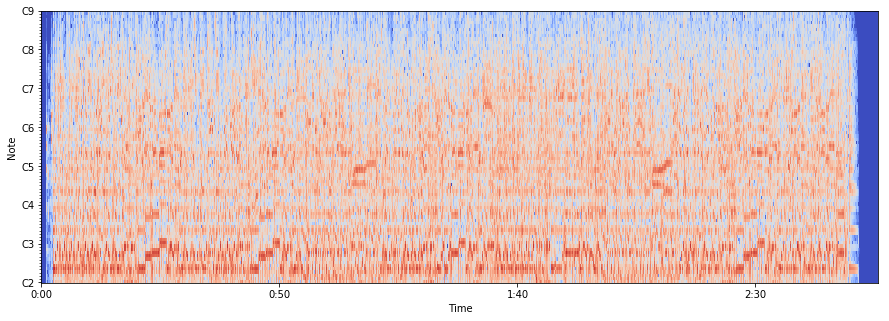}
 \caption{CQT spectrogram of \textit{I Saw Her Standing There} by The Beatles. In this spectrogram, the frequencies are placed logarithmically on the $y$-axis, because the octave numbers increase linearly.}
 \label{fig:cqt-spectrogram}
\end{figure}

In Figure~\ref{fig:cqt-spectrogram}, we see the spectrogram of the same Beatles song, but now calculated from the Constant-Q transform with $f_{min} = 65.4$Hz (the note C2) and $b=12$. Note that the frequencies are placed logarithmically on the $y$-axis because the octave numbers increase linearly, so it is easier to see which pitches sounded at each time instance in the song. Almost every pitch has a strictly positive intensity: the lion's share of the spectrogram's colors is not dark blue. Many of those pitches were not played intentionally by The Beatles' band members, but are caused by partials (any of the sinusoids of which a complex tone is composed, see Section~\ref{sec:pitchphysics}).

\begin{figure} 
 \centering
 \includegraphics[width=0.8\textwidth]{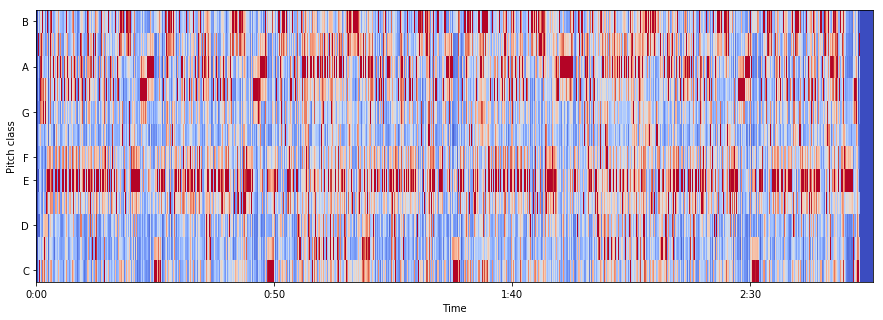}
 \caption{CQT chromagram of \textit{I Saw Her Standing There} by The Beatles. In this representation, octave information is discarded, but pitch class information is retained.}
 \label{fig:cqt-chromagram}
\end{figure}

For most \textsc{mir} tasks concerning pitches, octave information can be discarded. This is also the case for \textsc{ace}: the octave numbers are irrelevant in determining the chord from the pitches of which it is composed; we only need the pitch classes. \textbf{Chroma features} aggregate all spectral information that relates to a given pitch class into a single coefficient. There exist many variations of Chroma features, and they can be calculated from both the STFT and the Constant-Q transform. Basically, a Chroma feature is a 12-dimensional vector that can be obtained by summing all frequencies. Figure~\ref{fig:cqt-chromagram} shows a visualization of the chroma features of our example song. Here, we see for example that the pitch classes E and G\# have high chroma values in the beginning of the song. Indeed, the song starts with a E major chord.

\subsection{\textsc{Midi} representation}
\textsc{Midi} is an abbreviation for Musical Instrument Digital Interface. It is a protocol which allows electronic instruments and other digital musical tools to communicate with each other, by sending event messages. Such an event message can for example instruct a synthesizer to start playing a certain note (by a note on event), stop playing a note (note off event) or change to another instrument sound (program change event). A \textsc{midi} file is a sequence of \textsc{midi} messages, organized in a specific format. 

The information stored in \textsc{midi} files is therefore fundamentally different from the information stored in audio files: as we have seen in the previous section, audio files represent the waveform of a sound. Conversely, a \textsc{midi} file stores a list of instructions for a synthesizer, just like a musical score in a way stores instructions for a musician. The \textsc{midi} file itself does not contain any audio signals, but audio can be synthesized based on the \textsc{midi} event messages. \textsc{Midi} files can thus be considered as a compact way to store a musical score and form therefore a symbolic music representation. Compared to audio, it is way easier to extract note information from \textsc{midi}. This makes \textsc{midi} particularly interesting for research in \textsc{mir} and musicology.

Another difference between \textsc{midi} and audio is the file size: as \textsc{midi} stores music on a note level instead of on a sample level, \textsc{midi} files are typically much smaller than audio files. For instance, the .wav file of \textit{I Saw Her Standing Ther}e which we use in our data set has a size of 92.6 megabytes, while our \textsc{midi} files of the same song have file sizes ranging from just 27.3 to 28.8 kilobytes. This storage efficiency hugely contributed to the popularity of \textsc{midi} files before the advent of the compressed file format MP3. Up until now, \textsc{midi} files are still used in resource-scarce settings such as karaoke machines. This has lead to an abundance of \textsc{midi} files today. Recently, \cite{raffel2016learning} obtained as much as 178.561 \textsc{midi} files with unique MD5 checksums (i.e. a widely used 128-bit hash that is computed on a file) trough a large-scale web scrape, resulting in the ``Lakh \textsc{midi} Dataset''.

\begin{figure}
 \centering
 \includegraphics[width=0.8\textwidth]{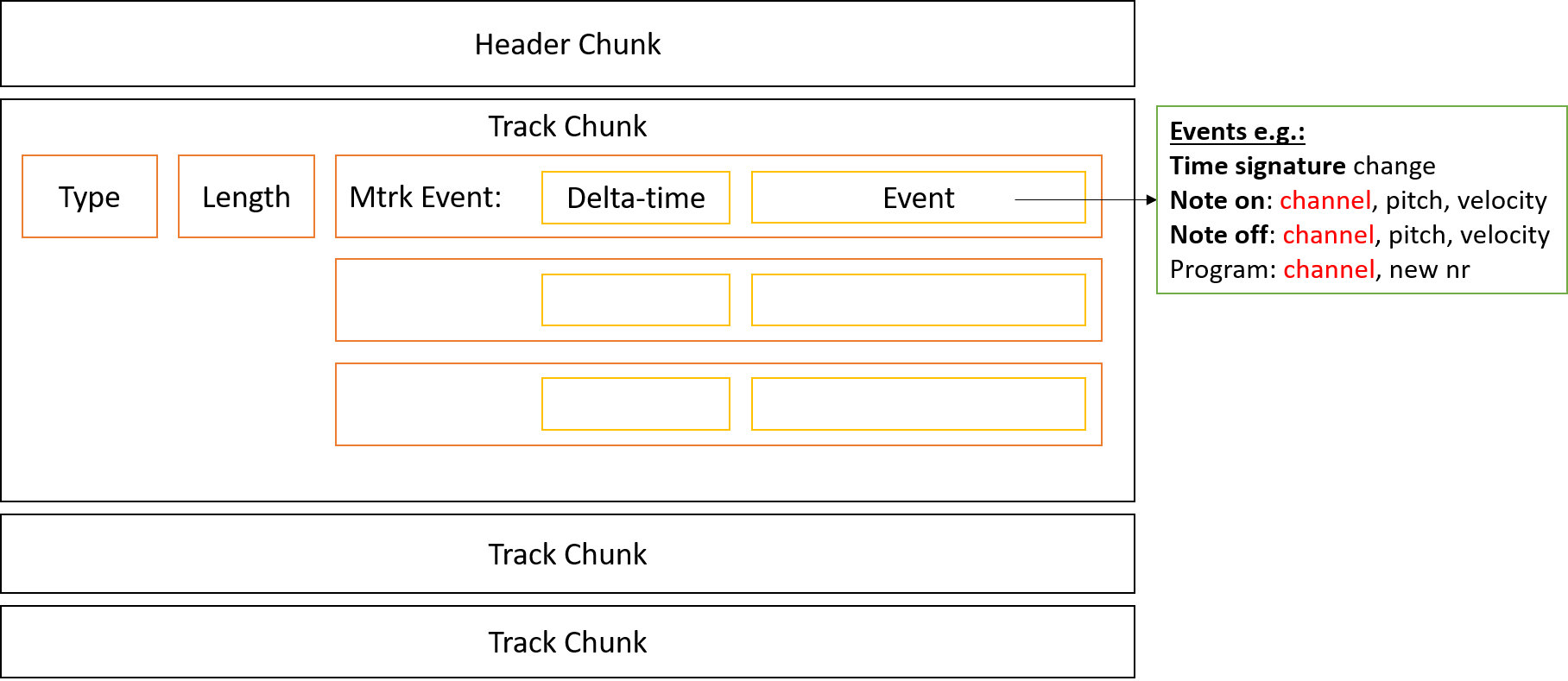}
 \caption{Structure of a \textsc{midi} file. A \textsc{midi} file has a header chunk and a number of track chunks. Each track chunk has a number of $\langle$delta-time, event$\rangle$ pairs. The most interesting events for our application are note on and note off events.}
 \label{fig:midi}
\end{figure}

The structure of a \textsc{midi} file is illustrated in Figure~\ref{fig:midi}. A \textsc{midi} file consists of a header chunk and a number of track chunks. The header chunk specifies the number of track chunks and the duration of a tick, which is the default time unit in a \textsc{midi} file. Each track chunk has a number of $\langle$delta-time, event$\rangle$ pairs. The delta-time is the time that has passed since the previous event, measured in ticks. An event is either a \textsc{midi} event, system exclusive (sysex) event or meta event. For a detailed description of these events and \textsc{midi} in general, we refer the reader to \cite{guerin2009midi}. In this section, we will only consider a subset of events. The most common \textsc{midi} events in an average \textsc{midi} file are note on, note off and program change events. Note on and note off events respectively specify the start and end of a note and have a channel, pitch and velocity. Both pitch and velocity are integers between 0 and 127. A larger pitch value results in a higher sound: a pitch of zero corresponds to  a C0, while the highest possible pitch (127) is the G10. If the velocity is high, a loud sound will be heard, while a velocity of zero only produces silence. A channel can be seen as a part in a full score: each channel is at any time mapped to a program number. These programs, or patches, determine the instrument sound. For example, program 1 is the acoustic grand piano, while a channel with program 59 will play the tuba for us. The program of a channel can be changed by program change events. 

\begin{figure}
 \centering
 \includegraphics[width=0.8\textwidth]{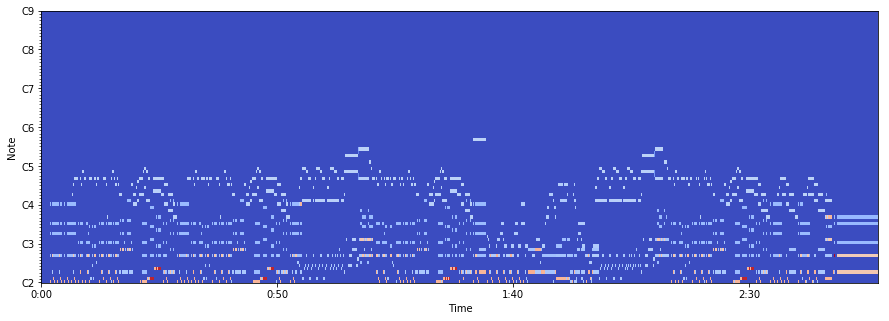}
 \caption{Piano-roll representation of \textit{I Saw Her Standing There} by The Beatles. Note that this representation is comparable to the CQT-specrogram, but way less noisy thanks to the absence of partials in \textsc{midi}.}
 \label{fig:piano-roll}
\end{figure}

Based on note on and note off events, we can easily extract a piano-roll representation of a \textsc{midi} file. This is a time-frequency visualization that strongly reminds of piano rolls for pianola or reproducing piano. The piano-roll representation of a \textsc{midi} for the Beatles song \textit{I Saw Her Standing There} is shown in Figure~\ref{fig:piano-roll}. When comparing this figure to the CQT-spectrogram in Figure~\ref{fig:cqt-spectrogram}, it becomes clear that the piano-roll representation is much ``cleaner'': thanks to the absence of partials, it is way easier to extract notes from \textsc{midi} than from audio.

\cite{raffel2016extracting} describe the sources of information available in a \textsc{midi} file. First, \textsc{midi} files are naturally suited to be used as transcriptions of pieces of music, thanks to the way they are specified. At each position in the file, we can infer exactly which instrument plays which note. To extract this kind of information, various software libraries were developed, for example the \textsc{midi} Toolbox \citep{eerola2004mir} and \lstinline{pretty_midi} \citep{raffel2014intuitive} Second, we can gather timing information from \textsc{midi} files, as there exist events for tempo changes and time signatures. Third, all 24 possible major and major keys can be specified in a key change event. Fourth, lyrics can be added to \textsc{midi} transcriptions by the use of lyrics meta-events. Finally, software libraries like \lstinline{jSymbolic} \citep{mckay2006jsymbolic} and \lstinline{music21} \citep{cuthbert2010music21} can be used to compute higher-level features.

To summarize, in this section we have seen that \textsc{midi} files are a symbolic music representation from which we can extract all kinds of musical information, including a note transcription, using one of the software libraries developed for this purpose. Thanks to their storage efficiency, \textsc{midi} files abound on the Internet. This makes \textsc{midi} files particularly interesting for \textsc{ace} research.

\subsection{Tab representation}\label{sec:tabs}
Guitar tablatures and chord sheets are collectively known as ``tabs''. In contrast to traditional musical scores, \textbf{(guitar) tablature} indicates the instrumental fingering rather than musical pitches. These tablatures are usually represented using an ASCII text notation, in which each line represents a string of the instrument. As reading tablature requires little musical training and tabs can be written and read without any specific software, they are very popular: millions of guitar tablature files can be found on websites like Ultimate Guitar\footnote{\url{https://www.ultimate-guitar.com/}} \citep{macrae2011guitar}. Figure~\ref{fig:guitar-tab} shows an example of guitar tablature. Chords can be extracted from tabs very easily. For example, above the word ``seventeen'' we see the fingering combination $\langle 0, 3, 1, 2, 2, 0 \rangle$, which means that a chord is played using open E strings, the third fret on the B string, the first fret on the G string and the second fret on the D and A string. This way, the notes E4, D4, G\#3, E3, B2 and E2 are played on the six strings. These notes form the seventh chord E7, as they consist of the notes E, G\#, B and D. In many tablatures, including the example in Figure~\ref{fig:guitar-tab}, chords are represented twice by also adding a chord. This makes it even easier to extract chord information from guitar tablature. However, note that tablature does not contain any timing information, in contrast to audio and \textsc{midi} representations. 

\begin{figure}
	\centering
    \begin{minipage}[b]{0.45\textwidth}
  	\tiny
 	\begin{verbatim}
e|-------------------------------|-------------------------------|
B|-------------------------------|-------------------------------|
G|-------------------------------|-------------------------------|
D|-------------------------------|-------------------------------|
A|2---2---2-------2---2-------2--|2---2---2---2-------2---2------|
E|0---0---0-------0---0-------0--|0---0---0---0-------0---0------|
                                              Well, she was   just                                             
                                              
          E7
e|--------0-------0---0----------|--------0---0-------0---0------|
B|--------3-------3---3----------|--------3---3-------3---3------|
G|--------1-------1---1----------|--------1---1-------1---1------|
D|--------2-------2---2----------|--------2---2-------2---2------|
A|--------2-------2---2----------|--------2---2-------2---2------|
E|--------0-------0---0----------|--------0---0-------0---0------|
                  seventeen,                              you know 

          A7                              E7
e|--------3-------3---3----------|--------0-------0---0----------|
B|--------2-------2---2----------|--------3-------3---3----------|
G|--------2-------2---2----------|--------1-------1---1----------|
D|--------2-------2---2----------|--------2-------2---2----------|
A|--------0-------0---0----------|--------2-------2---2----------|
E|--------x-------x---x----------|--------0-------0---0----------|
                  what I      mean                        and the
	\end{verbatim}
  	\caption{Guitar tablature excerpt}  
  	\label{fig:guitar-tab}
    \end{minipage}
    \hfill
    \begin{minipage}[b]{0.45\textwidth}
    \tiny
  	\begin{verbatim}
    
I SAW HER STANDING THERE     
THE BEATLES

[Verse]
             E7                     A7          E7
Well she was just seventeen and you know what I mean
                                      B7
And the way she looked was way beyond compare
   E           E7           A7    Am/C
So how could I dance with another oh,
       E7      B7       E7
when I saw her standing there


[Verse]
     E7                      A7      E7
Well she looked at me and I, I could see
                                           B7
That before too long I'd fall in love with her
E            E7           A7
She wouldn't dance with another
Am/C          E7      B7       E7
Oh,  when I saw her standing there

  	\end{verbatim}
  	\normalsize
  	\caption{Chord sheet excerpt}
  	\label{fig:tab-chord-sheet}
    \end{minipage}
\end{figure}

A \textbf{chord sheet} is a lyric sheet, in which chord symbols are placed above the lyric syllables with which they have to be timed. Chord sheets can also be very compactly represented in ASCII text notation. They can be found in abundance on the Internet. Figure~\ref{fig:tab-chord-sheet} is a fragment from a chord sheet for \textit{I Saw Her Standing There}. We see that chord information can be extracted directly from the chord sheet. However, similar to guitar tablature, there is no timing information available in chord sheets. In addition, it is quite common for chord sheets to represent only a single verse and chorus, as the chords of other verses and choruses in pop songs are often the same. 

In summary, tabs consist of guitar tablatures and chord sheets. Both types are created by music enthusiasts and can be found in abundance on websites like Ultimate Guitar. However, as there are no restrictions on the authorships of tabs, many tabs are erroneous or incomplete. Therefore, it is not trivial to select only the high-quality tabs.

\section{Summary of musical background}
This chapter provided an introduction to the concepts of pitch, music notation, intervals, chords and music representations. The \textbf{pitch} of musical tones refers to the degree of highness of sound and that the frequency is the number of vibrations in the sound wave. A tuning system discretises the space of all different pitches. In this study, we use twelve-tone equal-tempered tuning, in which the frequency ratio between of two subsequent scale steps equals $2^{\frac{1}{12}}$. A distance of twelve semitones is called an octave. We can group tones that are an octave apart in the same pitch class. In modern Western music notation, tones are given note names based on a pitch class name (i.e. one letter from \{A, B, C, D, E, F, G\}, possibly combined with an accidental) and an octave number. The reference note A4 has a frequency of 440 Hz. 
Music is \textbf{notated} on a staff. The pitch of each note is determined by its vertical position on the staff.
\textbf{Intervals} are defined as the distance between two pitches and are based on the notion of semitone. Interval names consist of a number and a quality.
\textbf{Chords} are groups of notes sounding at the same time, consisting of tones from at least different pitch classes. Most chords are built by stacking stacks of thirds on a root note. Triads are chords that consist of tones from three pitch classes and consist of the root, plus the third and fifth above it. Seventh chords, or tetrads, consist of the root, third, fifth and seventh. There exist multiple inversions for each chord, which are dependent on the relation of root note and bass note. We have seen various notations for chords. In this project, we use the chord grammar proposed by \citet{harte2005symbolic} as this grammar is unambiguous, context-independent and easy to use.

In Section~\ref{sec:musicrepresentations} we discussed the three different music representations which are used in this research project. The audio representation is an digitization of the waveform, obtained by sampling and quantization. Pitch information cannot be read from the audio directly, but can be estimated using Fourier analysis. \textsc{Midi} files are another music representation which, thanks to their storage efficiency, abound on the Internet. Interestingly, \textsc{midi} files are defined in such a way that we can easily extract all kinds of musical information, including pitch information. The third and final music representation we study are tabs, which is an umbrella term for guitar tablature and chord sheets. Tabs can be found in abundance on the Internet. Chords can be extracted almost directly from tabs, although tabs do not contain timing information and the chord labels are not always reliable.

\chapter{Framework of proposed system ``\textsc{decibel}''}\label{ch:framework}
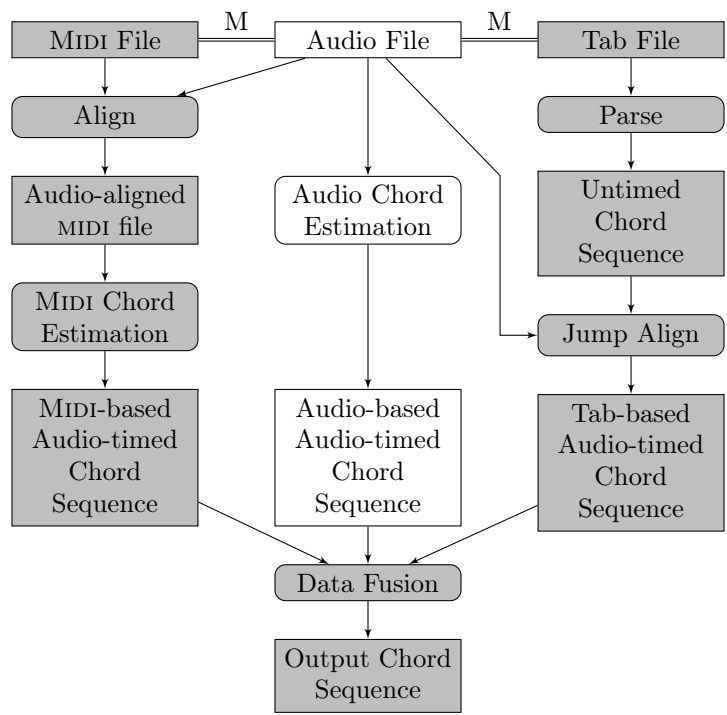
\begin{figure}
\centering
\begin{tikzpicture}[node distance = 0.5cm and 1cm, auto]
    \node [rect, fill=gray!50] (midi_file) {\textsc{Midi} File};
    \node [rect, right=of midi_file] (audio_file) {Audio File};
    \node [rect, right=of audio_file, fill=gray!50] (tab_file) {Tab File};
    
    \node [block, below=of midi_file, fill=gray!50] (align) {Align};
    \node [rect, below=of align, fill=gray!50] (audio_aligned_midi) {Audio-aligned \textsc{midi} file};
    \node [block, below=of audio_aligned_midi, fill=gray!50] (midi_ace) {\textsc{Midi} Chord Estimation};
    \node [rect, below=of midi_ace, fill=gray!50] (midi_cs) {\textsc{Midi}-based Audio-timed Chord Sequence};
    
    \node [block, right=of audio_aligned_midi] (audio_ace) {Audio Chord Estimation};
    \node [rect, right=of midi_cs] (audio_cs) {Audio-based Audio-timed Chord Sequence};
    
    \node [block, below=of tab_file, fill=gray!50] (tab_parse) {Parse};
    \node [rect, below=of tab_parse, fill=gray!50] (ucs) {Untimed Chord Sequence};
    \node [block, below=of ucs, fill=gray!50] (jump_align) {Jump Align};
    \node [rect, below=of jump_align, fill=gray!50] (tab_cs) {Tab-based Audio-timed Chord Sequence};
    
    \node [block, below=of audio_cs, fill=gray!50] (data_fusion) {Data Fusion};
    \node [rect, below=of data_fusion, fill=gray!50] (output) {Output Chord Sequence};
    
    \path [double_line] (audio_file) -- node[anchor=south] {M} (midi_file);
    \path [double_line] (audio_file) -- node[anchor=south] {M} (tab_file);
    
    \path [line] (midi_file) -- (align);
    \path [line] (audio_file) -- (align);
    \path [line] (align) -- (audio_aligned_midi);
    \path [line] (audio_aligned_midi) -- (midi_ace);
    \path [line] (midi_ace) -- (midi_cs);
    \path [line] (midi_cs) -- (data_fusion);
    
    \path[line] (audio_file) -- (audio_ace);
    \path[line] (audio_ace) -- (audio_cs);
    \path[line] (audio_cs) -- (data_fusion);
    
    \path[line] (tab_file) -- (tab_parse);
    \path[line] (tab_parse) -- (ucs);
    \path[line] (audio_file) -- (5.2, -1.8) -- (5.2, -3) |- (jump_align.west);
    \path[line] (ucs) -- (jump_align);
    \path[line] (jump_align) -- (tab_cs);
    \path[line] (tab_cs) -- (data_fusion);
    
    \path[line] (data_fusion) -- (output);
\end{tikzpicture}
\caption{Diagram of \textsc{decibel}'s framework. The \textsc{M} represents the matching between different representations of the same song. Data formats are depicted by rectangles; procedures are represented as rounded rectangles. The grey elements show how \textsc{decibel} extends existing audio \textsc{ace} methods.}
\label{fig:diagram}
\end{figure}

In this chapter, we present the framework of \textsc{decibel}: the proposed system for the Detection of Chords Improved By Exploiting Linked symbolic formats. In Chapter~\ref{ch:introduction} we have seen that existing chord estimation techniques that are only based on audio have some limitations: the performance of the \textsc{mirex} \textsc{ace} submissions seems to have reached a glass ceiling and some modern systems suffer from overfitting to subjective reference annotations. In Section~\ref{sec:musicrepresentations} we saw that the symbolic representations \textsc{midi} and tab have the convenient property that it is very easy to extract notes and chords from them. \textsc{Decibel} exploits this property by aligning \textsc{midi}s and tabs to the corresponding audio file, extracting chord sequences from each of the representations and using data fusion to combine the resulting chord sequences. \textsc{Decibel}'s framework is summarized in Figure~\ref{fig:diagram} and will be explained concisely in this section.

\section{\textsc{Decibel}'s framework in a nutshell}
The \textsc{decibel} system has a data set of audio, \textsc{midi} files and tabs at its disposal. \textsc{midi} files and tabs are obtained by a \textbf{web scrape}. These files are manually matched, based on meta-data. The dataset and matching process are described in Section~\ref{sec:data-set}. 

For each song, each of the three representations (audio, \textsc{midi} and tabs) is mapped to an \textbf{audio-timed chord sequence}, which is a sequence of chord events. Chord events are 3-tuples, consisting of a start time, end time and chord label. The possible chord label values are specified by the chosen chord vocabulary. The method for this chord estimation step depends on the representation: we used three different methods for audio, \textsc{midi} and tab representations, as specified below:

\begin{tabular}{p{0.1\textwidth} p{0.85\textwidth}}
\textbf{Audio}: &\textsc{Decibel} estimates chords from audio data using existing \textbf{audio \textsc{ace}} techniques.\\ 
\textbf{\textsc{Midi}}: &In order to estimate chords from \textsc{midi}, \textsc{decibel} first aligns each \textsc{midi} file to the audio recording using \textbf{audio-symbolic alignment} techniques. Then, chords are extracted from the audio-aligned \textsc{midi} file using a pattern-matching technique for \textbf{chord estimation in symbolic music}. This way, \textsc{decibel} obtains the chord sequences with the correct start and end times for the original audio file.\\
\textbf{Tabs}: &For the tabs, \textsc{decibel} can easily find the chord labels by \textbf{parsing} the ASCII text. Consequently, the system aligns them to the audio using \textbf{Jump Alignment}. 
\end{tabular}

Each of the three procedures for representation-dependent extraction of audio-timed chord sequences can be considered as a subsystem of \textsc{decibel}. In Chapters~\ref{ch:subsystem-audio}, \ref{ch:subsystem-midi} and \ref{ch:subsystem-tabs}, we will describe the audio, \textsc{midi} and tabs subsystems in detail. 

At this point, we have a rich harmonic representation, consisting of possible chord sequences for the song, obtained from symbolic and audio representations. As a final step, we use \textbf{data fusion} to combine these chord sequences into one final chord sequence. The data fusion method is treated in Chapter~\ref{ch:data-fusion}.

In the remainder of this section, we describe the collection of the data set in detail in Section~\ref{sec:data-set}. Also, we describe the performance evaluation measures, which are used to test both the representation-specific subsystems as well as the data fused result, in Section~\ref{sec:performance-evaluation}.

\section{Collection of data set}\label{sec:data-set}
\textsc{Decibel} uses a data set of audio, \textsc{midi} files and tabs. This data set is based on a subset of the Isophonics Reference Annotations \citep{mauch2009omras2}. The Isophonics data set contains chord annotations for 180 Beatles songs, 20 songs by Queen, 7 songs by Carole King and 18 songs by Zweieck. In this project, we only use the songs by the Beatles and Queen, as there were no \textsc{midi} or tabs for Zweieck available and there were some inconsistencies in the Carole King annotations. Using this 200 song data set has three advantages: 
(1) the music by The Beatles and Queen is popular music and therefore the correct genre for our dataset;
(2) thanks to the popularity of the two bands, it is easy to find \textsc{midi} and tab files for the songs in the data set; and
(3) the chord labels have been carefully checked and have been used for many years by the MIR community.

\begin{table}
\centering
\begin{tabular}{llll}
\textbf{Artist} & \textbf{Album} & \textbf{ID} & \textbf{\# songs} \\
The Beatles & Please Please Me & CDP 7 46435 2 & 14 \\
The Beatles & With the Beatles & CDP 7 46436 2 & 14 \\
The Beatles & A Hard Day’s Night & CDP 7 46437 2 & 13 \\
The Beatles & Beatles For Sale & CDP 7 46438 2 & 14 \\
The Beatles & Help! & CDP 7 46439 2 & 14 \\
The Beatles & Rubber Soul & CDP 7 46440 2 & 14 \\
The Beatles & Revolver & CDP 7 46441 2 & 14 \\
The Beatles & Sgt. Pepper’s Lonely Hearts & CDP 7 46442 2 & 13 \\
& Club Band &&\\
The Beatles & Magical Mystery Tour & CDP 7 48062 2 & 11 \\
The Beatles & The Beatles (the white album) & CDS 7 46443 8 & 30 \\ 
The Beatles & Abbey Road & CDP 7 46446 2 & 17 \\
The Beatles & Let It Be & CDP 7 46447 2 & 12 \\
Queen & Greatest Hits I & Parlophone, 0777 7 8950424 & 14 \\
Queen & Greatest Hits II & Parlophone, CDP 7979712 & 6 \\
\end{tabular}
\caption{Isophonics Reference Annotations}\label{table:isophonics}
\end{table}

We used the audio as provided on the CDs in Table~\ref{table:isophonics}. A complete list of all song names and the index that we assigned to them, is given in Appendix~\ref{appendix:data-set}. 
After collecting the audio files and annotations, we need to find \textsc{midi} and tab files and match them to the songs in our data set. First, we searched on the Internet for \textsc{midi} files of the aforementioned 200 songs. We downloaded \textsc{midi} files from 9 websites\footnote{\url{http://beatlesnumber9.com}}\footnote{\url{http://bmh.webzdarma.cz}}\footnote{\url{http://davidbmidi.com}}\footnote{\url{http://earlybeatles.com/midi}}\footnote{\url{http://en.midimelody.ru}}\footnote{\url{http://queen.wz.cz/midi}}\footnote{\url{http://www.angelfire.com}}\footnote{\url{http://www.dongrays.com}}\footnote{\url{http://www.rppmf.com}}. This way, we found 770 \textsc{midi} files with unique MD5-checksums, so multiple \textsc{midi} files (3.85 on average) map to a single audio file. we matched the \textsc{midi} and audio files by hand, based on the \textsc{midi} file name. Furthermore, we obtained tabs from Ultimate Guitar\footnote{\url{https://www.ultimate-guitar.com}}. We first automatically scraped all tabs from The Beatles and Queen from Ultimate Guitar's website. Then, we matched the tabs to the audio files by hand, based on song title. Tabs from songs that were not in the dataset, were discarded. This resulted in 1668 matched tabs (974 chords and 694 guitar tablature files).

\begin{figure}
  \centering
  \begin{subfigure}[b]{0.45\textwidth}
  	\includegraphics[width=\textwidth]{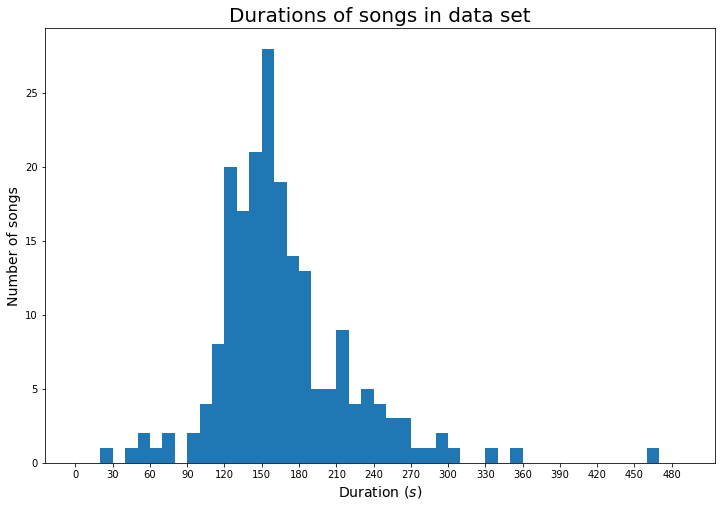}
    \caption{Song durations}
  \end{subfigure}
  \begin{subfigure}[b]{0.45\textwidth}
  	\includegraphics[width=\textwidth]{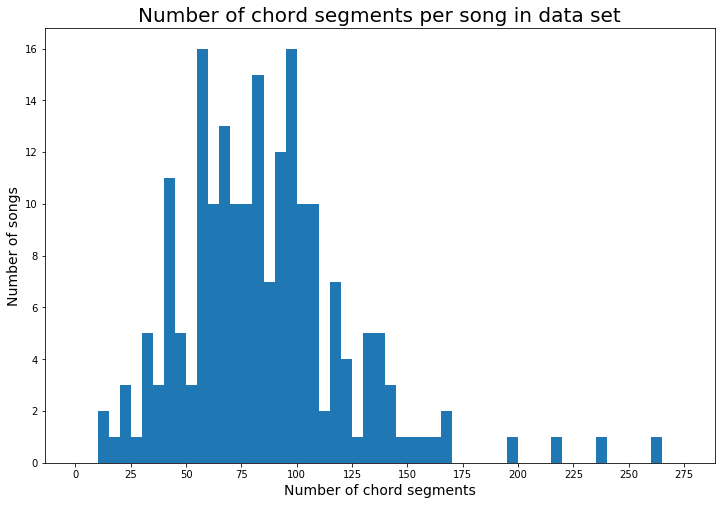}
    \caption{Number of chord segments per song}
  \end{subfigure}
  \begin{subfigure}[b]{\textwidth}
  	\includegraphics[width=\textwidth]{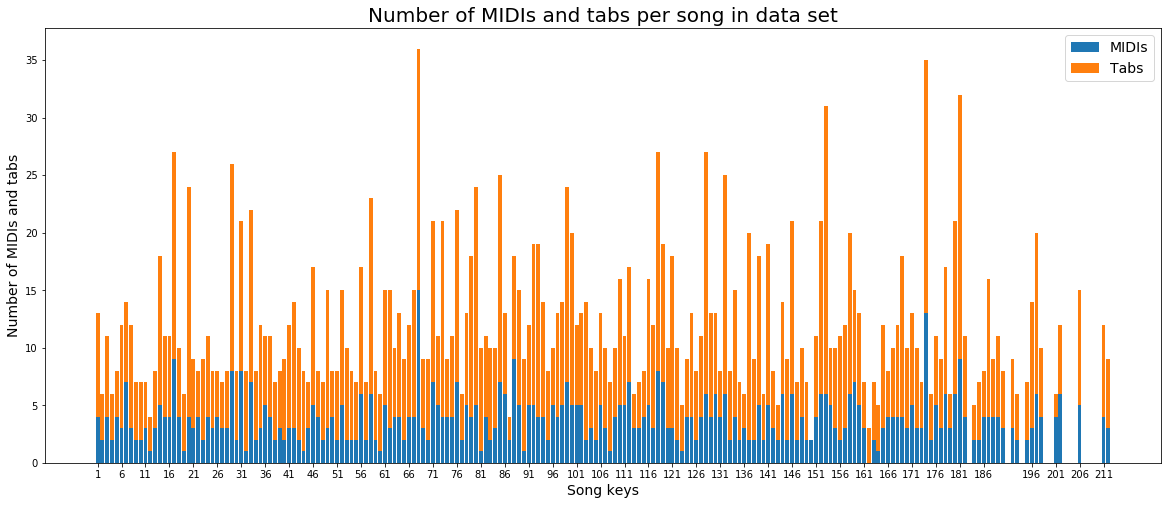}
    \caption{Number of \textsc{midi}s and tabs per song. Note that some songs (e.g. song number 208) are not matched to any \textsc{midi} or tab file. These are songs on the Queen CD's for which there were no reference annotations in the Isophonics data set.}
  \end{subfigure}
  \caption{Data set statistics. A typical song in our data set has a duration of 2 to 3 minutes, consists of 50-100 chord segments and is matched to 3 to 5 \textsc{midi} files and 5 to 10 tabs.}\label{fig:dataset_statistics}
\end{figure}

Some statistics on the data set are shown in Figure~\ref{fig:dataset_statistics}. A typical song in our data set has a duration of 2 to 3 minutes, consists of 50-100 chord segments and is matched to 3 to 5 \textsc{midi} files and 5 to 10 tabs.

\section{Performance evaluation}\label{sec:performance-evaluation}
In order to evaluate both the performance of \textsc{decibel}'s representation-specific subsystems and its final output chord sequence, we need evaluation measures. The quality of a chord sequence is usually determined by comparing it to a ground truth created by one or more human annotators. Commonly used \textbf{data sets} with chord annotations, which are also used in the \textsc{mirex} \textsc{ace} contest, are Isophonics\footnote{\url{http://isophonics.net/datasets}}, Billboard\footnote{\url{http://ddmal.music.mcgill.ca/research/billboard}}, RobbieWilliams\footnote{\url{https://www.researchgate.net/publication/260399240_Chord_and_Harmony_annotations_of_the_first_five_albums_by_Robbie_Williams}}, RWC-Popular\footnote{\url{https://github.com/tmc323/Chord-Annotations}}
and USPOP2002Chords\footnote{\url{https://labrosa.ee.columbia.edu/projects/musicsim/uspop2002.html}}. As stated before, \textsc{decibel} uses the Isophonics data set, augmented with matched \textsc{midi} and tab files.

The standard quality measure to evaluate the quality of an automatic transcription is \textbf{chord symbol recall (\textsc{csr})} \citep{harte2010towards}. This measure is also used in the \textsc{mirex} \textsc{ace} contest\footnote{\url{http://www.music-ir.org/mirex}}. \textsc{Csr} is the summed duration of time periods where the correct chord has been identified, normalized by the total duration of the song. Until 2013, \textsc{mirex} used an approximate, frame-based \textsc{csr} calculated by sampling both the ground-truth and the automatic annotations every 10 ms and dividing the number of correctly annotated samples by the total number of samples. Since 2013, \textsc{mirex} has used segment-based \textsc{csr}, which is more precise and computationally more efficient. The formula for segment-based \textsc{csr} is given in Equation~\ref{eq:csr}. We consider the ground-truth annotation $A$ as a sequence of segments $S_A$ and the estimated annotation $E$ as a sequence of segments $S_E$. The duration of a segment is notated as $|\cdot|$.

\begin{equation}\label{eq:csr}
CSR_T(S_E, S_A) = \frac{\sum_{S^j_A} \sum_{S^i_E}|S^i_E \cap S^j_A| \cdot \mathbb{M}_T(S^j_A, S^i_E)}{\sum_{S^j_A} |S^j_A|}
\end{equation}

$\mathbb{M}_T$ is a matching function as defined by Equation~\ref{eq:csr-matching-function}, in which $T$ denotes the comparison method used to evaluate the result of the matching function.

\begin{equation}\label{eq:csr-matching-function}
\mathbb{M}_T = \begin{cases}
    1 & \text{if } X \text{ matches } Y\\
    0 & \text{otherwise}
\end{cases}
\end{equation}

The comparison method is dependent of the \textbf{chord vocabulary}: both the chord labels ground truth annotations and the estimated annotations are mapped to a limited set of chord labels, as specified by the chord vocabulary. \textsc{mirex} uses the following five chord vocabularies:
\begin{enumerate}
\item Only the chord root note (\texttt{C, D, ..., B}), or no-chord (\texttt{N});
\item Major and minor: \texttt{\{N, maj, min\}};
\item Seventh chords: \texttt{\{N, maj, min, maj7, min7, 7\}};
\item Major and minor with inversions: \texttt{\{N, maj, min, maj/3, min/b3, maj/5, min/5\}};
\item Seventh chords with inversions: \texttt{\{N, maj, min, maj7, min7, 7, maj/3, min/b3, maj7/3, min7/b3, 7/3, maj/5, min/5, maj7/5, min7/5, 7/5, maj7/7, min7/b7, 7/b7\}}.
\end{enumerate}
Two chords match if and only if they are mapped to exactly the same chord label in the vocabulary. For example, the mapping to chord vocabulary 1 only preserves the root note. The chords \texttt{C:maj} and \texttt{C:min} will both be mapped to a \texttt{C} chord, so these chords would match in the first chord vocabulary. However, in all other vocabularies they would be mapped to \texttt{C:maj} and \texttt{C:min}, which do not match. 

For results that are calculated for the whole data set, we weigh the \textsc{csr} by the length of the song when computing an average for a given corpus. This final number is referred to as the \textbf{weighted chord symbol recall (\textsc{wcsr})}. Calculating the \textsc{wcsr} is basically the same as treating the data set as one big audio file, and calculating the \textsc{csr} between the concatenation of all ground-truth annotations and the concatenation of all estimated annotations.

The \textsc{csr} correctly indicates the accuracy of an \textsc{ace} algorithm in terms of whether the estimated chord for a given instant in the audio is correct. It it therefore widely used in the evaluation of \textsc{ace} systems. However, the annotation with the highest \textsc{csr} is not always the annotation that would be considered the best by human listeners. As an example, examine Figure~\ref{fig:csr-example}. Here we see two estimated annotations. Although Annotation A has the higher \textsc{csr}, most musicians would prefer Annotation B: Annotation A is clearly over-segmented, which makes it difficult to play and unnatural to listen to. On the other hand, human listeners would consider the 5 seconds with the wrong chord in Annotation B as just one major mistake. 

\begin{figure}
  \begin{tikzpicture}
    \draw[step=1cm,lightgray,very thin] (2,0) grid (15,3);
    \filldraw[fill=blue!25!white, draw=black] (2,2) rectangle (15,3);
    \filldraw[fill=blue!25!white, draw=black] (2,1) rectangle (3,2);
    \filldraw[fill=blue!25!white, draw=black] (4,1) rectangle (5,2);
    \filldraw[fill=blue!25!white, draw=black] (6,1) rectangle (8,2);
    \filldraw[fill=blue!25!white, draw=black] (9,1) rectangle (10,2);
    \filldraw[fill=blue!25!white, draw=black] (11, 1) rectangle (15,2);
    \filldraw[fill=blue!25!white, draw=black] (7, 0) rectangle (15,1);    
    \node[] at (0,2.5) {Ground Truth};
    \foreach \x in {2,7,10,12}
      \draw[thick] (\x cm,2) -- (\x cm,3);
    \node[] at (4.5,2.5) {C};
    \node[] at (8.5,2.5) {F};
    \node[] at (11,2.5) {G};
    \node[] at (13.5,2.5) {C};
    \draw[very thick] (2,2) rectangle (15,3);
    \node[] at (0,1.5) {Annotation A};
    \foreach \x in {2,3,4,5,6,7,8,9,10,11,12}
      \draw[thick] (\x cm,1) -- (\x cm,2);
    \node[] at (2.5,1.5) {C};
    \node[] at (3.5,1.5) {B};
    \node[] at (4.5,1.5) {C};
    \node[] at (5.5,1.5) {F\#};
    \node[] at (6.5,1.5) {C};
    \node[] at (7.5,1.5) {F};
    \node[] at (8.5,1.5) {B};
    \node[] at (9.5,1.5) {F};
    \node[] at (10.5,1.5) {B};
    \node[] at (11.5,1.5) {G};
    \node[] at (13.5,1.5) {C};
    \draw[very thick] (2,1) rectangle (15,2);
    \node[] at (0,0.5) {Annotation B};
    \foreach \x in {2,7,10,12}
      \draw[thick] (\x cm,0) -- (\x cm,1);
    \node[] at (4.5,0.5) {Am};
    \node[] at (8.5,0.5) {F};
    \node[] at (11,0.5) {G};
    \node[] at (13.5,0.5) {C};
    \draw[very thick] (2,0) rectangle (15,1);
    \node[] at (1,-0.5) {$t =$};
    \foreach \x in {0,1,2,3,4,5,6,7,8,9,10,11,12,13}
      \node[] at (\x + 2,-0.5) {\x};
  \end{tikzpicture}
  \caption{Annotation A has a \textsc{csr} of $9/13=69.2\%$ and Annotation B has a \textsc{csr} of $8/13=61.5\%$. However, Annotation B is the preferred chord sequence.}
  \label{fig:csr-example}
\end{figure}

Just measuring the (weighted) chord symbol recall is therefore not enough: we also need a metric for chord segmentation quality. For this purpose, \cite{mauch2010automatic} proposed the use of the directional hamming distance $h(S, S^0)$. It describes how fragmented segmentation $S$ is with respect to segmentation $S^0$, according to Equation~\ref{eq:hamming-distance}:

\begin{equation}\label{eq:hamming-distance}
h(S, S^0) = \sum_{i = 1}^{N_{S^0}} (|S_i^0| - \max_j |S_i^0 \cap S_j|)
\end{equation}

In our example, let Annotation A be $S^A$, let Annotation B be $S^B$ and let the ground truth be $S^G$. Then, $h(S^A, S^G)$ is a measure of over-segmentation of $S^A$ with respect to $S^G$. Indeed, 
\begin{equation*}
h(S^A, S^G) = \sum_{i = 1}^{N_{S^G}} (|S_i^G| - \max_j |S_i^G \cap S^A_j|) = (5 - 1) + (3 - 1) + (2 - 1) + (3 - 3) = 7
\end{equation*}
This is a high value, so the directional Hamming distance shows that $S^A$ is over-segmented. A high Hamming distance in the opposite direction indicates under-segmentation. 
\begin{equation*}
h(S^G, S^A) = \sum_{i = 1}^{N_{S^A}} (|S_i^A| - \max_j |S_i^A \cap S^G_j|) = 13 * (1 - 1) = 0
\end{equation*}
so $S^A$ is not under-segmented. The segmentation of Annotation B is identical to the ground truth:
\begin{equation*}
h(S^B, S^G) = h(S^G, S^B) = 0
\end{equation*}

We can easily transform the directional Hamming distance into a quality measures for over-segmentation and under-segmentation using the following equations:
\begin{equation}\label{eq:overseg}
\text{OverSegmentation}(S_E, S_A) = 1 - \frac{h(S_E, S_A)}{\sum_{i=1}^{N_{S_A}} |S_{A_i}|} \in [0,1]
\end{equation}
\begin{equation}\label{eq:underseg}
\text{UnderSegmentation}(S_E, S_A) = 1 - \frac{h(S_A, S_E)}{\sum_{i=1}^{N_{S_A}} |S_{A_i}|} \in [0,1]
\end{equation}

OverSegmentation or UnderSegmentation values near 0 correspond to highly under- and over-segmented annotations respectively. High values indicate good segmentation quality. Equations~\ref{eq:overseg} and \ref{eq:underseg} can be combined into an all-in-one segmentation measure using Equation~\ref{eq:seg}. We use the minimum, so an annotation only gets a high segmentation quality if neither under-segmentation nor over-segmentation is dominant.
\begin{equation}\label{eq:seg}
\text{Segmentation}(S_E, S_A) = \min \begin{cases}
\text{OverSegmentation}(S_E, S_A)\\ 
\text{UnderSegmentation}(S_E, S_A)
\end{cases}
\end{equation}

\chapter{Automatic Chord Estimation on audio}\label{ch:subsystem-audio}
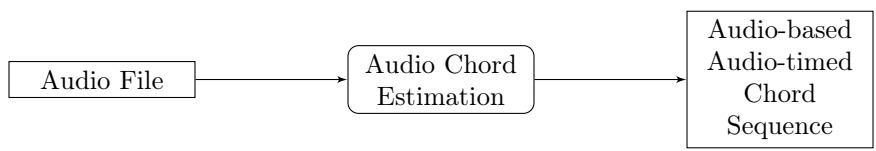
\begin{figure}[ht]
\centering
\begin{tikzpicture}[node distance = 2cm, auto]
    \node [rect] (audio_file) {Audio File};    
    \node [block, right=of audio_file] (audio_ace) {Audio Chord Estimation};
    \node [rect, right=of audio_ace] (audio_cs) {Audio-based Audio-timed Chord Sequence};
    
    \path[line] (audio_file) -- (audio_ace);
    \path[line] (audio_ace) -- (audio_cs);
\end{tikzpicture}
\caption{Diagram of \textsc{decibel}'s audio subsystem}
\label{fig:diagram-audio}
\end{figure}

In this chapter, we focus on the subsystem of \textsc{decibel} that extracts chord sequences from the audio representation, which is schematically summarized in Figure~\ref{fig:diagram-audio}. There exist multiple methods to extract chord labels from audio data. In Section~\ref{sec:aceaudio}, we give an overview of existing methods for \textsc{ace} in audio, showing a variety in implementations. Ten state-of-the-art systems are used in the audio \textsc{ace} subsystem of \textsc{decibel}, as described in Section~\ref{sec:audio-ace-selected-systems}. Finally, we evaluate these ten systems on our 200 song data set in Section~\ref{sec:audio-ace-evaluation} and conclude in Section~\ref{sec:audio-ace-conclusion}.

\section{Related work on audio \textsc{ace}}~\label{sec:aceaudio}
This section gives an overview of existing methods for \textsc{ace} on audio. \cite{fujishima1999realtime} was the first who considered \textsc{ace} as a task on its own. Since then, a lot of researchers have buckled down to this subject.

\begin{figure}[ht]
  \centering
  \begin{tikzpicture}[node distance = 0.8cm and 1.2cm]
    \node [rect] (training_audio) {Training Set Audio};
    \node [rect, right=of training_audio] (gt_annotations) {GT Training Annotations};
    \node [rect, right=of gt_annotations] (test_audio) {Test Set Audio};
    \node [block, below=of training_audio] (feature_extraction_training) {Feature \mbox{Extraction} \mbox{\textit{Section~\ref{sec:audio-ace-feature}}}};
    \node [block, below=of test_audio] (feature_extraction_test) {Feature \mbox{Extraction} \mbox{\textit{Section~\ref{sec:audio-ace-feature}}}};
    \node [rect, below=of feature_extraction_training] (training_features) {Training Set Features};
    \node [rect, below=of feature_extraction_test] (test_features) {Test Set Features};
    \node [block, right=of training_features] (model_training) {Model Training \mbox{\textit{Section~\ref{sec:audio-ace-models}}}};
    \node [rect, below=of model_training] (model) {Trained Model};
    \node [block, right=of model] (estimate_labels) {Estimate Chord Labels \mbox{\textit{Section~\ref{sec:audio-ace-models}}}};
    \node [rect, below=of estimate_labels] (estimate_annotation) {Estimated Annotations};
    \node [block, below=of estimate_annotation] (evaluate) {Evaluate \textit{Section \ref{sec:performance-evaluation} and \ref{sec:audio-ace-evaluation}}};
    \node [rect, left=of evaluate] (gt_test) {GT Test Annotations};
    \node [rect, right=of evaluate] (performance) {Performance};
    \path [line] (training_audio) -- (feature_extraction_training);
    \path [line] (feature_extraction_training) -- (training_features);
    \path [line] (training_features) -- (model_training);
    \path [line] (gt_annotations) -- (model_training);
    \path [line] (test_audio) -- (feature_extraction_test);
    \path [line] (feature_extraction_test) -- (test_features);
    \path [line] (model_training) -- (model);
    \path [line] (test_features) -- (estimate_labels);
    \path [line] (model) -- (estimate_labels);
    \path [line] (estimate_labels) -- (estimate_annotation);
    \path [line] (estimate_annotation) -- (evaluate);
    \path [line] (gt_test) -- (evaluate);
    \path [line] (evaluate) -- (performance);
\end{tikzpicture}
\caption{A prototypical audio \textsc{ace} pipeline, which estimates and evaluates chord label sequences based on a data set of audio files. There exist multiple methods for feature extraction and modeling. These are described in Sections~\ref{sec:audio-ace-feature} and \ref{sec:audio-ace-models}.}
\label{fig:audio-ace-pipeline}
\end{figure}
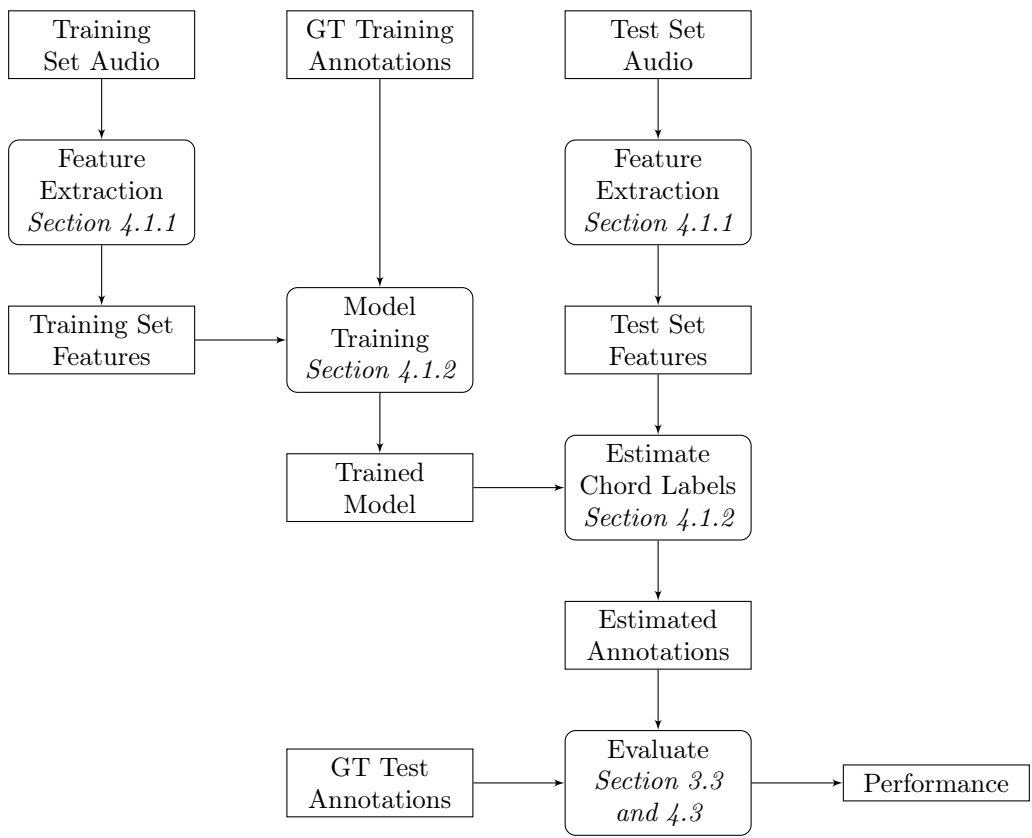

Most methods use the following pipeline~\citep{mcvicar2014automatic}: first, audio data is partitioned into a training set and a test set and features are calculated on both partitions. Subsequently, the features from the training set are used to train the parameters of a model. The chord labels for the test set are then estimated using this trained model. Finally, the performance of the system is evaluated by comparing the labels calculated by the model to the reference (ground truth) chord labels. Feature extraction and models are described in Sections~\ref{sec:audio-ace-feature} and \ref{sec:audio-ace-models} respectively. The standard pipeline is illustrated in Figure~\ref{fig:audio-ace-pipeline}.

\subsection{Feature extraction for audio \textsc{ace}}\label{sec:audio-ace-feature}
The first step in audio chord estimation is feature extraction. Features are extracted for both the audio in the training set and the audio in the test set, as shown in Figure~\ref{fig:audio-ace-pipeline}. 

The first operation most \textsc{ace} methods perform in feature extraction, is to calculate (a variation of) the chromagram \citep{wakefield1999mathematical}. As we have seen in Section~\ref{sec:musicrepresentations-audio}, the chromagram describes the pitch class salience over the duration of the audio, i.e. it represents which notes are present in each audio frame. For calculating the chromagram, the audio is first transformed to the \textbf{frequency domain} using either the Short Time Fourier Transform (STFT) or Constant-Q transform. 

In many \textsc{ace} systems, the spectrogram is \textbf{preprocessed} before calculating the chromagram. For example, \cite{mauch2010automatic} removes background noise by median filtering of the spectrogram; \cite{reed2009minimum} remove the percussive elements of the spectrum and \cite{pauws2004musical} removes harmonics. As some popular music songs are not tuned at the standard pitch of A4 = 440 Hz, many methods compensate for tuning issues by computing the spectrogram at a multiple of the required frequency resolution: \cite{sheh2003chord} doubled the frequency resolution of the spectrogram and \cite{harte2005automatic} even used the triple frequency resolution. As a next step, some authors remap the spectrogram, so it represents the human perception of pitch saliences more closely. \cite{ni2012end} for example use A-weighting, i.e. a technique to express the loudness as perceived by the human ear, to calculate loudness-based chromagrams. Finally, almost all methods discard octave information, as this is not relevant in chord estimation: the pitch A4 has the same function in a chord as the pitch A3 or A5.

Given the chromagram, some \textbf{post-processing} steps may help to prevent frequent chord changes in the predicted chord sequence. \cite{fujishima1999realtime} used smoothing techniques as post-processing, while \cite{bello2005robust} introduce beat-synchronous chromagrams, in which the chromagram is averaged between beat segments. 

Some methods however use \textbf{other representations} than traditional chromagrams: \cite{harte2006detecting} introduce the Tonal Centroid feature; \cite{mauch2008discrete} calculate a distinct bass chromagram and in more recent work, \cite{wu2017mirex} calculate a 36-dimensional binary acoustic feature using a Deep Residual Network, trained on \textsc{midi} data. \cite{sigtia2015audio} use Deep Neural Networks and \cite{korzeniowski2016feature, korzeniowski2016fully} train a Convolutional Neural Network to automatically learn musically interpretable features from the spectrogram. \cite{muller2010towards} introduce chroma DCT-reduced log pitch (CRP) features, which improve timbre invariance. These CRP features are implemented in the \textsc{ace} system by \cite{ohanlon2017improved}.

In this subsection, we saw various ways of extracting features from audio. These features are extracted from the audio of both the training set and the test set, and serve as an input for a model, as we will see in the next session.

\subsection{Audio \textsc{ace} models}\label{sec:audio-ace-models}
As a next step, we need to estimate chord labels, given the feature vectors of the test set. Although early methods for calculating the chord labels from feature vectors rely on template matching, recent works typically use statistical models or neural networks. Usually, these models are first trained on the training set, which is a subset of the data set for which ground truth chord labels are known. The model is then evaluated on the other partition of the data set: the test set. This section briefly indicates the variety of models used in audio \textsc{ace}.

\cite{sheh2003chord} were the first to use the \textbf{Hidden Markov Model (HMM)} and Expectation-Maximization algorithm to train a model for \textsc{ace}, in which the chord labels are represented by hidden states and the features correspond to the observed states. HMMs model the joint probability distribution $P(\textbf{X}, \textbf{y})$ over the feature vectors $X$ and the chord labels $y$. We will have a more detailed look at HMMs in Section~\ref{sec:jump-alignment-hmm}, as this type of model is also used in the Jump Alignment algorithm (with which \textsc{decibel} aligns tabs to audio). Some methods, for example the system proposed by \cite{shenoy2005key}, incorporate key information in a two-chain HMM. \cite{scholz2009robust} introduce higher-order HMMs, in order to better model the complexity of music.

An alternative model is the \textbf{Dynamic Bayesian Model (DBM)}, which is a Bayesian network that relates variables to each other over adjacent time steps. It was introduced for \textsc{ace} by \cite{mauch2010simultaneous}. This model has hidden nodes for metrical position, chord, bass note and musical key and observed nodes for bass and treble chromagrams. Thanks to its design, the DBM is able to integrate multiple pieces of musical context in a single model.

\cite{burgoyne2007cross} used different variations of \textbf{Conditional Random Fields (CRF)}. CRF is a discriminative model: it models $P(\textbf{y} \vert \textbf{X})$, which is the conditional distribution of chord labels given a sequence of feature vectors. Linear-chain CRFs were also used in the system of \cite{korzeniowski2016fully}.

In recent work, we see a trend towards using \textbf{Artificial Neural Network} models. This class of models is inspired by biological neural networks in our brains: they consist of a collection of connected nodes or neurons, often arranged in layers. Each of the neurons processes a signal that it receives from neurons of the previous layer, calculates some non-linear function on the sum of these inputs and passes the result on to the neurons in the next layer. The neurons and edges between them typically have weights that are adjusted in the training phase. This way, neural networks can learn a function that maps feature vectors to chord labels without modeling a probability distribution - provided that there is sufficient training data. Neural networks come in many flavors. For example, \cite{sigtia2015audio} use a hybrid Recurrent Neural Network (RNN) model and \cite{wu2017mirex} combine a Bi-direction LSTM Network and CRFs.

\section{Selected systems}\label{sec:audio-ace-selected-systems}
In order to evaluate \textsc{decibel}, we experiment with ten different audio \textsc{ace} systems: the nine submissions for the \textsc{mirex} 2017 and/or 2018 \textsc{ace} competition, together with the Chordify algorithm. We selected these ten systems as they are state of the art. It is therefore, in contrast to inferior audio \textsc{ace} systems, not trivial that we can improve these systems by incorporating information from \textsc{midi} files and tabs. The implementation of the selected systems is summarized below.

\begin{longtable}{p{0.21\textwidth} p{0.75\textwidth}}
\textbf{\textsc{chf}}: &The exact details of the Chordify algorithm are not public, but the current implementation is based on \cite{koops2017chord}. \textsc{Chf} uses a deep convolutional-recurrent model for automatic chord recognition from the CQT spectrogram.\\ 
\textbf{\textsc{cm2} (2017) / \textsc{cm1} (2018)}: &\textsc{Cm2} is the algorithm implemented in the Chordino plugin in Sonic Visualiser \citep{cannam2017mirex}. First, NNLS chroma features are extracted from the CQT spectrogram. Then, a fixed dictionary of chord profiles is used to calculate frame-wise chord similarities. Finally, the chord labels are computed using a HMM. \textsc{Cm2} was resubmitted in 2018 as \textsc{cm1}.\\ 
\textbf{\textsc{jlw1, jlw2} (2017)}: &Both \textsc{jlw} algorithms are based on the random forest model, which is a collection of decision trees \citep{jiang2017extended}. First, the signal is separated into a harmonic and percussive signal, using HPSS. Then the harmonic part is transformed to the frequency domain and NNLS chroma features are calculated on the result. Consequently, a random forest model is trained on these features. This results in a chordogram, which is a matrix that represents the emission probability for chord $c$ at frame $n$. As a final step, the result is smoothed, and this is where versions \textsc{jlw1} and \textsc{jlw2} differ. \textsc{Jlw1} uses a HMM, similar to the one used by Chordino. \textsc{Jlw2} uses a beat tracking CRF model that only allows chord changes on beats or at half-beat positions.\\ 
\textbf{\textsc{kbk1, kbk2} (2017) / \textsc{fk2} (2018)}: &Each of the two \textsc{kbk} algorithms is based on neural networks and only recognizes chords in the Major/Minor alphabet. \textsc{Kbk1} uses a deep neural network to extract chroma vectors, as shown in \citep{korzeniowski2016feature}, while \textsc{kbk2} learns features automatically by a fully convolutional neural network \citep{korzeniowski2016fully}. Then, in both versions, the chord sequence is decoded using a linear-chain Conditional Random Field (CRF) as described in \cite{korzeniowski2016fully} and chords segments are aligned to beats using a beat tracker. \textsc{Kbk2} was resubmitted in 2018 as \textsc{fk2}.\\ 
\textbf{\textsc{wl1} (2017)}: &In \textsc{wl1}, the acoustic features are first calculated from the  spectrogram of each music  signal with a deep residual network.  Then,  the  feature  vectors  are  fed  as a sequence into  a  Bidirectional Long Short Term Memory network, and a class likelihood vector  is  calculated  for  each  frame.  Finally,  the  class  likelihood sequence is passed on to the trained CRF layer to decode the optimal label sequence. \citep{wu2017mirex}\\
\textbf{\textsc{jlcx1}, \textsc{jlcx2} (2018)}&
Both \textsc{jlcx1} and \textsc{jlcx2} are based on a model that represents a chord label by its root, triad and bass. Then, a classifier based on a Recurrent Convolutional Neural Network (RCNN) is trained to recognise all subparts given the audio feature. Finally, labels of these subparts are reassembled to form the final chord label. The difference between \textsc{jlcx1} and \textsc{jlcx2} is that \textsc{jlcx2} restricts the chord transition to align with beat positions, while \textsc{jlcx1} allows chord transitions at any location. \citep{jiang2018mirex}\\
\textbf{\textsc{sg1} (2018)}&The approach of \textsc{sg1} \citep{gasser2018mirex} is based on previous year's \textsc{mirex} submission \textsc{kbk2}. \textsc{sg1} uses a convolutional neural network as well, and augments the training data by applying pitch shifts and detuning. However, they train the network on three objectives: root note, chord description and inversion.\\
\end{longtable}

\section{Evaluation of audio \textsc{ace}}\label{sec:audio-ace-evaluation}

\begin{figure}
  \centering
  \begin{subfigure}[b]{0.65\textwidth}
  	\includegraphics[width=\textwidth]{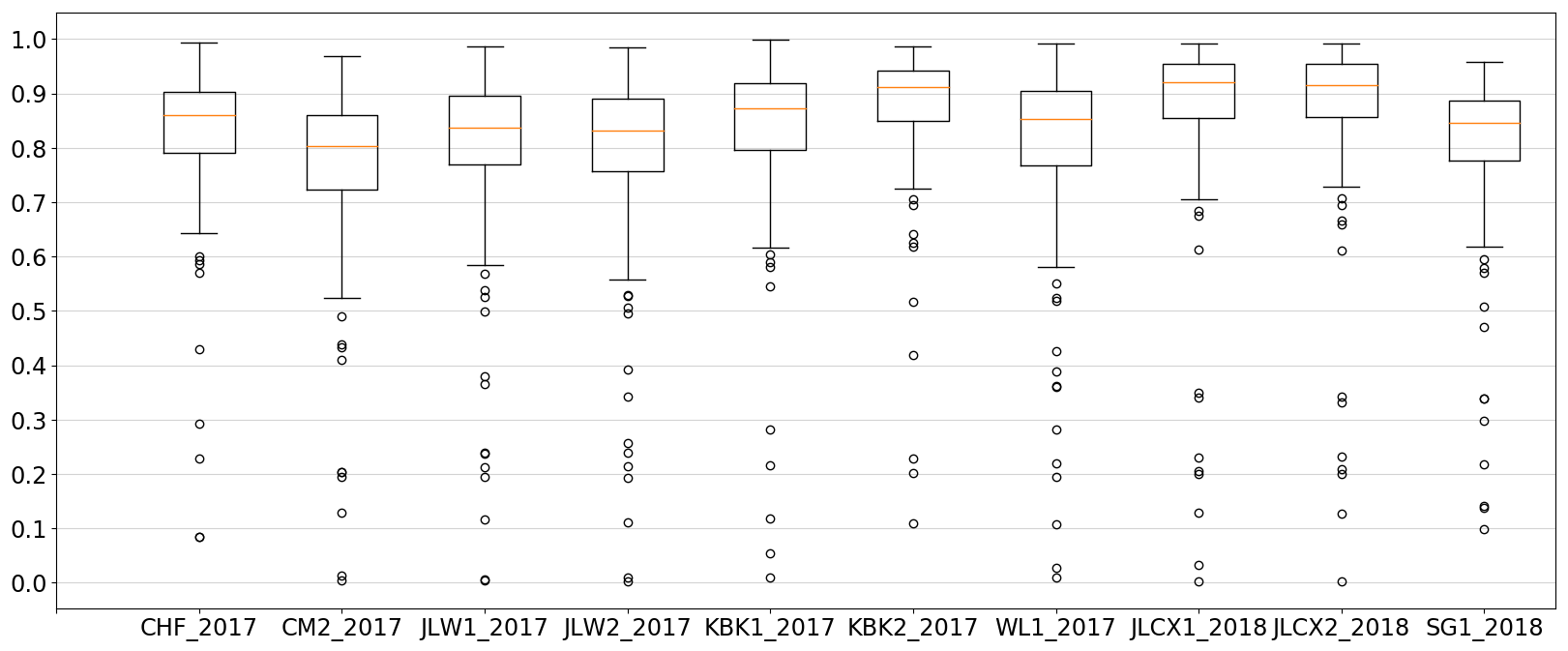}
    \caption{Chord Symbol Recall}
  \end{subfigure}
  \begin{subfigure}[b]{0.65\textwidth}
  	\includegraphics[width=\textwidth]{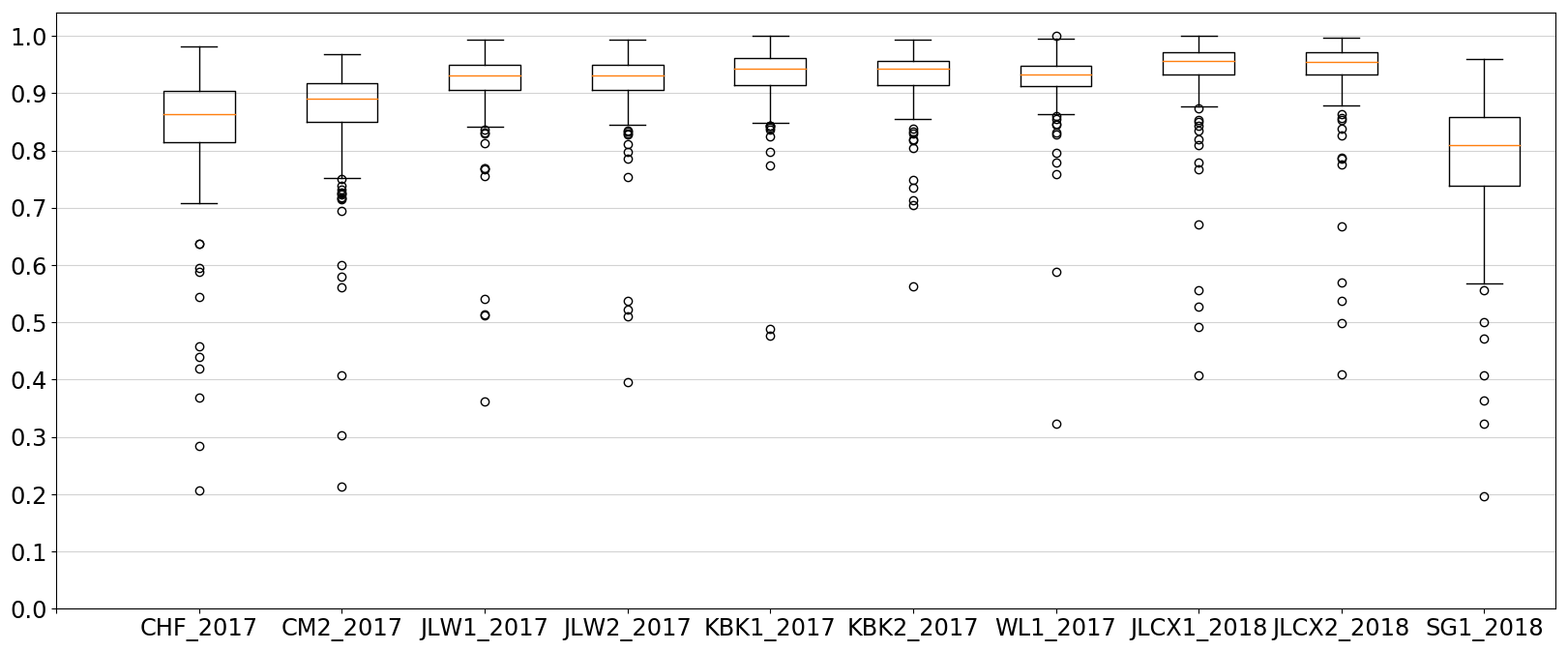}
    \caption{Oversegmentation}
  \end{subfigure}
  \begin{subfigure}[b]{0.65\textwidth}
  	\includegraphics[width=\textwidth]{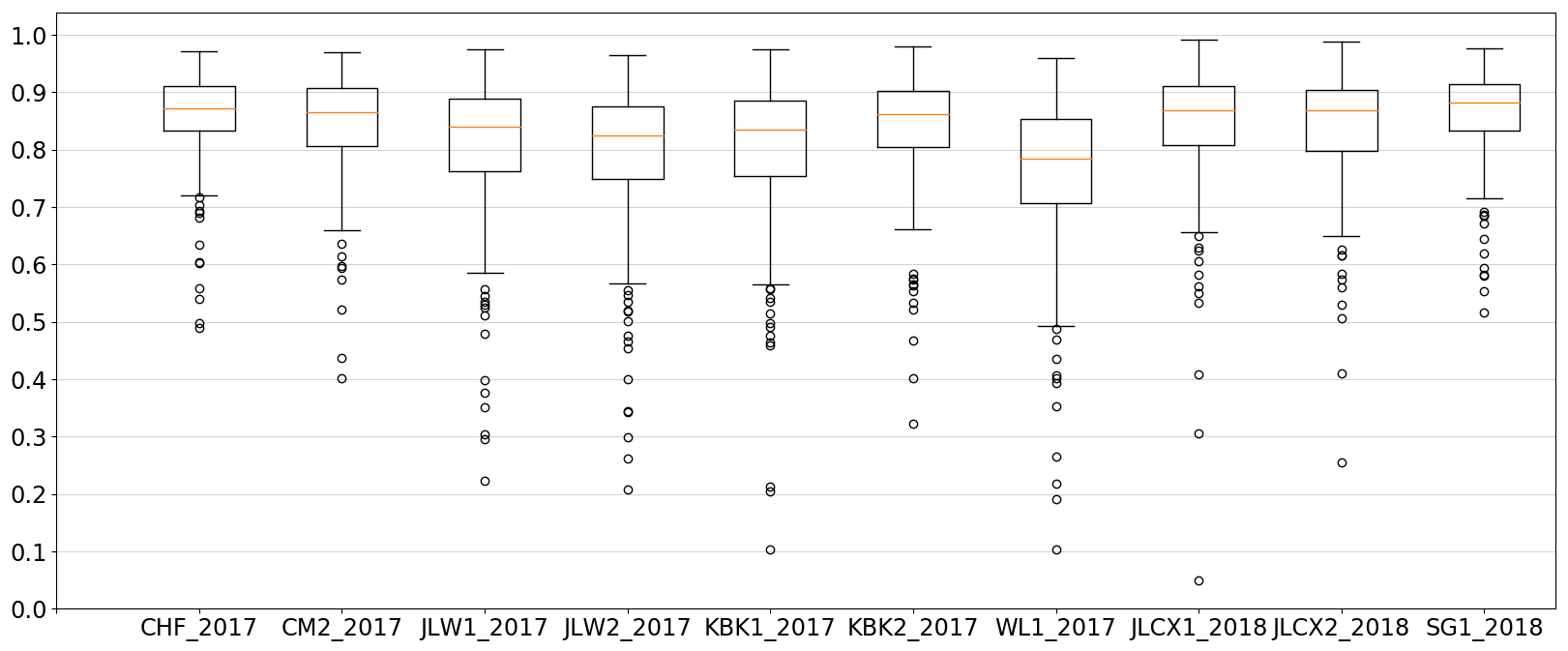}
    \caption{Undersegmentation}
  \end{subfigure}
  \begin{subfigure}[b]{0.65\textwidth}
  	\includegraphics[width=\textwidth]{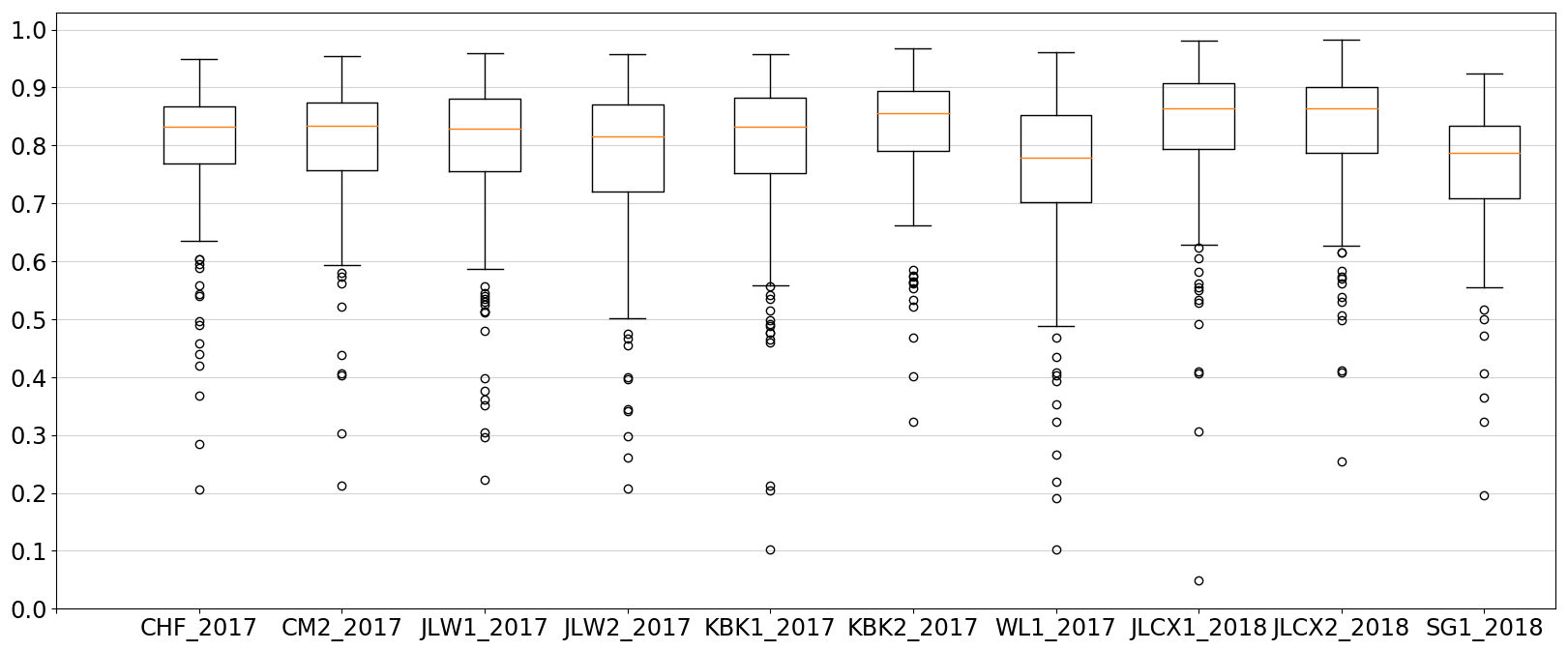}
    \caption{Segmentation}
  \end{subfigure}
  \caption{Comparison of performance measures for Chordify (\textsc{chf}) and \textsc{mirex} 2017/2018 \textsc{ace} submissions. For a description of these performance measures, see Section~\ref{sec:performance-evaluation}. \textsc{Kbk2} is the best performing system in terms of \textsc{csr}. \textsc{Chf} and \textsc{cm2} tend to oversegment the song. On the other hand, \textsc{wl1} inclines to undersegmentation.}\label{fig:audio-ace-compare}
\end{figure}

In this section, we evaluate the ten audio \textsc{ace} systems, of which we described the implementation in the previous section, on our data set\footnote{Note that the results are not exactly the same as the \textsc{mirex} evaluation on the Isophonics data set, as we use a subset of this data set, consisting of the Beatles and Queen songs.}. Figure~\ref{fig:audio-ace-compare} shows the performance of all ten systems in terms of \textsc{csr} and segmentation measures. From the figure we see that the \textsc{csr} of all ten systems is quite high, with a median value of at least 80\%. \textsc{jlcx2} is the best performing system, but given these high \textsc{csr} scores we can already conclude that all ten systems are state of the art. We also observe some outliers for all systems, which are songs for which the \textsc{csr} is very low. Some outliers are song-specific, that is: (almost) all audio \textsc{ace} systems perform badly on this song. Other outliers are specific for an \textsc{ace} system.

\begin{figure}[H]
	\centering
    \includegraphics[width=\textwidth]{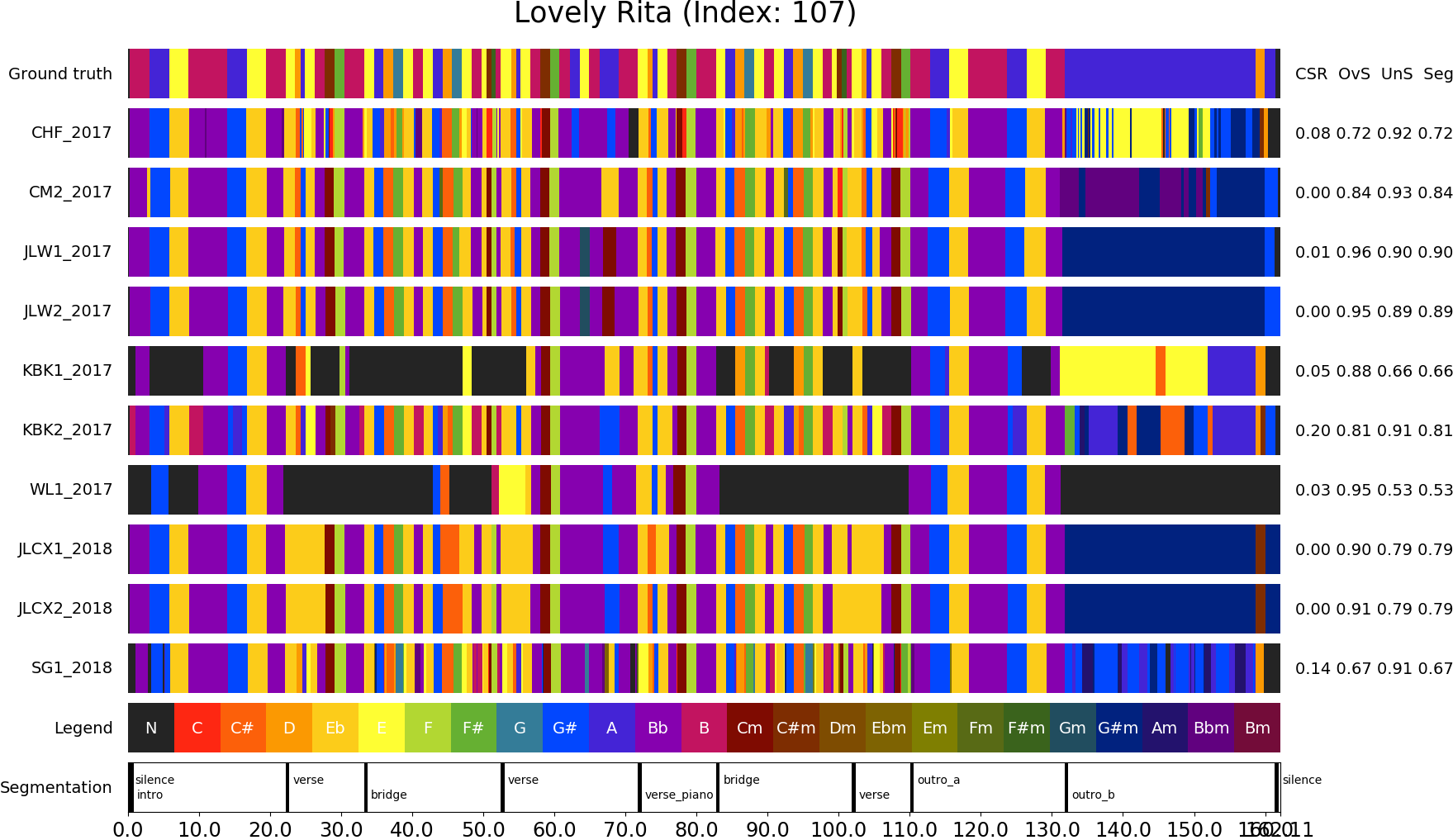}
    \caption{Chord sequences from ten different audio \textsc{ace} systems on the Beatles song \textit{Lovely Rita}, which has tuning issues. \textsc{Csr} = Chord Symbol Recall; OvS = oversegmentation; UnS = undersegmentation; and Seg = segmentation.}\label{fig:ace_lovely_rita}
\end{figure}

First, we will look at song-specific outliers, which have low \textsc{csr}s due to problems in the audio recording or in the ground truth labels. Some songs have \textbf{tuning issues}. This is for example the case in the song \textit{Lovely Rita} from the Beatles. The song was originally performed in E major, but during mixdown the tape machine ran at a lower frequency, resulting in a pitch drop of a quarter tone \citep{lewisohn1989beatles}. Without this background information, it would be very hard to decide if the song is in E major or in E$\flat$ major. From Figure~\ref{fig:ace_lovely_rita} it becomes clear that all ten audio \textsc{ace} systems choose the E$\flat$ major key: the chord labels of the \textsc{ace} systems are quite consistently one semitone lower than the ground truth chord labels. This results in very low \textsc{csr}s for all ten audio \textsc{ace} systems, ranging from 0\% (e.g.\ \textsc{jlw2}) to 20\% (e.g.\ \textsc{kbk2}). The Beatles song\textit{ Wild Honey Pie} has tuning issues as well, resulting in \textsc{csr}s from 0\% (\textsc{jlw1}) to 42\% (\textsc{kbk2}). \textit{Wild Honey Pie} is a short, experimental song, characterized by Indian influences, played with a lot of vibrato. The Queen song \textit{Another One Bites The Dust} has some tuning issues too, albeit to a lower extent than \textit{Lovely Rita} and \textit{Wild Honey Pie}.

\begin{figure}[H]
	\centering
    \includegraphics[width=\textwidth]{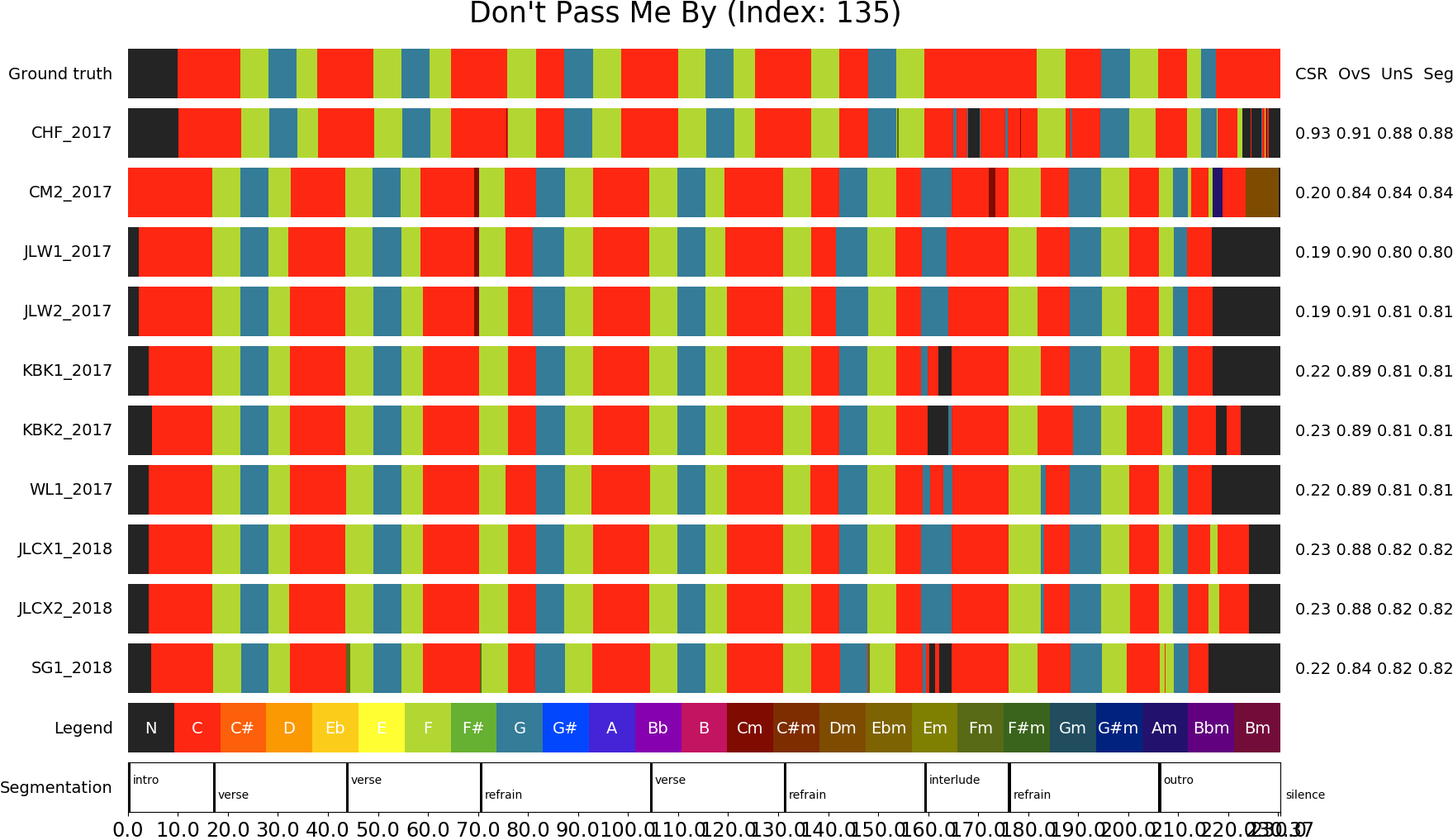}
    \caption{Chord sequences from ten different audio \textsc{ace} systems on the Beatles song \textit{Don't Pass Me By}, in which the chord labels are shifted in time. \textsc{Csr} = Chord Symbol Recall; OvS = oversegmentation; UnS = undersegmentation; and Seg = segmentation.}\label{fig:ace_dont_pass_me_by}
\end{figure}

In two other songs (\textit{The Continuing Story of Buffalo Bill} and \textit{Don't Pass Me By}), the chord labels of the nine \textsc{mirex} chord \textsc{ace} systems are \textbf{shifted} in time compared to the ground truth data, as shown in Figure~\ref{fig:ace_dont_pass_me_by}. This is probably due to the use of different audio versions in chord annotation and evaluation.

\begin{figure}[H]
	\centering
    \includegraphics[width=\textwidth]{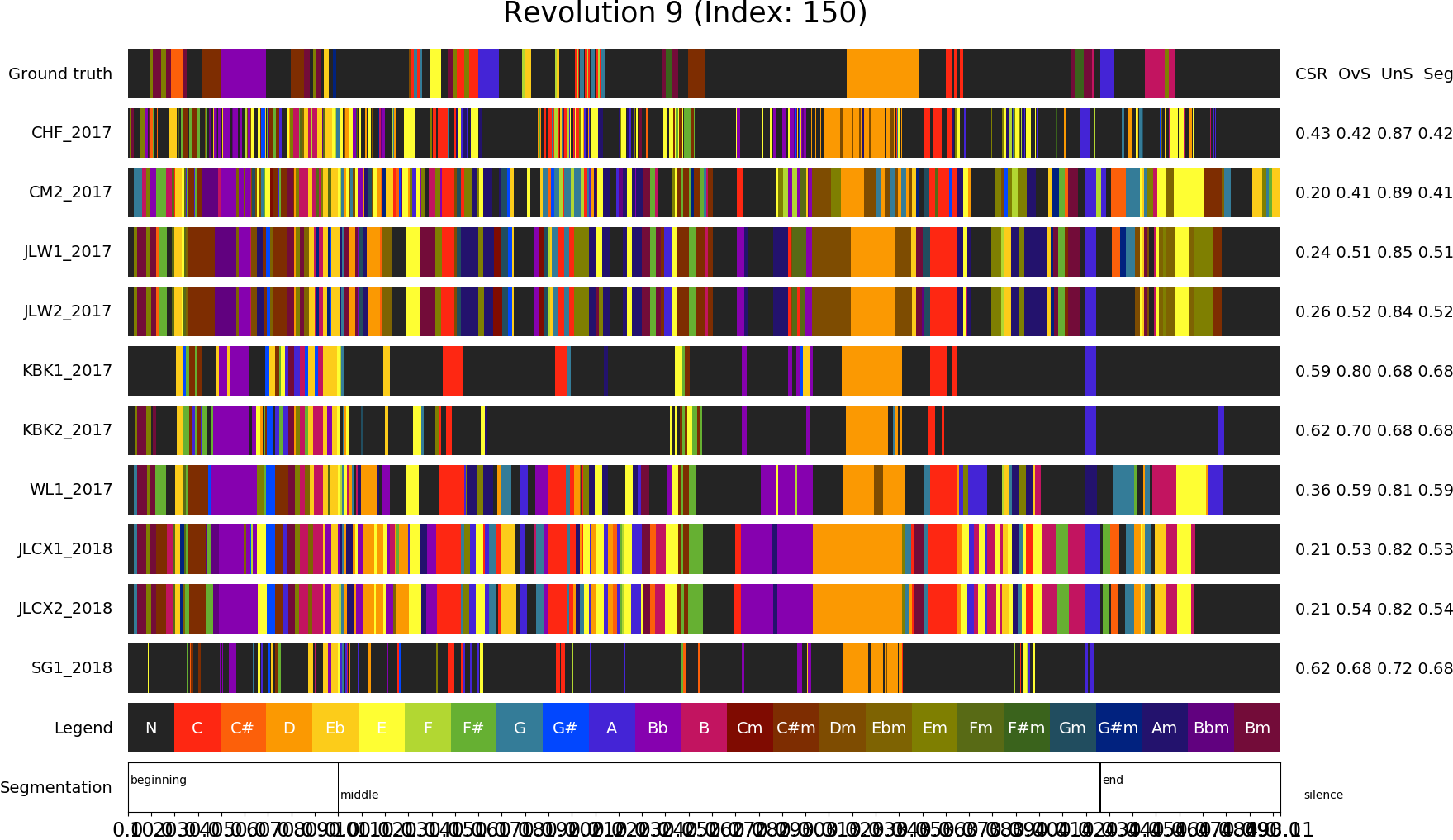}
    \caption{Revolution 9 is a sound collage, consisting of many non-harmonic sounds. \textsc{Csr} = Chord Symbol Recall; OvS = oversegmentation; UnS = undersegmentation; and Seg = segmentation.}\label{fig:revolution-9}
\end{figure}

The Beatles song \textit{Revolution} has another song-specific issue. This song is a sound collage: an experimental recording that is glued together from sound fragments, many of which are \textbf{non-harmonic}. Accordingly, we see in Figure~\ref{fig:revolution-9} that a large part of the ground truth is represented with black, corresponding to the no-chord symbol. However, most audio \textsc{ace} systems do not correctly recognize non-harmonic content as they are trained on tonal music.

\begin{figure}[H]
	\centering
    \includegraphics[width=\textwidth]{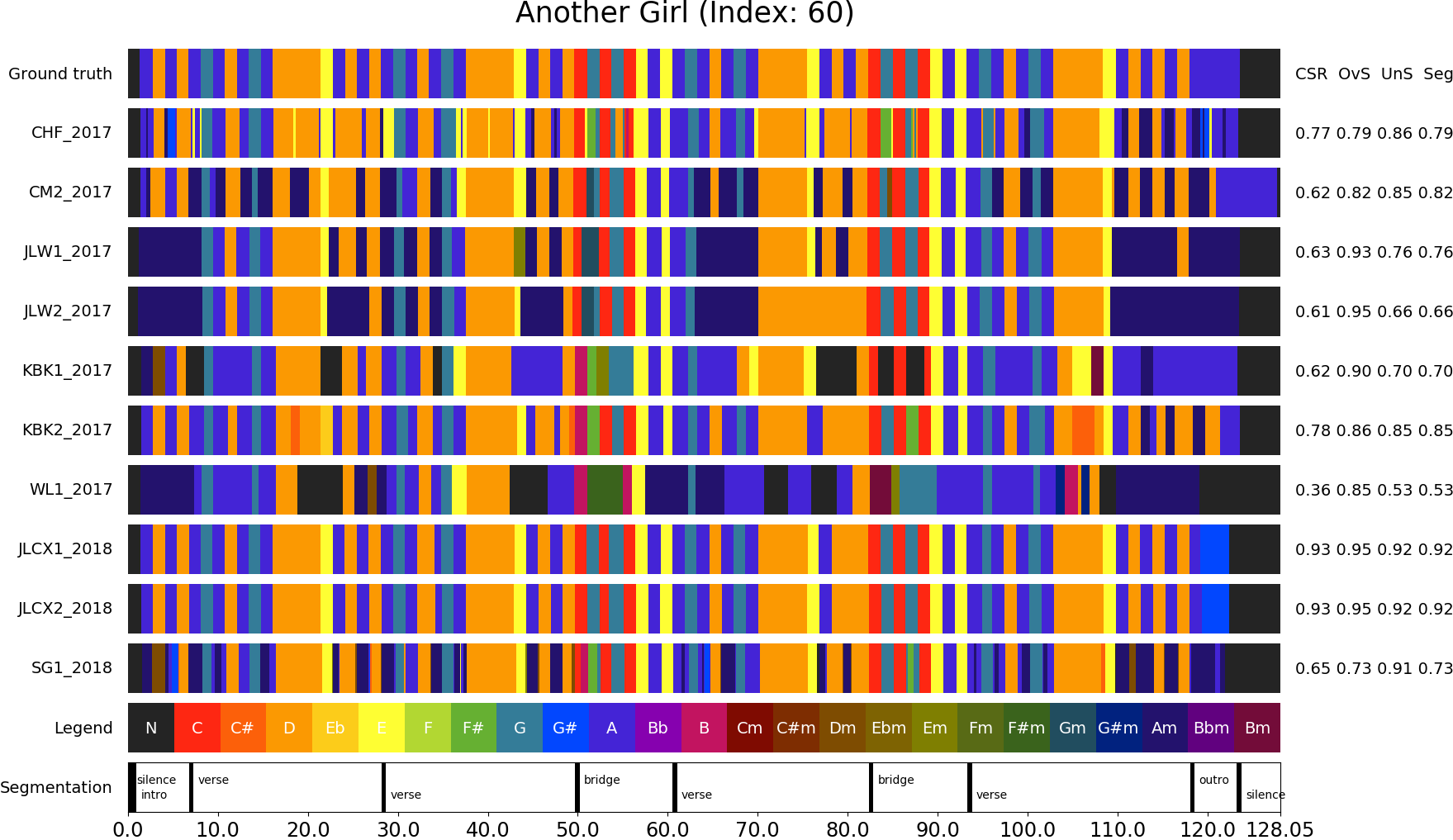}
    \caption{Chord sequences from ten different audio \textsc{ace} systems on the Beatles song \textit{Another Girl}. \textsc{Csr} = Chord Symbol Recall; OvS = oversegmentation; UnS = undersegmentation; and Seg = segmentation.}\label{fig:ace_audio_labels}
\end{figure}

Apart from song-specific problems (like tuning issues, shifted recordings and non-harmonic notes), there are also audio \textsc{ace}-specific issues. From Figure~\ref{fig:audio-ace-compare}a, we can see that \textsc{kbk2} performs the best, while \textsc{cm2} performs the worst in terms of \textsc{csr}. \textsc{Chf} and \textsc{cm2} tend to oversegment the song, considering the relatively low values for oversegmentation in Figure~\ref{fig:audio-ace-compare}b. On the other hand, \textsc{wl1} inclines to undersegmentation. We see this for example in Figure~\ref{fig:ace_audio_labels}: in this example we see that \textsc{chf} and \textsc{cm2} create too many segments and correspondingly get relatively low oversegmentation scores. On the other hand, \textsc{wl1} creates too few segments, resulting in undersegmentation: \textsc{wl1} here has a undersegmentation score of only 0.53.

\section{Conclusion}\label{sec:audio-ace-conclusion}
In this chapter, we have described the implementation and performance characteristics of ten systems for audio \textsc{ace}. By evaluating the audio \textsc{ace} systems on our 200 song data set, we learned that all ten systems are state of the art, as they have high chord symbol recalls. However, different \textsc{ace} systems behave differently: some systems tend to undersegment the chord sequence, while others tend to oversegmentation. In this study, we will use of each of these systems: we will compare the performance of the original audio \textsc{ace} system to the data fused result, which incorporates not only the audio \textsc{ace} system but also information from \textsc{midi} and tab files.

\chapter{Automatic Chord Estimation on \textsc{midi}}\label{ch:subsystem-midi}
In this chapter, we discuss the subsystem of \textsc{decibel} that detects an audio-timed chord sequence based on a \textsc{midi} file. This subsystem is illustrated below, in Figure~\ref{fig:diagram-midi}. In order to receive audio-timed chord labels from a \textsc{midi} file, \textsc{decibel} first finds an optimal alignment from the \textsc{midi} file to the audio file, realigns the \textsc{midi} file using this alignment and then uses a \textsc{midi} chord recognizer to estimate the chord labels on the realigned \textsc{midi} file. In Section~\ref{sec:midi-audio-align} we will explain the alignment step and in Section~\ref{sec:chord-est-midi} the \textsc{midi} chord recognition step. We then have audio-timed chord labels for each \textsc{midi} file. However, there may be \textsc{midi} files that are unsuitable to use, because they are bad transcriptions (i.e. the notes in the \textsc{midi} file are not consistent with the notes in the audio file), because they are in a wrong transposition or because they are a transcription of just one part of the song. It may help to exclude these \textsc{midi} files from the data fusion step. We experiment with \textsc{midi} selection methods in Section~\ref{sec:midi-selection} and conclude in Section~\ref{sec:midi-conclusion}. 

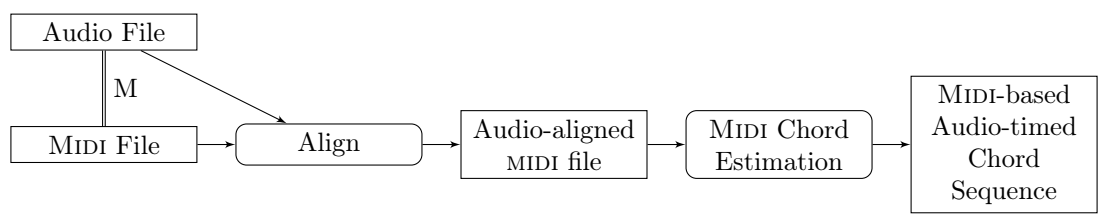
\begin{figure}   
\centering
\begin{tikzpicture}[node distance = 1cm and 0.5cm, auto]
    \node [rect] (audio_file) {Audio File};
    \node [rect, below=of audio_file] (midi_file) {\textsc{Midi} File};
    
    \node [block, right=of midi_file] (align) {Align};
    \node [rect, right=of align] (audio_aligned_midi) {Audio-aligned \textsc{midi} file};
    \node [block, right=of audio_aligned_midi] (midi_ace) {\textsc{Midi} Chord Estimation};
    \node [rect, right=of midi_ace] (midi_cs) {\textsc{Midi}-based Audio-timed Chord Sequence};
    
    \path [double_line] (audio_file) -- node[] {M} (midi_file);
    
    \path [line] (midi_file) -- (align);
    \path [line] (audio_file) -- (align);
    \path [line] (align) -- (audio_aligned_midi);
    \path [line] (audio_aligned_midi) -- (midi_ace);
    \path [line] (midi_ace) -- (midi_cs);
\end{tikzpicture}
\caption{Diagram of \textsc{decibel}'s \textsc{midi} subsystem; the M between Audio File and \textsc{midi} file indicates that they are manually matched.}
\label{fig:diagram-midi}
\end{figure}

\section{\textsc{Midi}-to-audio alignment}\label{sec:midi-audio-align}
Music alignment is the procedure which, given any position in one representation of a piece of music, determines the corresponding position within another representation. It is also called \textit{music synchronization}. In the \textsc{midi} subsystem, \textsc{decibel} aligns each of the \textsc{midi} representations to the corresponding audio representation. This section explains the \textsc{midi}-to-audio alignment procedure. \textsc{Midi}-to-audio alignment is an important task in \textsc{mir}: a lot of research has already been conducted on this subject. Therefore, we will first study related work on alignment in Section~\ref{sec:midialign}. We can then make a well-grounded choice for the alignment system implemented in \textsc{decibel}. This alignment system is described in Section~\ref{sec:alignment-selected-system} and evaluated in Section~\ref{sec:alignment-evaluation}. 

\subsection{Related work on \textsc{midi}-to-audio alignment}\label{sec:midialign}

There are two variants of \textsc{midi}-to-audio alignment: on-line and off-line alignment. In \textbf{off-line alignment}, the full performance is accessible for the alignment process. By contrast, in \textbf{on-line alignment}, the aligner processes the data in real-time as the signal is acquired. On-line alignment is also known as score following and has applications such as automatic accompaniment, audio editing and automatic turning of score pages. In general, off-line techniques work better than on-line systems, because they do not need to calculate the alignment in real-time and they can access the full performance at all times. For our project, the full performance is accessible from the beginning of the alignment process and the alignment does not have to be calculated in real-time. Therefore, we use off-line audio-symbolic alignment techniques.

The alignment method is typically based on either \textbf{statistical approaches} (for example HMMs, as used by e.g. \citet{cuvillier2014coherent}) or \textbf{Dynamic Time Warping} \citep{carabias2015audio, raffel2016optimizing, arzt2016flexible, lajugie2016weakly}. In this project, we focus on Dynamic Time Warping (\textsc{dtw}) approaches, as the \textsc{dtw} algorithm is efficient, is conceptually simple and calculates an easily interpretable alignment confidence score.

In the remainder of this section, we first give a general introduction to \textsc{dtw}. Subsequently, we see how \textsc{dtw} can be applied to alignment of music.

\subsubsection{An introduction to Dynamic Time Warping}
In \textsc{dtw} \citep{sakoe1978dynamic} we try to find the optimal alignment path between two sequences of feature vectors $X \in \mathbb{R}^{M \times D}$ and $Y \in \mathbb{R}^{N \times D}$, in which $M$ and $N$ are the lengths of vectors $X$ and $Y$ respectively and $D$ is the dimension of the feature vector. For example: for chroma vectors, $D = 12$. An alignment path is defined as two nondecreasing sequences $p \in \mathbb{N}^L$ and $q \in \mathbb{N}^L$, that satisfy the following three conditions:
\begin{itemize}
\item Boundary condition: $p[1] = 1$, $q[1] = 1$, $p[L] = N$ and $q[L] = M$;
\item Monotonicity condition: $p[1] \leq p[2] \leq \ldots \leq p[L]$ and $q[1] \leq q[2] \leq \ldots \leq q[L]$;
\item Step size condition: $(p[l + 1] - p[l], q[l+1] - q[l]) \in \{(1,1), (1,0), (0,1)\}$ for $l \in [1, L - 1]$.
\end{itemize}

In the alignment path $p, q$, $p[i] = n$ and $q[i] = m$ implies that the $n$\textsuperscript{th} feature vector in $X$ is matched to the $m$\textsuperscript{th} feature vector in $Y$. The optimal alignment path is the path with the lowest total cost, in which the total cost is defined as the sum of local costs $c$ between each pair of features $(X[p[i]], Y[q[i]])$ on the alignment path. Finding the optimal path $p, q$ can thus be defined as the following minimization problem:
\begin{equation}\label{eq:dtw}
p, q = \operatorname*{argmin}_{p,q} \sum_{i=1}^L c(X[p[i]], Y[q[i]])
\end{equation}

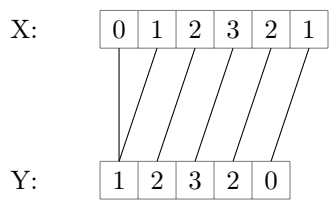
\begin{figure}[ht]
\centering
\begin{tikzpicture}
\draw[step=.5cm, gray, thin] (-0.0001,1.9999) grid (3, 2.5);
\draw[step=.5cm, gray, thin] (-0.0001,-0.0001) grid (2.5, 0.5);
\node at(-1, 2.25) {X:};
\node at(-1, 0.25) {Y:};
\node at (0.25, 2.25) {0}; \node at (0.75, 2.25) {1}; \node at (1.25, 2.25) {2}; \node at (1.75, 2.25) {3}; \node at (2.25, 2.25) {2}; \node at (2.75, 2.25) {1};
\node at (0.25, 0.25) {1}; \node at (0.75, 0.25) {2}; \node at (1.25, 0.25) {3}; \node at (1.75, 0.25) {2}; \node at (2.25, 0.25) {0};
\draw (0.25, 2) -- (0.25, 0.5); \draw (0.75, 2) -- (0.25, 0.5); \draw (1.25, 2) -- (0.75, 0.5); \draw (1.75, 2) -- (1.25, 0.5); \draw (2.25, 2) -- (1.75, 0.5); \draw (2.75, 2) -- (2.25, 0.5); 
\end{tikzpicture}
\caption{Alignment of X~=~{[0,~1,~2,~3,~2,~1]} and Y~=~{[1,~2,~3,~2,~0]}}\label{fig:dtw-toy-example}
\end{figure}

As an example, consider two sequences X = {[0, 1, 2, 3, 2, 1]} and Y = {[1, 2, 3, 2, 0]} and define the cost function as the absolute distance, i.e. $c(X[p[i]], Y[q[i]]) = \vert X[p[i]] - Y[q[i]] \vert$. The optimal alignment path is illustrated in Figure~\ref{fig:dtw-toy-example}. In this path, the first two elements of X are aligned to the first element of Y; the third element of X is aligned to the second element of Y; the fourth element of X is aligned to the third element of Y, etcetera. Therefore, the optimal path is $p, q = {[1, 2, 3, 4, 5, 6]}, {[1, 1, 2, 3, 4, 5]}$. This path has length $\sum_{i=1}^6 c(X[p[i]], Y[q[i]]) = c(0, 0) + c(1, 0) + c(2, 2) + c(3, 3) + c(2, 2) + c(1, 0) = 0 + 1 + 0 + 0 + 0 + 1 = 2$.

To determine the optimal alignment path, one could compute the total cost of all possible warping paths and then return the path with the lowest cost. Unfortunately this is not feasible for large sequences, because the number of possible warping paths is exponential in the length of the sequences. Luckily, there exist other methods to find the optimal alignment path. 

Consider a path $p_n, q_m$ with length $k$ that is a prefix of the alignment path. That is: $p_n, q_m$ fulfills all requirements of the alignment path, except for the last part of the boundary criterion. So:
\begin{itemize}
\item $p_n[1] = 1$, $q_m[1] = 1$, $p_n[k] = \textbf{n} \in [1, N]$ and $q_m[k] = \textbf{m} \in [1, M]$; 
\item $p_n[1] \leq p_n[2] \leq \ldots \leq p_n[k]$ and $q_m[1] \leq q_m[2] \leq \ldots \leq q_m[k]$; 
\item $(p_n[l + 1] - p_n[l], q_m[l+1] - q_m[l]) \in \{(1,1), (1,0), (0,1)\}$ for $l \in [1, k - 1]$.
\end{itemize}

Now note that, thanks to the step size condition, any path $p_n, q_m$ starts with either $p_{n - 1}, q_{m - 1}$; $p_{n}, q_{m - 1}$ or $p_{n - 1}, q_{m}$, followed by one additional step. Therefore, if we know the costs of the optimal $p_{n - 1}, q_{m - 1}$; $p_{n}, q_{m - 1}$ and $p_{n - 1}, q_{m}$, we can derive the cost of the optimal path $p_n, q_m$ in constant time, using the following equation:
\begin{equation}\label{eq:dtw-1}
\begin{split}
\text{cost}(p_n, q_m) & = \operatorname*{min}_{p_n,q_m} \sum_{i=1}^k c(X[p_n[i]], Y[q_m[i]])\\
 & = c(X[p_n[l]], Y[q_m[l]]) + \operatorname*{min}\begin{cases}
\sum_{i=1}^{k-1} c(X[p_{n-1}[i]], Y[q_{m-1}[i]])\\
\sum_{i=1}^{k-1} c(X[p_n[i]], Y[q_{m-1}[i]])\\
\sum_{i=1}^{k-1} c(X[p_{n-1}[i]], Y[q_m[i]])
\end{cases}
\end{split}
\end{equation}

All paths $p_n, q_1$ have the special property that they must exclusively consist of horizontal steps, i.e. $(p_n[l + 1] - p_n[l], q_1[l+1] - q_1[l]) = (1, 0)$ for $l \in [1, k - 1]$. Therefore, we can calculate their  cost without minimization: 
\begin{equation}\label{eq:dtw-2}
\begin{split}
\text{cost}(p_n, q_1) & = \operatorname*{min}_{p_n,q_1} \sum_{i=1}^k c(X[p_n[i]], Y[q_1[i]])\\
 & = \sum_{i=1}^{n} c(X[p[i]], Y[q[1]])
\end{split}
\end{equation}

Likewise, we do need to minimize for calculating the cost of any path $p_1, q_m$:
\begin{equation}\label{eq:dtw-3}
\begin{split}
\text{cost}(p_1, q_m) & = \operatorname*{min}_{p_1,q_m} \sum_{i=1}^k c(X[p_1[i]], Y[q_m[i]])\\
 & = \sum_{i=1}^{m} c(X[p[1]], Y[q[i]])
\end{split}
\end{equation}

Also, note that the optimal path $p_n, q_m$ with $n = N$ and $m = M$ equals the optimal full alignment path $p, q$. 

The Dynamic Time Warping algorithm uses Equations~\ref{eq:dtw-1}, \ref{eq:dtw-2} and \ref{eq:dtw-3} to efficiently compute the optimal full alignments path $p, q$ and its cost. Basically, the algorithm computes the cost of any path $p_n, p_m$ with $n \in [1, N]$ and $m \in [1, M]$ in a convenient order and stores the result. This way, each of these $(N \times M)$ costs is computed exactly once, so the algorithm runs in $\mathcal{O}(N \times M)$ time.

The pseudocode of the \textsc{dtw} algorithm is given in Algorithm~\ref{alg:dtw-pseudocode}. The algorithm takes a cost matrix $C$ with $C[n,m] = c(X[n], Y[m])$. Any entry $D[n, m]$ contains the cost of the optimal path $p_n, q_m$, so after running the algorithm $D[N, M]$ contains the minimum cost of $p, q$. We can find the optimal alignment path $p, q$ by following the arrows backwards from $P[N, M]$.

\begin{algorithm}
\begin{algorithmic}[1]
\Function{\textsc{dtw}}{Cost Matrix $C$ of size $N \times M$}
  \State Initialize Accumulated cost Matrix $D$ of size $N \times M$
  \State Initialize Path Matrix $P$ of size $N \times M$.
  \State $D[1, 1] = C[1, 1]$
  \State $P[1, 1] = \cdot$
  \ForAll{$n \in [2, N]$}
  	\State $D[n, 1] = C[n, 1] + D[n - 1, 1]$
    \State $P[n, 1] = \leftarrow$
  \EndFor
  \ForAll{$m \in [2, M]$}
    \State $D[1, m] = C[1, m] + D[1, m - 1]$
    \State $P[1, m] = \downarrow$
  \EndFor
  
  \ForAll{$n \in [2, N]$}
      \ForAll{$m \in [2, M]$}
      	\State $s = \operatorname*{min}{D[n - 1, m - 1], D[n, m - 1], D[n - 1, m]}$
        \If{$s \equiv D[n - 1, m - 1]$}
          \State $D[n, m] = C[n, m] + D[n - 1, m - 1]$
          \State $P[n, m] = \swarrow$
        \ElsIf{$s \equiv D[n, m - 1]$}
          \State $D[n, m] = C[n, m] + D[n, m - 1]$
          \State $P[n, m] = \downarrow$
        \Else
          \State $D[n, m] = C[n, m] + D[n - 1, m]$
          \State $P[n, m] = \leftarrow$
        \EndIf
     \EndFor
  \EndFor
\EndFunction
\end{algorithmic}
\caption{Pseudocode of the \textsc{dtw} algorithm}\label{alg:dtw-pseudocode}
\end{algorithm}

Let us now test the \textsc{dtw} algorithm on our example X~=~{[0,~1,~2,~3,~2,~1]} and Y~=~{[1,~2,~3,~2,~0]} from Figure~\ref{fig:dtw-toy-example}. First, we calculate the cost matrix $C$. Remember that in our example, $c$ is defined as the absolute difference. For example, $C[4, 1] = c(X[4], Y[1]) = \vert X[4] - Y[1] \vert = \vert 3 - 1 \vert = 2$. The resulting cost matrix $C$ is given in Figure~\ref{fig:dtw-output-toy-example}a. Figures~\ref{fig:dtw-output-toy-example}b and c show the accumulated cost matrix $D$ and the path matrix $P$ respectively, which are calculated by the algorithm. We indeed find the optimal alignment path ($p, q = {[1, 2, 3, 4, 5, 6]}, {[1, 1, 2, 3, 4, 5]}$) by following the arrows back from $P[N, M]$ and read from $D[N, M]$ that its  total cost truly equals 2.

\begin{figure}
\centering
\begin{subfigure}[b]{0.3\textwidth}
  \centering
  \begin{tikzpicture}
  \matrix [matrix of math nodes, every node/.style={anchor=center,text depth=.5ex,text height=2ex,text width=1em, draw=lightgray}]
  {
  \textbf{0} & 0 & 1 & 2 & 3 & 2 & 1\\
  \textbf{2} & 2 & 1 & 0 & 1 & 0 & 1\\
  \textbf{3} & 3 & 2 & 1 & 0 & 1 & 2\\
  \textbf{2} & 2 & 1 & 0 & 1 & 0 & 1\\
  \textbf{1} & 1 & 0 & 1 & 2 & 1 & 0\\
    & \textbf{0} & \textbf{1} & \textbf{2} & \textbf{3} & \textbf{2} & \textbf{1}\\
  };
  \end{tikzpicture}
  \caption{Cost matrix $C$}
\end{subfigure}
\begin{subfigure}[b]{0.35\textwidth}
  \centering
  \begin{tikzpicture}
  \matrix [matrix of math nodes, every node/.style={anchor=center,text depth=.5ex,text height=2ex,text width=1em, draw=lightgray}]
  {
  \textbf{0} & 8 & 6 & 4 & 5 & 3 & 2\\
  \textbf{2} & 8 & 5 & 2 & 2 & 1 & 2\\
  \textbf{3} & 6 & 4 & 2 & 1 & 2 & 4\\
  \textbf{2} & 3 & 2 & 1 & 2 & 2 & 3\\
  \textbf{1} & 1 & 1 & 2 & 4 & 5 & 5\\
    & \textbf{0} & \textbf{1} & \textbf{2} & \textbf{3} & \textbf{2} & \textbf{1}\\
  };
  \end{tikzpicture}
  \caption{Accumulated cost matrix $D$}
\end{subfigure}
\begin{subfigure}[b]{0.3\textwidth}
  \centering
  \begin{tikzpicture}
  \matrix [matrix of math nodes, every node/.style={anchor=center,text depth=.5ex,text height=2ex,text width=1em, draw=lightgray}]
  {
  \textbf{0} & $\downarrow$ & $\downarrow$ & $\downarrow$ & $\swarrow$ & $\downarrow$ & $\swarrow$\\
  \textbf{2} & $\downarrow$ & $\downarrow$ & $\downarrow$ & $\downarrow$ & $\swarrow$ & $\leftarrow$\\
  \textbf{3} & $\downarrow$ & $\downarrow$ & $\downarrow$ & $\swarrow$ & $\leftarrow$ & $\leftarrow$\\
  \textbf{2} & $\downarrow$ & $\swarrow$ & $\swarrow$ & $\leftarrow$ & $\leftarrow$ & $\leftarrow$\\
  \textbf{1} & $\cdot$ & $\leftarrow$ & $\leftarrow$ & $\leftarrow$ & $\leftarrow$ & $\leftarrow$\\
    & \textbf{0} & \textbf{1} & \textbf{2} & \textbf{3} & \textbf{2} & \textbf{1}\\
  };
  \end{tikzpicture}
  \caption{Path matrix $P$}
\end{subfigure}
\caption{Cost matrix and \textsc{dtw} output for the alignment of our toy example X~=~{[0,~1,~2,~3,~2,~1]} and Y~=~{[1,~2,~3,~2,~0]}}\label{fig:dtw-output-toy-example}
\end{figure}

\subsubsection{Dynamic Time Warping in music alignment}
In the previous subsection, we saw an example of \textsc{dtw} on a one-dimensional signal. When aligning two music representations, the \textbf{features} typically have more dimensions. For example, chroma features would be a sensible choice: they can be calculated from both the audio and the synthesized \textsc{midi}. In that case, the features are twelve-dimensional. Chroma features are for example used for alignment by \citet{hu2003polyphonic}. \citet{pratzlich2016triple} and \citet{wang2014robust} also use chroma based-features, but those are combined with features capturing onset information. Other methods, those designed by \citet{dixon2005match} and \citet{raffel2015large}, use the result from a Constant-Q transform.

As we have seen in Section~\ref{sec:musicrepresentations-audio}, we need to set a \textbf{time scale} for the feature vectors. Some methods, for example \citet{dixon2005match}, compute feature vectors over short, overlapping frames of audio. Other methods, like \citet{raffel2015large} use beat-synchronous feature vectors, obtained by aggregating the original feature vectors between two beats.

Multidimensional features require an appropriate \textbf{cost function}. A common choice (used by e.g. \citet{turetsky2003ground,wang2014robust,raffel2015large}) is the cosine distance. The cosine distance between two $d$-dimensional features $x$ and $y$ is defined as:
\begin{equation}
c(x,y) = 1 - \frac{\left\langle x \vert y \right\rangle}{\vert \vert x \vert \vert \cdot \vert \vert y \vert \vert} = 1 - \frac{\sum_{i=1}^d x_i \cdot y_i}{\sqrt[]{\sum_{i=1}^d x_i^2} \cdot \sqrt[]{\sum_{i=1}^d y_i^2}}
\end{equation}

Some methods use a \textbf{penalty} to discourage non-diagonal moves \citep{raffel2015large}. That is, Equation~\ref{eq:dtw} is replaced by Equation~\ref{eq:dtw-penalty}, in which $\Phi(i) \leq 0$ if $p[i] = p[i-1]$ or $q[i] = q[i-1]$.
\begin{equation}\label{eq:dtw-penalty}
p, q = \operatorname*{argmin}_{p,q} \sum_{i=1}^L c(X[p[i]], Y[q[i]]) + \Phi(i)
\end{equation}

In music alignment, it is not always necessary that the alignment path $p, q$ spans the entirety of feature vectors $X$ and $Y$. For instance, sometimes we want to match only a part of a \textsc{midi} file to the audio recording. In order to enable subsequence matching, we drop the boundary condition ($p[1] = 1$, $q[1] = 1$, $p[L] = N$ and $q[L] = M$) and relax it to the more flexible condition that either $g \cdot N \leq p[L] \leq N$ or $g \cdot M \leq q[L] \leq M$. $g$ is the \textbf{gully} parameter and determines the proportion of the subsequence that must be successfully matched. \cite{raffel2015large} for example choose a gully of 0.95, allowing for some tolerance that the beginning or ending of a \textsc{midi} file is incorrect.

\begin{figure}[ht]
  \centering
  \begin{subfigure}[b]{0.3\textwidth}
  \centering
  \begin{tikzpicture}
  \draw[step=0.25,black,very thin] (0,0) grid (3,3);
  \foreach \x in {0,0.25,...,2.75}
  { \filldraw[fill=blue!25!white, draw=black] (\x,\x) rectangle (\x + 0.25,\x + 0.25); }
  \foreach \x in {0,0.25,...,2.5}
  { \filldraw[fill=blue!25!white, draw=black] (\x,\x + 0.25) rectangle (\x + 0.25,\x + 0.5);
    \filldraw[fill=blue!25!white, draw=black] (\x + 0.25,\x) rectangle (\x + 0.5,\x + 0.25);}
  \foreach \x in {0,0.25,...,2.25}
  { \filldraw[fill=blue!25!white, draw=black] (\x,\x + 0.5) rectangle (\x + 0.25,\x + 0.75);
    \filldraw[fill=blue!25!white, draw=black] (\x + 0.5,\x) rectangle (\x + 0.75,\x + 0.25);}
  \foreach \x in {0,0.25,...,2}
  { \filldraw[fill=blue!25!white, draw=black] (\x,\x + 0.75) rectangle (\x + 0.25,\x + 1);
    \filldraw[fill=blue!25!white, draw=black] (\x + 0.75,\x) rectangle (\x + 1,\x + 0.25);} 
  \end{tikzpicture}
  \caption{Sakoe-Chiba band}\label{fig:sakoe-chiba-band}
  \end{subfigure}
  \begin{subfigure}[b]{0.3\textwidth}
  \centering
  \begin{tikzpicture}
  \draw[step=0.25,black,very thin] (0,0) grid (3,3);
  \foreach \x in {0,0.25,...,2.75}
  { \filldraw[fill=blue!25!white, draw=black] (\x,\x) rectangle (\x + 0.25,\x + 0.25); }
  \foreach \x in {0.25,0.5,...,2.25}
  { \filldraw[fill=blue!25!white, draw=black] (\x,\x + 0.25) rectangle (\x + 0.25,\x + 0.5);
    \filldraw[fill=blue!25!white, draw=black] (\x + 0.25,\x) rectangle (\x + 0.5,\x + 0.25);}
  \foreach \x in {0.5,0.75,...,1.75}
  { \filldraw[fill=blue!25!white, draw=black] (\x,\x + 0.5) rectangle (\x + 0.25,\x + 0.75);
    \filldraw[fill=blue!25!white, draw=black] (\x + 0.5,\x) rectangle (\x + 0.75,\x + 0.25);}
  \foreach \x in {0.75,1,...,1.25}
  { \filldraw[fill=blue!25!white, draw=black] (\x,\x + 0.75) rectangle (\x + 0.25,\x + 1);
    \filldraw[fill=blue!25!white, draw=black] (\x + 0.75,\x) rectangle (\x + 1,\x + 0.25);} 
  \end{tikzpicture}
  \caption{Itakura parallelogram}\label{fig:itakura-parallelogram}
  \end{subfigure}
\caption{Global constraint regions}
\end{figure}
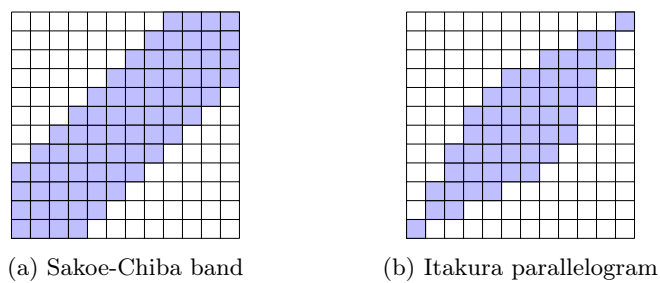

To conclude, some methods place \textbf{global constraints} \citep[Chapter~3]{muller2015fundamentals} on the alignment path. These global constraints reduce complexity of the \textsc{dtw} algorithm and prevent paths that diverge too much from the diagonal line. A global constraint only allows points on the warping path inside the global constraint region $R \subseteq [1 : N] \times [1 : M]$. Two well-known constraints are the Sakoe-Chiba band \citep{sakoe1978dynamic} and the Itakura parallelogram \citep{itakura1975minimum}. The Sakoe-Chiba band constrains the path to lie within a fixed distance of the diagonal, as shown in Figure~\ref{fig:sakoe-chiba-band}. The Itakura parallelogram is illustrated in Figure~\ref{fig:itakura-parallelogram} and constrains the path to lie within a parallelogram around the diagonal of the matrix.

\subsection{Selected system}\label{sec:alignment-selected-system}
As we have seen in Section~\ref{sec:midialign}, Dynamic Time Warping (\textsc{dtw}) is a common technique to align two feature vectors, for example two representations of the same song. We also learned that there are many variations of the \textsc{dtw} algorithm: one can use different features, time scales, cost functions, penalties, gully parameters and global constraints. In this section, we will explain the \textsc{midi}-to-audio alignment method we selected for \textsc{decibel}. 

For alignment between \textsc{midi} files and audio recordings, \textsc{decibel} uses a \textsc{dtw} algorithm by \citet{raffel2016optimizing}. We selected this algorithm for a couple of reasons. In the first place, it benefits from the two advantages of \textsc{dtw}: the algorithm calculates an easily interpretable alignment confidence score, which gives a good indication of the alignment quality. Furthermore, the algorithm is conceptually simple and easy to implement. Typically, the performance of \textsc{dtw} heavily relies on the chosen parameters. An advantage of this specific system is that the optimal parameter setting is trained on a synthetically trained data set of 1000 \textsc{midi} files which were transcriptions of Western popular music songs, using Bayesian optimization. 

\begin{figure}[p!]
 \centering
 \begin{subfigure}[b]{0.49\textwidth}
 	\includegraphics[width=\textwidth]{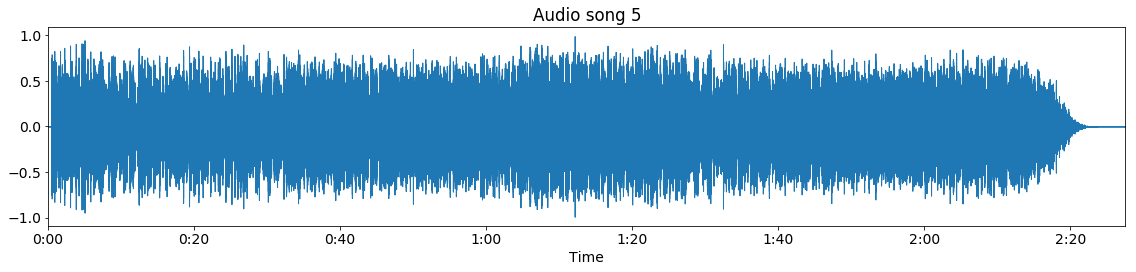}
    \caption{Audio waveform}\label{fig:alignment-procedure-a}
 \end{subfigure}
 \begin{subfigure}[b]{0.49\textwidth}
 	\includegraphics[width=\textwidth]{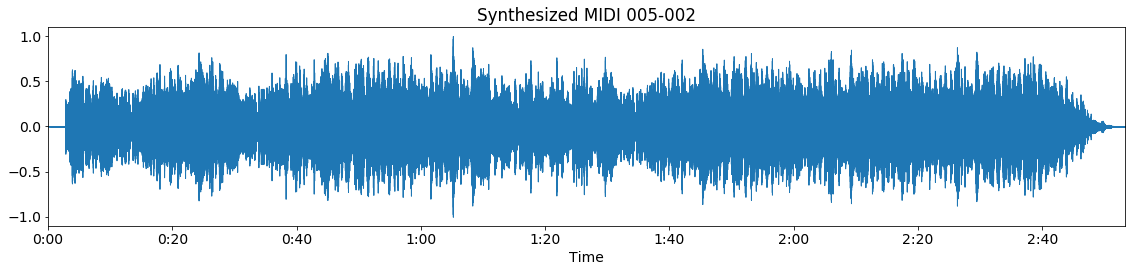}
    \caption{Synthesized \textsc{midi} waveform}\label{fig:alignment-procedure-b}
 \end{subfigure}
 \begin{subfigure}[b]{0.49\textwidth}
 	\includegraphics[width=\textwidth]{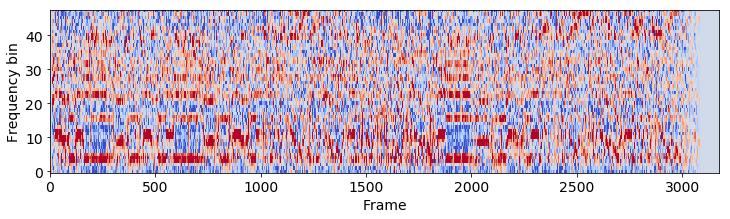}
    \caption{CQT of the audio waveform}\label{fig:alignment-procedure-c}
 \end{subfigure}
 \begin{subfigure}[b]{0.49\textwidth}
 	\includegraphics[width=\textwidth]{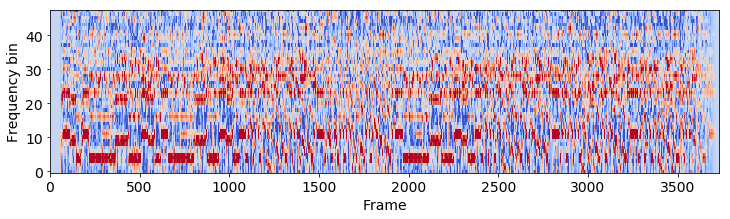}
    \caption{CQT of the synthesized \textsc{midi} waveform}\label{fig:alignment-procedure-d}
 \end{subfigure}
 \begin{subfigure}[b]{0.6\textwidth}
 	\includegraphics[width=\textwidth]{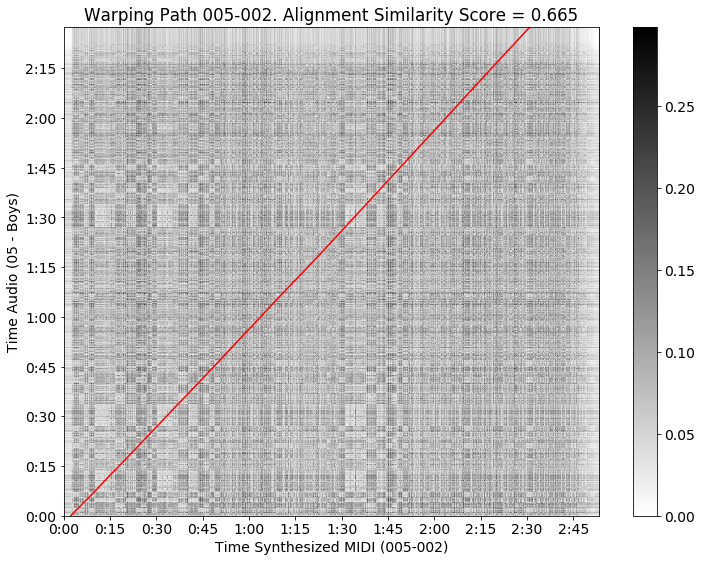}
    \caption{Alignment obtained by \textsc{dtw}}\label{fig:alignment-procedure-e}
 \end{subfigure}
 \begin{subfigure}[b]{0.7\textwidth}
 	\includegraphics[width=\textwidth]{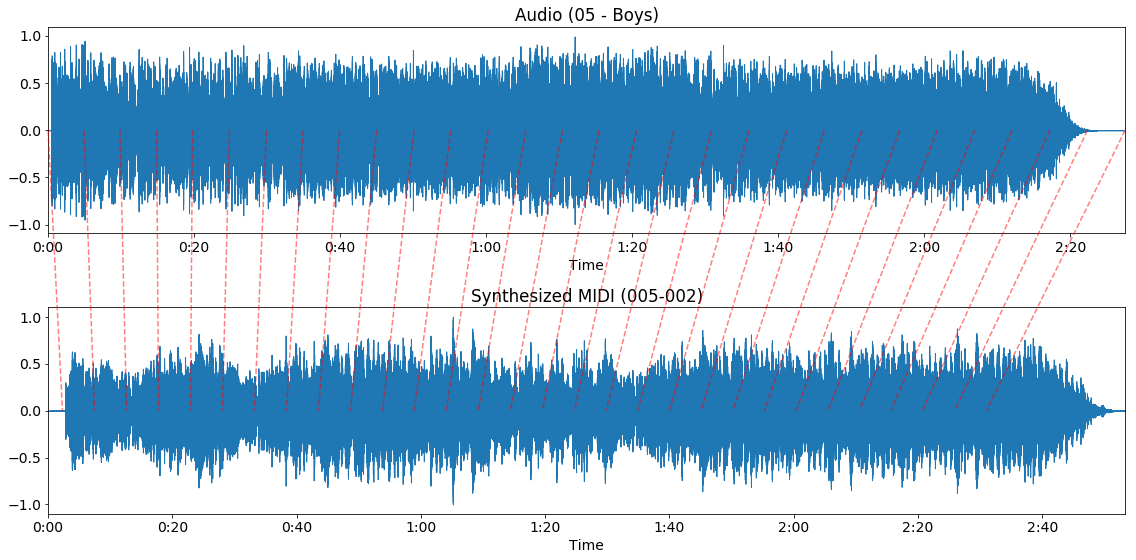}
    \caption{Alignment remapping}\label{fig:alignment-procedure-f}
 \end{subfigure}
 \caption{Illustration of the alignment procedure}
 \label{fig:alignment-procedure}
\end{figure}

Before we look at the specific parameter settings, let us first consider the \textbf{outline} of the algorithm. The outline is illustrated with an example in Figure~\ref{fig:alignment-procedure}. First, all \textsc{midi} files are synthesized using the fluidsynth\footnote{\url{http://www.fluidsynth.org/}} software synthesizer with the FluidR3\_GM soundfont. Now we have a waveform representation for both the audio (Figure~\ref{fig:alignment-procedure-a}) and the \textsc{midi} file (Figure~\ref{fig:alignment-procedure-b}). Note that our example \textsc{midi} file starts with silence, while in the audio recording the music starts immediately. Also, the \textsc{midi} file has a longer duration, as the \textsc{midi} file repeats the chorus an additional time, compared to the audio file. Then, the Constant-Q transform is calculated for both the audio (Figure~\ref{fig:alignment-procedure-c}) and the synthesized \textsc{midi} waveform (Figure~\ref{fig:alignment-procedure-d}). Features are found by aggregation over the Constant-Q transform vectors. Then, the optimal path between the audio file and the synthesized \textsc{midi} is calculated using \textsc{dtw}. This results in an optimal path and the alignment confidence score, as illustrated in Figure~\ref{fig:alignment-procedure-e}. In this figure, we see that the alignment path starts not in the coordinate (0, 0), but a bit to the right: the silence at the start of the \textsc{midi} file is not mapped to any position in the audio file. The same goes for the end of the \textsc{midi} file, which is a superfluous repetition of the chorus. Finally, this alignment path is used to remap the \textsc{midi} file to the audio recording (see Figure~\ref{fig:alignment-procedure-f}).

\begin{table}
\centering
\begin{tabular}{ll}
\textbf{Parameter}     & \textbf{Setting}                                        \\
Feature representation & log-magnitude Constant-Q transform                      \\
Time scale             & every 46 milliseconds                                   \\
Cost function          & cosine distance										 \\
Penalty                & median distance of all pairs of frames                  \\
Gully                  & 0.96                                                    \\
Band path constraint   & none                                                   
\end{tabular}
\caption{Parameter settings of audio-\textsc{midi} \textsc{dtw} alignment algorithm}\label{table:alignmentpars}
\end{table}

\textsc{Decibel} uses the unchanged \textbf{parameter setting} reported in the paper by \citet{raffel2016optimizing}. The parameters are listed in Table~\ref{table:alignmentpars}. We did not experiment with alternative parameter settings as the parameters were already optimized for a dataset in the same genre as mine. In the optimal parameter setting, the features are represented by log-magnitude Constant-Q transform. This is the Constant-Q transform we have seen in Section~\ref{sec:musicrepresentations-audio}, but the log of the features is calculated to mimic human perception more closely. An optimal hop size of 1024 samples is found, which corresponds to a time scale of 46 milliseconds, given the sampling rate of 22050 Hz: $\frac{1024}{22040} = 0.046$s. Feature vectors are normalized by L2 norm before calculating the local distances. This is equivalent to using the cosine distance. An penalty of the median distance between all pairs of frames is found to give the best results. In the optimal system, the gully parameter is 0.96 and there is no band path constraint.

The \textbf{output} of the \textsc{dtw} system is an optimal path and its alignment confidence score. The path specifies which time point, measured in seconds, in the \textsc{midi} file is aligned to which time point in the audio file. We use this path to recompute the times in the \textsc{midi} file using the pretty\_midi package~\citep{raffel2014intuitive}. This gives us the ``audio-aligned \textsc{midi} file'' we will use in the chord estimation step. The alignment confidence score is the mean  distance  between  all  pairs  of  frames  over  the  entire aligned portion of both feature sequences. A qualitative evaluation on 500 real word \textsc{midi}/audio pairs by \cite{raffel2016optimizing} shows that the alignment confidence score is a reliable measure for the quality of the alignment in most cases, although there were a few outliers which had a low (good) alignment confidence score despite being aligned badly. In general, \textsc{midi}/audio pairs with an alignment confidence score below 0.85 are aligned well. In the next section, we will do a similar evaluation for to verify if the alignment confidence score is a good indicator for the alignment quality in our data set.

\subsection{Evaluation of \textsc{midi}-to-audio alignment}\label{sec:alignment-evaluation}
\begin{figure}
\centering
\includegraphics[width=0.8\textwidth]{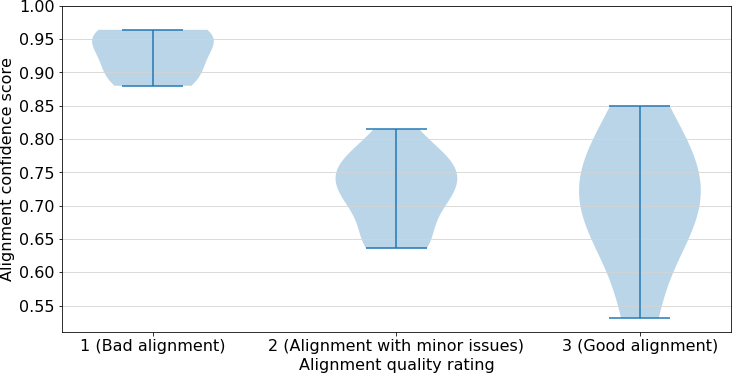}
\caption{Violin plot showing the distribution of alignment confidence scores for each rating in our qualitative evaluation.}\label{fig:violin-plot}
\end{figure}

After aligning all \textsc{midi} files in the data set to the corresponding audio file, we found an average alignment confidence score of 0.768. In order to verify if the results found by \citep{raffel2016extracting} are applicable to our data set, we evaluated the performance of the \textsc{dtw} system on a random sample of 25 \textsc{midi}s. For each \textsc{midi} file, we synthesized the realigned \textsc{midi} version and played it simultaneously with the original audio file in Sonic Visualiser, listening to the realigned \textsc{midi} file on the left earphone and the original audio on the right earphone. In this listening test, we classified each \textsc{midi} into one of three alignment quality categories: bad alignments; alignments with minor issues; or good alignments. The results of this evaluation can be found in the Appendix, in Table~\ref{tab:alignment-listening-test} and is also depicted in a violin plot in Figure~\ref{fig:violin-plot}. A violin plot \citep{hintze1998violin} is an alternative for the box plot that reveals density information from the data. Our evaluation confirmed Raffel's observation that alignments with a low alignment confidence score are good, while high alignment score (above 85\%) have major issues. These problems are mostly due to \textsc{midi} files in a wrong transposition or \textsc{midi} files that were bad transcriptions, for example because they only represented one part of the song.

\section{Chord estimation on \textsc{midi}}\label{sec:chord-est-midi}
In the previous section we saw how \textsc{decibel} aligns \textsc{midi} files to audio recordings. As a next step, we need a chord estimator to calculate the chord sequences from this realigned \textsc{midi} file. In this section, we give an overview of related work on \textsc{midi ace} in Subsection~\ref{sec:chordMIDI} and describe \textsc{decibel}'s \textsc{midi} chord estimation method in Subsection~\ref{sec:midi-ace-implementation}. This method will be briefly evaluated in Subsection~\ref{sec:midi-ace-evaluation}.

\subsection{Related work on chord estimation on \textsc{midi}}\label{sec:chordMIDI}
Chord estimation in \textsc{midi} files\footnote{also referred to as \textit{symbolic chord recognition} or \textsc{midi ace}} is the task of dividing a \textsc{midi} file into segments, in such a way that each segment boundary corresponds to a chord change, and assigning a chord label to each segment. In contrast to audio chord estimation, only few methods have been proposed that extract chords from symbolic music like \textsc{midi}.

In a pioneering paper, \cite{winograd1968linguistics} proposes a grammar-based approach to perform automatic tonal (roman numeral) analysis. This type of analysis identifies chords in their harmonic context. \cite{maxwell1992expert} developed a rule-based expert system that performs harmonic chord function analysis for tonal music, for example Bach keyboard pieces. \cite{temperley1999modeling} present a computational system for analyzing both metrical and harmonic structure. The system is based on eight preference rules, which are criteria for selecting the best analysis of a piece out of many possible ones. Their algorithm is implemented in the first version of Melisma Music Analyzer \citep{sleator2001melisma}.

\cite{pardo2001chordal,pardo2002algorithms} propose another method for segmentation and chord labeling. They segment the score using partition points (the set of all start and end points, excluding duplicates). Then, the score for each segment is calculated by computing the minimum of the distance between the segment and any of the 180 (template, root) pairs. An example of a template is $\langle 0,3,7 \rangle$, corresponding to a major chord. The authors give 15 different templates and 12 possible roots - one for each pitch class. In case of a tie, the best (template, root) pair is found using tie breaking rules. Inspired by aforementioned research by \citeauthor{pardo2001chordal}, \cite{scholz2008cochonut} propose COCHONUT: a new way to recognize chords from symbolic \textsc{midi} guitar data in bossa nova music, which is also dealing with complex chords like ninth and suspended chords. Chord estimation is split into three steps: (1) run a segmentation algorithm; (2) apply a utility function to identify the most probable chords for each segment and build a graph representing them; and (3) choose the best chord label for each segment, considering contextual information. 

\cite{raphael2003harmonic, raphael2004functional} use probabilistic models to perform functional harmonic analysis on \textsc{midi} data. The analysis is performed on a fixed musical period, for example a measure. They use 12 pitch classes. A chord is  specified by the tonic, mode and chord. For example, $(t, m, c) = (2, \text{major}, \text{II})$ would represent a triad in the key of D major built on the second scale degree. So this chord contains the pitches E, G and B (e minor). The most likely chord sequence is computed with a HMM using the Baum-Welch algorithm.

\cite{radicioni2010breve} perform chord estimation using a HMPerceptron model, in which the domain knowledge is modeled in Boolean features. As chord vocabulary, they consider chords with 12 possible root nodes, 3 possible modes and 3 possible added notes, resulting in 108 possible chord labels.

In more recent research, \cite{masada2017chord} propose a machine learning model for chord estimation that uses semi-Markov Conditional Random Fields (semi-CRFs) to perform a joint segmentation and labeling of symbolic music.

In this section, we have seen that the number of algorithms for \textsc{midi} chord recognition is limited.
Each of these algorithms would require modification and/or labelled training data in order to be used in \textsc{decibel}: (1) rule-based systems \citep{winograd1968linguistics, maxwell1992expert, temperley1999modeling} are designed for a specific music genre (usually classical music); (2) some algorithms \citep{winograd1968linguistics, maxwell1992expert, temperley1999modeling,raphael2004functional} recognise functional harmony instead of chord labels; and (3) other systems \citep{radicioni2010breve, masada2017chord} are based on machine learning, which requires a lot of labelled training data that is not available for \textsc{midi} files of popular music. Therefore, we designed \textsc{cassette} (Chord estimation Applied to Symbolic music by Segmentation, Extraction and Tie-breaking TEmplate matching). \textsc{Cassette} is a template-matching based algorithm for \textsc{midi} chord recognition that is easy to implement and understand and does not require any training. It is based on \cite{pardo2002algorithms}, but we adapted the segmentation method and use alternative tie breaking rules.

\subsection{Implementation of \textsc{cassette}}\label{sec:midi-ace-implementation}
In Section~\ref{sec:chordMIDI} we have seen that the existing algorithms for \textsc{midi} chord recognition are not suitable for direct use in our \textsc{decibel} system. Therefore, we designed \textsc{cassette} (Chord estimation Applied to Symbolic music by Segmentation, Extraction and Tie-breaking TEmplate matching). \textsc{Cassette} is a template-matching based algorithm for \textsc{midi} chord recognition that is easy to implement and understand. Similar to the good old cassette tapes, this algorithm is certainly not state of the art. However, it is simple to implement and does not require any training.

\textsc{Cassette} recognizes chords in a three-step procedure: first, it segments each audio-aligned \textsc{midi} file. Then, it calculates a weighted chroma feature for each of the segments, based on the notes that are present within the segment. Finally, \textsc{cassette} matches the features of each segment to the features of a predefined chord vocabulary and assigns each segment to the most similar chord. In the remainder of this section, we zoom in on each of these three steps.

\textsc{Cassette} \textbf{segments} the \textsc{midi} both on the bar and on the beat level. In many cases, segmentation on the bar level is sufficient, as chord changes in popular music often are placed on the downbeat, i.e. on the first beat of a bar. An advantage on segmentation on the bar level is that non-harmony notes, which are typically short, are less problematic in the template matching step as they have relatively lower weights than the (typically longer) harmony notes. On the other hand, segmentation on the bar level does not work well for songs that have chord changes at other positions than the start of a bar. That is why \textsc{cassette} segments the \textsc{midi} file on the beat level as well. This typically results in a chord sequence with more chord changes, although some of which are based on non-harmony notes. For segmentation, \textsc{cassette} uses the pretty\_midi package by~\cite{raffel2014intuitive}. Bar segments are found with the get\_downbeats function, while the algorithm finds beat segments with the get\_beats function. 

After segmentation, the next step is \textbf{feature extraction} on each of the segments of the \textsc{midi} file. For each segment, \textsc{cassette} extracts the notes sounding between its start and end time. From these notes, we calculate weighted chroma, a feature that is similar to A-weighted chroma as used by~\citet[Section~4.1]{bonvini2014automatic}. In order to calculate the weighted chroma feature for a given segment, \textsc{cassette} computes for each note the product of the \textsc{midi} velocity and the proportion of the bar during which this note sounds, and sums this product over all notes in the same pitch class. For example, the weighted chroma of a quarter note C in a 4/4 bar with \textsc{midi} velocity of 100 is $[25, 0, 0, 0, 0, 0, 0, 0, 0, 0, 0, 0]$. In the resulting vector, both the louder notes (with high velocity) and the longer notes (with higher duration) in the \textsc{midi} file are relatively more important, compared to softer or shorter notes. \textsc{Cassette} normalizes the weighted chroma vector by dividing each element by the total sum of all its elements. This makes the feature invariant to the total loudness and duration of the notes in the segment.

As a third and final step, we need to find the \textbf{best matching chords} for each segment. Therefore, \textsc{cassette} calculates the template similarity score between the feature of the segment and the feature of each template in our chord vocabulary. In this project, we use a vocabulary of 24 chords, consisting of all 12 major chords and all 12 minor chords, and the no-chord. The chroma-like feature of each template is a 12-dimensional vector, in which each value is 1 if the corresponding note occurs in that chord and the value is 0 otherwise. For example: the D minor chord consists of a D, F and an A. The corresponding chroma would be $[0, 0, 1, 0, 0, 1, 0, 0, 0, 1, 0, 0]$. The score function that measures the similarity between the chroma of the segment and the chroma template is based on earlier work by~\cite{pardo2002algorithms}. The template similarity score $S$ is calculated with the formula $S = P - (N + M)$. $P$ is the positive evidence: the sum of the weights of the chroma of the bar which match a template element. $N$ is the negative evidence: the sum of the weights of the chroma of the bar which does not match a template element. $M$ stands for misses: the count of template elements not matched by any note. High scores correspond to well-matched templates. For each bar, \textsc{cassette} finds the chord with the highest template similarity score. If the score is -3 or lower, the algorithm assigns a no-chord, as there is no evidence for any chord if three notes or more from the template are missing. Furthermore, \textsc{cassette} applies a single tie-breaking rule: if multiple templates have the same template similarity score, it selects the template whose root pitch has the greatest weight in the segment's chroma.

\subsection{Evaluation of \textsc{cassette}}\label{sec:midi-ace-evaluation}
In the previous section, we have seen how \textsc{cassette} estimates chords from \textsc{midi} files. For the evaluation of the system, we run into a minor issue: a suitable data set with chord annotations for \textsc{midi} files of popular music does not yet exist. As a solution, we selected 50 well-aligned \textsc{midi} files from our data set and tested them against the Isophonics annotations, as described below.

First, we needed a subset of \textsc{midi}s for which the timing information was consistent with the Isophonic annotations. In Section~\ref{sec:alignment-evaluation} we observed that \textsc{midi}/audio pairs that are aligned well typically have a low alignment error. Therefore, we selected the 50 \textsc{midi} files with the lowest alignment error. This subset had alignment confidence scores ranging from 0.459 to 0.602, so we can reasonably expect that these \textsc{midi} files are well aligned to the audio. 

Then, we ran the \textsc{cassette} algorithm on the audio-aligned versions of these 50 \textsc{midi} files and calculated the \textsc{wcsr} for each of the resulting chord label sequences. The results are shown in Table~\ref{tab:midi-ace-evaluation}. Note that particularly the beat-based \textsc{midi} chord recognition method performs quite well in terms of \textsc{wcsr} with a 80.0\% score. Compared to the bar-based \textsc{midi} chord recognition method, beat-based chord recognition has minor oversegmentation issues, while the bar-based method is undersegmenting. This also explains the lower \textsc{wcsr} score of 70.9\% of the bar-based method. We can conclude that \textsc{cassette}, despite its simplicity, performs quite well on our data set with popular music and a limited chord vocabulary. 


\begin{table}[tbp]
\centering
\begin{tabular}{lllll}
\hline
\textbf{Sementation method} & \textbf{\textsc{wcsr}} & \textbf{OvS} & \textbf{UnS} & \textbf{Seg} \\ \hline
Beat & 80.0\% & 83.0\% & 89.1\% & 80.8\% \\
Bar & 70.9\% & 95.4\% & 67.1\% & 67.1\% \\ \hline \\
\end{tabular}
\caption{Results of \textsc{midi} chord recognition for the 50 \textsc{midi}s with the lowest alignment error.}\label{tab:midi-ace-evaluation}
\end{table}

Another interesting property of \textsc{cassette} is that it computes template similarity scores, which we can use to calculate an approximation score for the quality of the output chord sequences. The template similarity score for each segment gives an indication of the fit of the chord label to this segment: a high score indicates a well-fitting chord label while a low-score segment probably has a less suitable chord label. This may be the case when the actual chord label is more complex than the best fitting chord in our simplified vocabulary. When we average the scores over all segments, we get an approximation score for the quality of the chord label. We call this score the Average Template Similarity (\textsc{ats}) and we will see its advantages in the next section.

\section{\textsc{Midi} file selection}\label{sec:midi-selection}
In the previous section we showed that \textsc{cassette}, our \textsc{midi} chord recognition method, was quite successful for a 50-song subset of well-aligned \textsc{midi} files. 
Yet extending the aforementioned experiment to the full data set showed a considerably worse performance: when comparing the chord sequences found by \textsc{midi} file alignment and chord recognition system on all \textsc{midi} files, we found a \textsc{wcsr} of 62.9\% for beat-based chord recognition and 60.3\% for bar-based chord recognition. 
These results are not competitive to the performance of other (audio) chord recognition systems. 
However, these poor results are partly due to low-quality \textsc{midi} files. Table~\ref{tab:midi-selection-performance} shows that \textsc{midi} quality highly influences the \textsc{ace} performance. If the worst \textsc{midi} file for each song is chosen, then the \textsc{wcsr} for our data set is 45.2\% (beat) and 43.9\% (bar); if the best \textsc{midi} file for each song is chosen, then the \textsc{wcsr} for our data set is 78.9\% (beat) and 75.0\% (bar).

In this section, we describe a method to select the estimated best \textsc{midi} file for each song. Since we cannot calculate the \textsc{csr} for unlabeled data, we use a proxy measure based on the \textsc{aes} and \textsc{ats} scores.

As reported in Section~\ref{sec:midi-audio-align},  some \textsc{midi} files are badly aligned. Accordingly, these files will typically not yield good chord labels. Therefore, we discarded all \textsc{midi} files with an \textsc{aes} higher than 85\%. This leads to a great shift in performance: the \textsc{wcsr} on all \textsc{midi} files that are sufficiently aligned, is 72.7\% (beat) and 69.3\% (bar).

\begin{figure}
 \centering
 \includegraphics[width=\columnwidth]{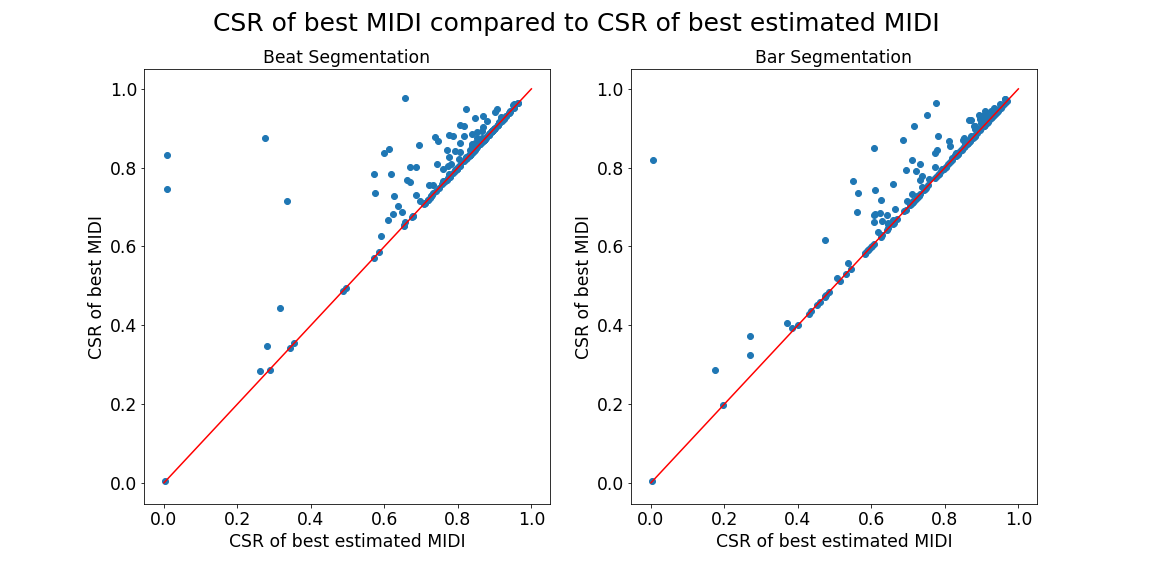}
 \caption{\textsc{Csr} of the real best \textsc{midi} compared to the \textsc{csr} of the estimated best \textsc{midi} for both beat and bar segmentation. Points on the diagonal line (i.e. $x=y$) correspond to songs for which the best \textsc{midi} file was estimated correctly. The distance in $y$ direction between each point and the line is the difference between the \textsc{csr} of the best \textsc{midi} and the \textsc{csr} of the estimated best \textsc{midi} file.}\label{fig:csr-ats-diff}
\end{figure}

After this \textsc{midi} selection step based on alignment quality, we have 592 out of our initial 770 \textsc{midi} files left. 
Since we still have multiple \textsc{midi} files per song, we do an additional selection step.
For this, we use the \textsc{ats} score that is calculated by \textsc{cassette}. 
The \textsc{csr} correlates with the \textsc{ats}: the Pearson correlation coefficient is 0.483 for beat segmentation and 0.597 for bar segmentation. 
For most songs, the \textsc{csr} of the \textsc{midi} with the highest \textsc{ats} is (almost) as good as the \textsc{csr} of the actual best \textsc{midi}, as shown in Figure~\ref{fig:csr-ats-diff}. 
In this scatter plot, each point belongs to a song from our data set. 
The \textsc{csr} of the estimated best \textsc{midi} of the song corresponds with its position on the $x$-axis and the \textsc{csr} of the actual best \textsc{midi} corresponds with its position on the $y$-axis. 
Points on the diagonal line (where $x = y$) are songs for which the best \textsc{midi} file was estimated correctly. 
The plot shows that there are very few songs for which there is a big difference between \textsc{csr} of the best \textsc{midi} and \textsc{csr} of the estimated best \textsc{midi}: there are only three songs for which the difference is higher than 0.5. In two of these songs, the estimated best \textsc{midi} was a semitone transposed, compared to the original audio. In the third song, the estimated best \textsc{midi} is only a transcription of part of the song. However, in the vast majority of songs, the difference in \textsc{csr} between the estimated and actual best \textsc{midi} files is small: most points are close to the diagonal line. We conclude that the \textsc{ats} is a suitable score measure to select the estimated best \textsc{midi} file for each song. 
This is also reflected in the performance shown in Table~\ref{tab:midi-selection-performance}: the chord sequences from \textsc{midi}s selected with our proposed selection method (Estimated best) have \textsc{wcsr}s of 75.7\% and 72.9\% and thereby outperform the performance of all well-aligned \textsc{midi} files with three percentage points. 

\begin{table}[t]
\centering
\resizebox{\columnwidth}{!}{
\begin{tabular}{lllllllll}
\hline
 & \multicolumn{2}{c}{\textbf{\textsc{wcsr}}} & \multicolumn{2}{c}{\textbf{OvSeg}} & \multicolumn{2}{c}{\textbf{UnSeg}} & \multicolumn{2}{c}{\textbf{Seg}} \\
\textbf{} & \textbf{Beat} & \textbf{Bar} & \textbf{Beat} & \textbf{Bar} & \textbf{Beat} & \textbf{Bar} & \textbf{Beat} & \textbf{Bar} \\ \hline
\textbf{Min \textsc{csr}} & 45.2\% & 43.9\% & 75.5\% & 88.6\% & 75.4\% & 62.2\% & 65.9\% & 60.8\% \\
\textbf{All} & 62.9\% & 60.3\% & 78.9\% & 91.4\% & 81.7\% & 66.9\% & 72.6\% & 66.0\% \\
\textbf{Well-aligned} & 72.7\% & 69.3\% & 80.2\% & 93.1\% & 86.6\% & 71.0\% & 76.8\% & 70.4\% \\
\textbf{Estimated best} & \textbf{75.7}\% & \textbf{72.9}\% & \textbf{81.3}\% & \textbf{93.2}\% & \textbf{87.2}\% & \textbf{73.3}\% & \textbf{77.6}\% & \textbf{72.5}\% \\
\textbf{Max \textsc{csr}} & 78.9\% & 75.0\% & 82.3\% & 93.2\% & 87.2\% & 72.9\% & 78.6\% & 71.9\% \\ \hline \\
\end{tabular}}
\caption{Performance comparison of five \textsc{midi} selection methods in terms of \textsc{wcsr}, oversegmentation, undersegmentation and segmentation as defined in \cite{harte2010towards}.}\label{tab:midi-selection-performance}
\end{table}

\begin{figure*}
 \centering
 \includegraphics[width=\textwidth]{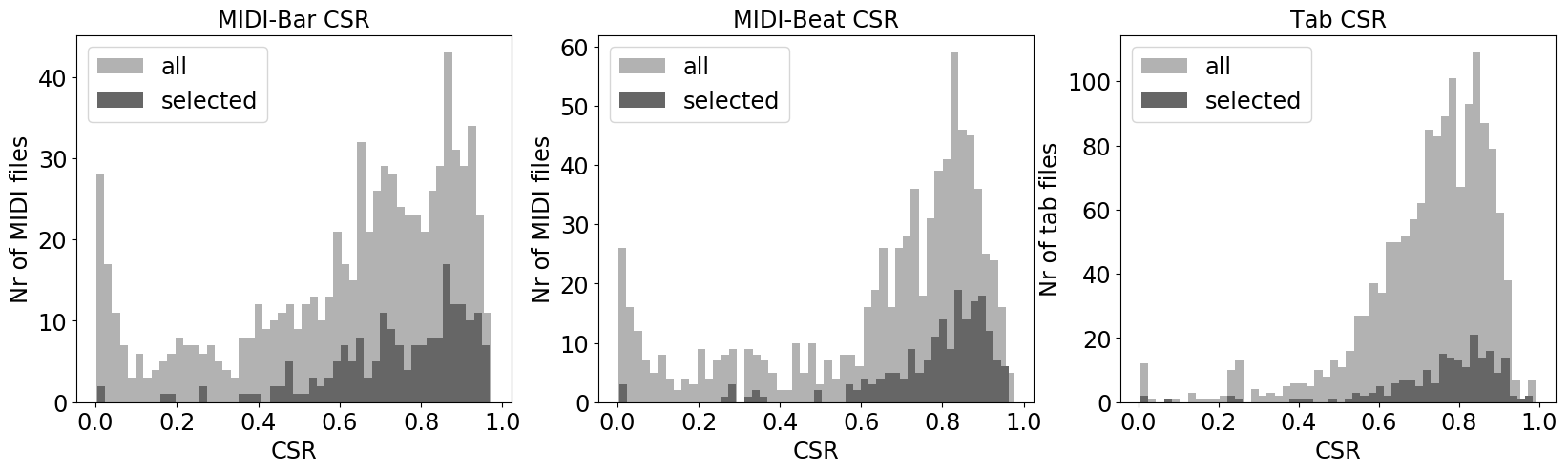}
 \caption{Histograms showing the distribution of CSR for (A) \textsc{midi} files with bar segmentation; (B) \textsc{midi} files with beat segmentation; and (C) tab files.}\label{fig:hist-symbolic}
\end{figure*}

Figure~\ref{fig:hist-symbolic} shows the \textsc{csr} distribution for all \textsc{midi} files in grey and all estimated best \textsc{midi} for each song in a darker shade. It shows that \textsc{midi} files corresponding to the poorest chord estimates (grey peaks at the left) are typically not selected.

\section{Conclusion}\label{sec:midi-conclusion}
In this section, we have examined the \textsc{midi} subsystem of \textsc{decibel}. This subsystem extracts chord sequences from \textsc{midi} files by first aligning the \textsc{midi} file to the audio file and then running a simple chord estimation method on the re-aligned \textsc{midi} file. In our data set, we have collected multiple \textsc{midi} files for each song. The subsystem computes a chord sequence for each \textsc{midi} file, so we obtain multiple chord sequences per song. We have seen that we can select ``good'' (and in many cases the best) \textsc{midi}s for each song by (1) ignoring \textsc{midi}s with a high alignment error and (2) selection of the \textsc{midi} with the highest Average Template Similarity (\textsc{ats}). This resulted in \textsc{wcsr}s of 75.7\% for the beat segmentation and a 72.9\% for bar segmentation.  

\chapter{Automatic Chord Estimation on tabs}\label{ch:subsystem-tabs}
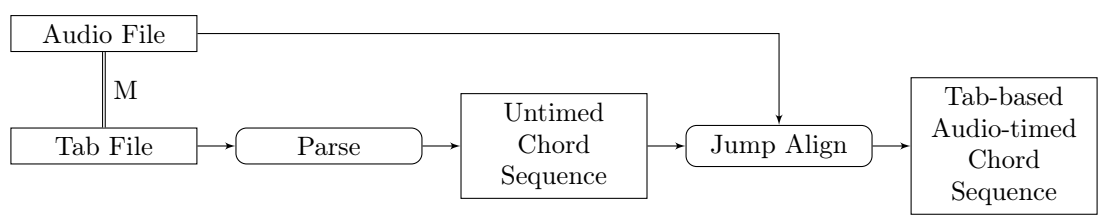
\begin{figure}   
\centering
\begin{tikzpicture}[node distance = 1cm and 0.5cm, auto]
    \node [rect, right=of midi_file] (audio_file) {Audio File};
    \node [rect, below=of audio_file] (tab_file) {Tab File};
    
    \node [block, right=of tab_file] (tab_parse) {Parse};
    \node [rect, right=of tab_parse] (ucs) {Untimed Chord Sequence};
    \node [block, right=of ucs] (jump_align) {Jump Align};
    \node [rect, right=of jump_align] (tab_cs) {Tab-based Audio-timed Chord Sequence};
    
    \path [double_line] (audio_file) -- node[] {M} (tab_file);
    
    \path[line] (tab_file) -- (tab_parse);
    \path[line] (tab_parse) -- (ucs);
    \path[line] (audio_file) -| (jump_align);
    \path[line] (ucs) -- (jump_align);
    \path[line] (jump_align) -- (tab_cs);
\end{tikzpicture}
\caption{Diagram of \textsc{decibel}'s tab subsystem; the M between Audio File and Tab File indicates that they are manually matched.}
\label{fig:diagram-tabs}
\end{figure}

In the previous chapters we have seen how our system estimates audio-timed chord labels from audio representations and \textsc{midi} files. In this chapter, we will look at \textsc{decibel}s third subsystem, which uses tab files for estimating chord labels. As we have seen in Section~\ref{sec:tabs}, there exist two types of tabs and we can easily extract chord symbols from both of them: guitar tablature indicates the instrumental fingering, which directly maps to the notes of a chord; in a chord sheet, the chord symbols are represented explicitly, together with the lyric syllables with which they have to be timed. In this chapter, we will explain how \textsc{decibel} uses tabs from either of these two types in order to estimate audio-timed chord labels. The outline of this subsystem is also illustrated in Figure~\ref{fig:diagram-tabs}. First, the tabs are parsed and the chord information is extracted. This step is described in Section~\ref{sec:tab-parse}. As a next step, \textsc{decibel} aligns the chord information to the audio file, as described in Section~\ref{sec:jump-alignment}. This gives us multiple chord estimates: one for each tab file of the corresponding song. In Section~\ref{sec:tab-selection}, we explain how \textsc{decibel} selects the expected best tab file for each song. Finally, we summarize the chapter in Section~\ref{sec:tab-conclusion}.

\section{Tab parsing}\label{sec:tab-parse}
Before \textsc{decibel} can align the tab files to the audio, it first needs to parse them and extract the chord information. The used tab parsing method is explained in Section~\ref{sec:tab-parse-methods} and evaluated in Section~\ref{sec:tab-parse-evaluation}.

\subsection{Tab parsing methodology}\label{sec:tab-parse-methods}
\textsc{Decibel} parses the tabs in a similar way to the parser proposed by \cite{macrae2012linking}. First, it classifies each line in the tab file to a line type. Then, it segments the tab by splitting on empty lines. As a next step, all systems in each segment are identified. We define a system as a set of subsequent lines that belong to each other. For example: a tab system is very common in guitar tablature files and consists of exactly six subsequent tab files. In chord sheets, a common system is the alteration between chord lines and lyrics lines. From these systems, \textsc{decibel} can then extract the chord labels. Thereby, the system retains line information (i.e. the line of the chord in the text file), as this will be used in the tab-audio alignment step. The steps of line type classification, segmentation and system and chord extraction are explained more thoroughly in the remainder of this section.

\subsubsection{Line type classification}
For each line in the tab file, \textsc{decibel} estimates its line type using a set of heuristics. We distinguish the following line types: \textit{empty}, \textit{chords}, \textit{tuning definition}, \textit{capo change}, \textit{structural marker}, \textit{chord definition}, \textit{tablature}, \textit{lyrics} and \textit{combined chord and lyrics}.

\begin{itemize}

\item A line has the \textit{empty} line type if it is empty or consists only of a space.

\item To determine if a line has the \textit{chords} line type, the system first splits the line by spaces. Then it checks if each element of the line matches the chord pattern. An element is a chord if:
\begin{itemize}
\item It consists of at most 10 characters;
\item The first character is a note letter (a, b, c, d, e, f or g);
\item The element does not contain sequences of four digits (as this would indicate a chord definition);
\item The element does not contain symbols; and
\item Any three letter sequence is in the following list: {\textit{min}, \textit{add}, \textit{aug}, \textit{dim}, \textit{maj}, \textit{sus}, \textit{flat}}.
\end{itemize}

\item A line has the \textit{tuning} line type if it contains the word ``tuning''.

\item To find out if a line is a \textit{structural marker}, our parser searches for words like ``verse'', ``chorus'' or ``bridge''.

\item A line is a \textit{chord definition} if it contains a sequence of exactly 6 digits.

\item A line is a \textit{tablature line} if it contains at least 10 characters which are a digit, hyphen, vertical bar, slash, letter `h', `b' or `p' or a space and the number of hyphens is larger than the number of spaces.

\item A \textit{lyrics line} fulfills each of the following conditions:
\begin{itemize}
\item It does not contain square brackets, the equals sign or the at sign;
\item It contains at most 10 hyphens;
\item Either:
\begin{itemize}
\item It consists of just one word, which has only letter characters and contains at least three of the same letters after each other, like `ooooh' or `aaah'; or
\item It contains at least two words.
\end{itemize}
\end{itemize}

\item To find out if a line is a \textit{combined chords and lyrics} line, \textsc{decicbel} first extracts all characters between square brackets ([ and ]). If these characters form a chord, it removes all characters between each pair of square brackets. If the remaining line is a lyrics line, we can conclude that our original line was a combined chords and lyrics file.
\item If a line does not satisfy any of these requirements, the parser assigns the line type \textit{undefined}.
\end{itemize}

\subsubsection{Segmentation}
At this point, we have line types for each line in the tab. As a next step, our parser divides the tab file into segments. A \textit{segment} is simply defined as the lines between lines of the empty line type. In a typical tab, verses and choruses are separated by empty lines and therefore are designated to their own segment.

\subsubsection{System and chord extraction}
Now we have our tab file divided into segments, and we know the line type of each of the lines in a segment. We want to extract the chord labels, but we only extract them from a suitable system. A \textit{system} consists of all of lines that should be played or sung together at the same time. For example, in a typical chord sheet, we will find a lot of systems consisting of a \textit{chord line} above a \textit{lyrics line}, as the chords in the chord line should be played at the same time as the word in the lyrics line should be sung. The six different system types are listed in the first column of  Table~\ref{tab:system-chord-extraction}. 

\begin{table}[tbp]
\centering
\begin{tabular}{ll}
\hline
\textbf{System type} & \textbf{Chord Extraction} \\ \hline
Combined chords and lyrics line & Parse chords from combined chords and lyrics line \\\hline
Chord line & Parse chords from chord line \\\hline
\begin{tabular}[c]{@{}l@{}}Chord line\\ Exactly six tablature lines\end{tabular} & Parse chords from chord line \\\hline
\begin{tabular}[c]{@{}l@{}}Chord line\\ Exactly six tablature lines\\ 1 to 3 lyrics lines\end{tabular} & Parse chords from chord line \\\hline
Exactly six tablature lines & Parse chords from six tablature lines \\\hline
\begin{tabular}[c]{@{}l@{}}Exactly six tablature lines\\ 1 to 3 lyrics lines\end{tabular} & Parse chords from six tablature lines \\ \hline
\end{tabular}
\caption{System types and their chord extraction methods}\label{tab:system-chord-extraction}
\end{table}

\textsc{Decibel} assigns each line to the largest possible system. Consider for example a segment consisting of three chord definition lines, followed by a chord line, six tablature lines and one lyrics line. Our parser finds only one system, consisting of a chord line, six tablature lines and a lyrics line. In this case, the six tablature lines do not form a system on their own, as the chord line and lyrics line are also played and sung at the same time. This has to do with the very reason why we consider systems: we only extract one chord sequence from a system, to prevent duplicate chord sequences. This is particularly important in guitar tablatures, as they often have a chord line directly followed by six tab lines with exactly the same chord information. In that case, the parser ignores the six tab lines and extract the chords from the chord line only. For each of the systems, the chord extraction method is displayed in the second column of Table~\ref{tab:system-chord-extraction}. We see that there are basically three chord extraction methods:\begin{itemize}
\item \textbf{Parse chords from combined chords and lyrics line}: our parser extracts all characters between square brackets ([ and ]), so we have a list of strings, in which each string represents a chord. Then it parses each string by splitting it into a root note, chord type and optional bass note using regular expressions. For example, the chord string ``D7/F\#'' has root note ``D'', chord type ``7'' (dominant seventh chord) and bass note ``F\#''. Based on these three substrings, the parser now converts the chord string to Harte's chord notation~\citep{harte2005symbolic}. It also saves the line number and character index of the start of the chord string.
\item \textbf{Parse chords from chord line}: the parser extracts chord strings from the chord line by splitting on spaces and ``|'' characters. Then it parses each string and converts it to Harte's chord notation, line numbers and character indices, as described above.
\item \textbf{Parse chords from six tablature lines}: for each character index ($x$-value), the parser extracts the six characters that are notated above each other. If all characters are either a digit of an `x', a chord is played at this position. The parser then derives the notes played at the same time based on these characters and their string - for example: if a digit `1' appears on the B string line, the corresponding chord is a C; the digit `2' on the B string corresponds to a C\#. Finally it finds the nearest chord from our chord vocabulary (based on cosine distance). Again, it also saves line number (of the first tab line) and character index.
\end{itemize}

\subsection{Tab parsing evaluation}\label{sec:tab-parse-evaluation}
In this section, we describe the evaluation of the tab parsing algorithm. For the evaluation of the tab parsing algorithm, we used a random sample of 25 tabs from the data set, as listed in Appendix~\ref{appendix:tabs-evaluation-set}. For these 25 tabs, we annotated the untimed chord sequence by hand. Then, we ran \textsc{decibel}'s tab parsing algorithm on these 25 tabs and compared the result to the annotation. 

\begin{figure}
 \centering
 \includegraphics[width=0.75\textwidth]{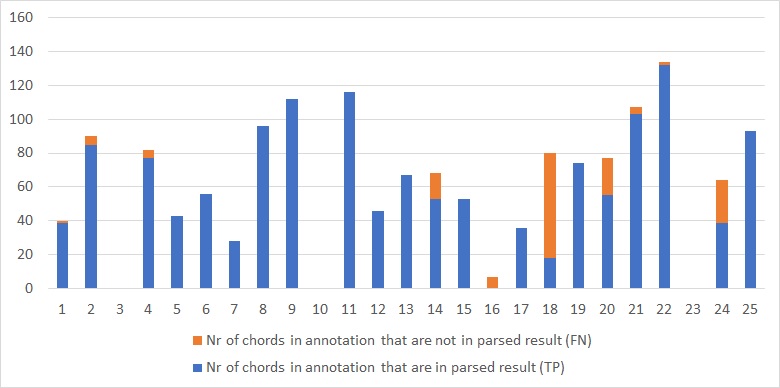}
 \caption{Recall of chords in our sample of 25 tab files.}
 \label{fig:tab-parse-eval}
\end{figure}

The results are shown in Figure~\ref{fig:tab-parse-eval}, in which the height of each bar corresponds to the number of chords in the annotation of a tab file. The True Positives (TP) are the chords in the annotation which are also in the parsed results. In the figure, we see them as the blue part of the bar. The orange part of the bar corresponds to the False Negatives (FN): the chords that are in the annotation of a tab file, but not found by our parser. As we see in this figure, the parser finds (almost) all chords in most of the songs, as the orange part is very small. There were no chords in either the annotation or the parser output of the third, tenth and 23rd song. For the other songs, chords were missing in the tab output for the following reasons:
\begin{itemize}
 \item The parser does not recognize chords that are glued after each other (without a separator) in the tab file;
 \item The parser does not recognize chords in a line with more text or strange symbols, as such a line is not classified as a chord line;
 \item The parser does not recognize chords in guitar tablature systems when there are no digits or 'x' in all six lines.
\end{itemize}

In some songs, text that was actually not a chord was misinterpreted as chords. This happened in one tab file which added some rhythm information by for example putting a 'Q' above a chord symbol for quarter note durations and an 'E' for eighth note durations. Our parser falsely detected a line of eighth notes as a sequence of E chords. In another line, chord vocabulary information was falsely detected as a chord line.

Still, in most songs we see that the tab parser performs quite well, as there are no missing chords in 15 of the 25 songs.

\section{Jump Alignment}\label{sec:jump-alignment}
Having completed the tab parsing step, we have extracted the chord labels and their corresponding line and word numbers from the tab file. However, tab files retain no timing information, so we need an additional step to align the chord labels to the audio file. As we have seen in Section~\ref{sec:introduction-related-work}, there already exist four different algorithms by \cite{mcvicar2011using} that incorporate tab information into a HMM-based system for audio chord estimation. To the best of our knowledge, these are the only algorithms that use tabs in audio \textsc{ace}. The most promising of these four algorithms is Jump Alignment. In this section, we describe and evaluate our implementation of Jump Alignment. 

Jump Alignment is based on a Hidden Markov Model (HMM). A HMM models the joint probability distribution $P(X, y | \Theta)$ over the feature vectors $X$ and the chord labels $y$, where $\Theta$ are the parameters of the model. In Section~\ref{sec:jump-alignment-preprocessing}, we describe the preprocessing steps that are used for extraction of the features $X$ and the chord labels $y$. Section~\ref{sec:jump-alignment-hmm} gives an introduction to the HMM. The details on the Jump Alignment algorithm are treated in Section~\ref{sec:jump-alignment-algorithm}.

\subsection{Preprocessing}\label{sec:jump-alignment-preprocessing}
First, the audio file needs to be \textbf{preprocessed}. For this purpose, we use the python package librosa~\citep{brian_mcfee_2018_1252297}. First, we convert the audio file to mono. Then, we use the HPSS function to separate the harmonic and percussive elements of the audio. Then, we extract chroma from the harmonic part, using constant-Q transform with a sampling rate of 22050 and a hop length of 256 samples. Now we have chroma features for each sample, but we expect that the great majority of chord changes occurs on a beat. Therefore, we beat-synchronize the features: we run a beat-extraction function on the percussive part of the audio and average the chroma features between the consecutive beat positions. The chord annotations need to be beat-synchronized as well. We do this by taking the most prevalent chord label between beats. 
Each mean feature vector with the corresponding beat-synchronized chord label is regarded as one frame. Now we have the feature vectors $X$ and chord labels $y$ for each song, which we feed to our HMM.

\subsection{Hidden Markov Model}\label{sec:jump-alignment-hmm}
As we have seen in Section~\ref{sec:audio-ace-models}, a HMM models the joint probability distribution $P(X, y | \Theta)$ over the feature vectors $X$ and the chord labels $y$, where $\Theta$ are the parameters of the model. The HMM is an extension of the Markov chain in which the states are not directly observable (but hidden). In this section, we give a summary of the Hidden Markov Model, based on the tutorial by~\cite{rabiner1989tutorial} and Section 5.3.2 from \cite{muller2015fundamentals}. 

In a \textbf{Markov chain}, we make the assumption that the probability from the current state $s_t$ to the next state $s_{t+1}$ only depends on the current state, and not on any of the previous states. Formally: 
\begin{equation}
P(Y_{t + 1} = s_{t + 1} | Y_1 = s_1, Y_2 = s_2, \ldots, Y_t = s_t) = P(Y_{t + 1} = s_{t + 1} | Y_t = s_t)
\end{equation}
This assumption is called the \textbf{Markov property}. In our implementation, states are represented by chord labels. Any chord label is an element from the discrete set of chords from our chord vocabulary. Although we could use any chord vocabulary, we use the vocabulary with all major and minor chords and the no-chord symbol. Our vocabulary thus consists of 25 items. Thanks to the Markov property, we can efficiently model the probability that chord $j$ occurs after chord $i$ by the \textbf{transition probability} $P_{tr}(y_t = j | y_{t-1} = i)$. Given a chord vocabulary of 25 items, we can model all transition probabilities in $25 \cdot 25 = 625$ probability values, that can be efficiently stored in the matrix $\mathbf{P_{tr}}$.

The first chord of a chord progression cannot be computed from its previous chord. This information is specified by the \textbf{initial state probabilities}. There are 25 initial state probability values, stored in the vector $\mathbf{P_{ini}}$, in which $P_{ini}(y_1 = c)$ is the probability that the first chord is chord $c$. 

Based on the Markov chain, we could now compute the probability for a given chord label given a sequence of chord labels, or compute the probability of any chord label sequence. However, in a \textsc{ace} scenario, we cannot directly observe the chord labels, but we do observe a sequence of chroma features that is related to the chord labels. We thus need to expand our model, by adding \textbf{emission probabilities}. The resulting model is a Hidden Markov Model. The \textbf{emission probability} $P_{obs}(X_t | y_t)$ equals the probability of an observation (chroma feature vector) $X$ given a chord label $y$ at time $t$. In contrast to transition and initial state probabilities, the emission probabilities cannot be represented by a matrix, as chroma features are continuous. Instead, they are modeled by continuous probability density functions: the observation probability distribution for each of the 25 chord labels $a_i$ in the vocabulary as a 12-dimensional Gaussian with 12-dimensional \textbf{mean vector} $\mu_i$ and $12 \times 12$ \textbf{covariance matrix} $\Sigma_i$. We can now find our emission probabilities for any observed chroma feature vector $x_t$ and any chord label $a_i$, using the Gaussian with parameters $\mu_i$ and $\Sigma_i$:

\begin{equation}
P_{obs}(x_t | \mu_i, \Sigma_i) = \frac{1}{(2 \pi)^6 |\Sigma_i|^{1/2}} \exp [-\frac{1}{2} (x_t - \mu_i)^T \Sigma^{-1} (x_t - \mu)]
\end{equation}

We now have seen that a HMM is parametrized by transition, initial state and emission probabilities. We define $\Theta$ as the set of these parameters:
\begin{equation}
\Theta = \{\mathbf{P_{tr}}, \mathbf{P_{ini}}, \{\mu_i\}_{i=1}^{25}, \{\Sigma_i\}_{i=1}^{25}\}
\end{equation}

There exist different ways of estimating $\Theta$. We adopt the machine learning approach, that is: we \textbf{train} the probabilities on a subset of our data set. In our case, the training set is a 100-song random sample of our data set. The initial state and transition probability matrices are calculated simply by frequency counting and normalization of all chord labels in the training set. For calculating $\mu_i$ and $\Sigma_i$, we take the set of all feature vectors that are labeled with chord $i$ in our data set, and then calculate the mean vector and covariance matrix respectively.

Now we know all elements of the HMM, we can represent the joint representation for the chroma feature vectors $X$ and chord labels $y$ mathematically in the following equation:
\begin{equation}
P(X, y | \Theta) = P_{ini}(y_1 | \Theta) \cdot P_{obs} (x_1 | y_1, \Theta) \cdot \prod_{t=2}^{|y|} P_{tr} (y_t | t_{t-1}, \Theta) \cdot P_{obs} (x_t | y_t, \Theta)
\end{equation}

HMMs can be used to solve three basic problems, as pointed out in \cite{rabiner1989tutorial}. One of these problems, which is called the \textbf{uncovering problem} or \textbf{decoding problem}, is the relevant problem for our purpose: automatic chord estimation. The goal of the uncovering problem is to find the state (chord) sequence that best explains the observed (chroma) feature sequence. Formally: we want to find the chord label sequence $y^*$ where $y^* = \operatorname*{argmax}_y P(y | X, \Theta)$. Note that this is equivalent to finding $y^* = \operatorname*{argmax}_y P(y, X | \Theta)$. 

Note that the optimal chord sequence for an observation sequence of $n$ elements starts with the optimal chord sequence for the observation sequence of the first $n - 1$ elements. The \textbf{Viterbi algorithm} efficiently solves the uncovering problem by exploiting this property. The pseudocode of Viterbi is given in Algorithm~\ref{alg:viterbi-pseudocode}. In each iteration, the algorithm computes the indices $i_1, i_2, \ldots, i_t$ to the optimal chord sequence for $X_1, X_2, \ldots, X_t$ (i.e. the first $t$ steps of the observation sequence $X$) ending with chord label $y_{i_t}$. The result is then stored in the matrices $V$ and $B$: in $V$, the algorithm keeps the likelihood of the current path, while the chord index is stored in $B$. When calculating a chord sequence of length $n$, the likelihoods of the optimal chord sequences of length $n - 1$ can be efficiently retrieved by reading the corresponding value from matrix $V$. In the termination phase, we can read the final chord index $i_T$ by maximizing over the last column of $B$ and find the full optimal chord sequence by following the backpointers in $B$ back in time. We can also read the likelihood of the full optimal chord sequence from $V[i_T, T]$.

\begin{algorithm}
\begin{algorithmic}[1]
\Function{Viterbi}{Observation Sequence $X_1, X_2, \ldots, X_T$, HMM specified by $\Theta$, Chord alphabet $y_1, y_2, \ldots, y_{25}$}
  \State Initialize Viterbi Matrix $V$ of size $25 \times T$
  \Comment{Initialization}
  \State Initialize Backpointer Matrix $B$ of size $25 \times (T - 1)$.
  \ForAll{$i \in [1, 25]$}
  	\State $V[i, 1] = \mathbf{P_{ini}}[y_i] \cdot P_{obs}(y_i | X_1)$
  \EndFor  
  \ForAll{$t \in [2, T]$}
      \Comment{Recursion step}
      \ForAll{$i \in [1, 25]$}
        \State $V[i, t] = \operatorname*{max}_{j=1}^{25} V[j, t - 1] \cdot \mathbf{P_{tr}}[y_j, y_i] \cdot P_{obs}(y_j | X_t)$
        \State $B[i, t - 1] = \operatorname*{argmax}_{j=1}^{25} V[j, t - 1] \cdot \mathbf{P_{tr}}[y_j, y_i]$
     \EndFor
  \EndFor
  \State{$i_T = \operatorname*{argmax}_{j=1}^{25} V[j, T]$}
  \Comment{Termination}
  \ForAll{$t \in [T - 1, 1]$}
    \State{$i_t = B[i_{t+1}, t]$}
  \EndFor
  \State{\Return{Chord Sequence $y_{i_1}, y_{i_2}, \ldots, y_{i_T}$ and Likelihood Value $V[i_T, T]$}}
\EndFunction
\end{algorithmic}
\caption{Pseudocode of the Viterbi algorithm}\label{alg:viterbi-pseudocode}
\end{algorithm}

\subsection{Jump Alignment}\label{sec:jump-alignment-algorithm}
Jump Alignment is an extension to the HMM described in the previous section, which utilizes the chords that are parsed from tabs. Following \cite{mcvicar2011using}, we refer to these chords parsed from tab files as Untimed Chord Sequences (UCSs). Compared to the original HMM, in the Jump Alignment algorithm the state space and transition probabilities are altered in such a way that it can align the UCSs to audio, while allowing for jumps to the start of other lines. We will explain how this works in the remainder of this section.

Let us assume that we have an audio file and tab file from the same song. Let $\mathbf{e} = e_1, e_2, \ldots, e_L$ be the UCS we extracted from the tab file, applying the parsing algorithm described in Section~\ref{sec:tab-parse}. Because we want an alignment between the audio file and this UCS $\mathbf{e}$, we define a new set of hidden states: $Y' = \{1, 2, \ldots, L\}$ corresponds to the indices of the chords in $\mathbf{e}$. For example, given a (very small) chord vocabulary $[C, D, E]$ and the UCS $[C, E, D, E, D, C, E]$, our hidden state set $Y'$ would be $[1, 3, 2, 3, 2, 1, 3]$. Note that different hidden states can now refer to different occurrences of the same chord, in contrast to our original state space.

We also need to alter the transition probability distribution $P'_{tr}$, as we only want to allow chord sequences that are alignments of $\mathbf{e}$, our Untimed Chord Sequence. However, \cite{mcvicar2011using} showed that exact alignment of $\mathbf{e}$ to audio results in a decrease in performance compared to the original HMM, because repetition cues (like ``play this verse twice'') are not interpreted by the tab parser. In addition, some tabs do not specify the chords of all verses. Therefore, Jump Alignment allows jumps from the end of a line in the tab file to the beginning of any line (the current one, of any of the previous or subsequent lines). The new transition probability distribution is expressed as:

\begin{equation}
P'_{tr} (j | i, \Theta, e) = \begin{cases}
\frac{1}{Z_i} P_{tr} (e_i, e_j) & \text{if } j \in \{i, i+1\},\\
\frac{p_f}{Z_i} P_{tr} (e_i, e_j) & \text{if } i < j \text{ and } i \text{ is the end and } j \text{ is the beginning of a line},\\
\frac{p_b}{Z_i} P_{tr} (e_i, e_j) & \text{if } i > j \text{ and } i \text{ is the end and } j \text{ is the beginning of a line},\\
0 & \text{otherwise}.
\end{cases}
\end{equation}
$Z_i$ is a normalization factor, which is used to re-normalize $P'_{tr}$ so that the transition probabilities $P'_{tr}(y_j | y_i, \Theta, e)$ sum to one. $p_f$ is the probability of jumping forward when this is allowed. Likewise, $p_b$ is the probability of a jump backwards, provided that this is allowed.

The forward probability $p_f$ and backward probability $p_b$ need to be specified. We estimate the optimal $p_f$ and $p_b$ as follows: first, we manually select a subset of 22 songs for which we are sure that they are in the right key (so we do not need to consider transpositions). Appendix~\ref{appendix:tabs-pf-pb-training-set} lists these songs. Then, we run the Jump Alignment algorithm with all combinations of forward and backward probabilities between 0 and 1 in steps of 0.05 and evaluate the resulting chord label sequences in terms of \textsc{csr}. We will use the $p_f$ and $p_b$ with the highest average \textsc{csr}.

\begin{figure}
 \centering
 \includegraphics[width=0.7\textwidth]{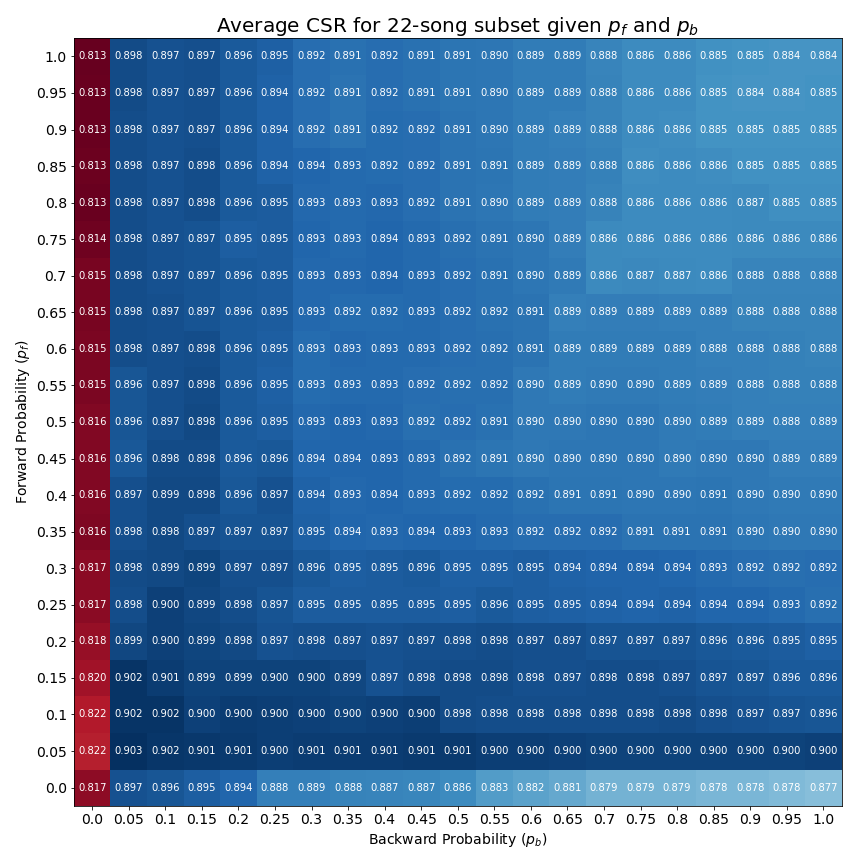}
 \caption{Average \textsc{csr} for 22-song subset given $p_f$ and $p_b$. We find an optimal parameter setting of $p_f = 0.05$ and $p_b = 0.05$.}\label{fig:forward-backward-probabilities}
\end{figure}

Figure~\ref{fig:forward-backward-probabilities} shows the average \textsc{csr} for our 22-song subset, given a $(p_f, p_b)$ parameter setting. We see that positive forward or backward probabilities improve \textsc{csr} compared to a strict alignment (in which $p_f = 0$ and $p_b = 0$): particularly non-zero backward probabilities yield considerably better chord label sequences than $p_b$'s of zero. However, if the forward and backward probabilities become too large, the average \textsc{csr} decreases as the importance of the order of lines in the tab file decreases. We find an optimal parameter setting of $p_f = 0.05$ and $p_b = 0.05$. This is the parameter setting that we use in our implementation of Jump Alignment.

Tabs are often notated in a transposed key compared to the original audio file, because some keys are easier to play on the guitar than others. In order to correct for such ``simplified'' tabs, Jump Alignment considers the tab files in all 12 transpositions and chooses the transposition with the highest likelihood.

\section{Tab file selection}\label{sec:tab-selection}
The performance of the jump alignment algorithm is dependent on the quality of the tab file. Following \citet{mcvicar2011using}, we select for each song the tab file for which the log-likelihood was the highest. The results for our full 200 song dataset are shown in Table~\ref{tab:tab-file-selection-evaluation}. If we would take the average \textsc{csr} of all tabs for a song, the \textsc{wcsr} is 72.4\%. The upper limit for the selection method is a \textsc{wcsr} of 78.7\%: this is what we would get if we selected the best tab. 
Since we do not know the \textsc{csr} for unlabelled data, we select the expected best tab-file for each song based on the log-likelihood, following \cite{mcvicar2011using}. 
By choosing the tab file with the highest log-likelihood for each song, the resulting \textsc{wcsr} improves to 75.3\%. The distribution of \textsc{csr}s of selected tab files compared to all tab files is also shown in Figure~\ref{fig:hist-symbolic}.

\begin{table}
\centering
\begin{tabular}{ll}
\hline
\textbf{} & \textbf{\textsc{wcsr}} \\ \hline
\textbf{Worst \textsc{csr} of all tabs} & 59.1\% \\
\textbf{Average \textsc{csr} of all tabs} & 72.4\% \\
\textbf{Best log-likelihood of all tabs} & 75.3\% \\
\textbf{Best \textsc{csr} of all tabs} & 78.7\% \\ \hline \\
\end{tabular}
\caption{\textsc{wcsr} of all songs, with different tab file selection methods}\label{tab:tab-file-selection-evaluation}
\end{table}

\section{Conclusion}\label{sec:tab-conclusion}
In this section, we have examined the last of the three representation-specific subsystems of \textsc{decibel}: the tab subsystem. This subsystem consists of a parser that extracts Untimed Chord Sequences (UCSs) from guitar tablature or chord sheets, thereby retaining line information. This is then input for the Jump Alignment algorithm, which aligns the UCSs to the audio recording. When selecting the estimated best tabs for each song, based on log-likelihood, this subsystem reaches a \textsc{wcsr} of 75.3\% on the Isophonics data set.

\chapter{Data fusion}\label{ch:data-fusion}
\textsc{Decibel} estimates chord label sequences from different music representations, i.e. audio, \textsc{midi} and tab files, as we have seen in the previous three chapters. This results in a set of chord label sequences for each song in our data set. This set of chord labels forms a rich harmonic representation that is already interesting in itself, as we will see in Section~\ref{sec:harmonic-representation}. However, in order to answer our research question we need to combine these chord label sequences into one final sequence (and compare the resulting chord sequence to the sequence obtained by using only an audio \textsc{ace} method). \textsc{Decibel} achieves this using a data fusion step. In this chapter, we motivate the chosen data fusion method by a literature study and an evaluation of various methods.

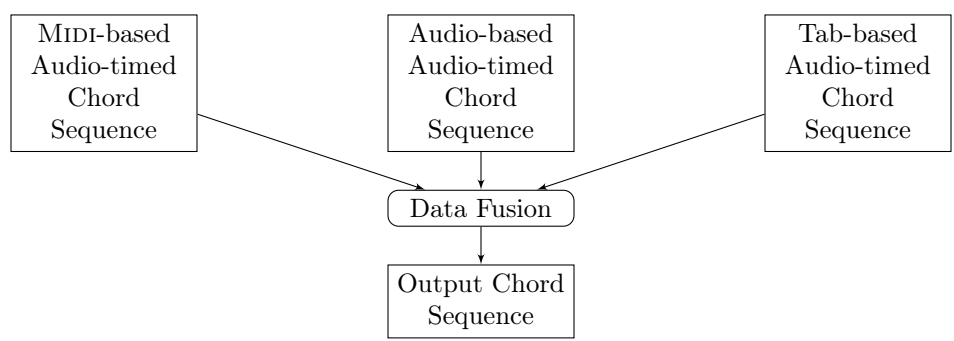
\begin{figure}[ht]
\centering
\begin{tikzpicture}[node distance = 0.5cm and 2.5cm, auto]
    \node [rect] (midi_cs) {\textsc{Midi}-based Audio-timed Chord Sequence};
    \node [rect, right=of midi_cs] (audio_cs) {Audio-based Audio-timed Chord Sequence};
    \node [rect, right=of audio_cs] (tab_cs) {Tab-based Audio-timed Chord Sequence};
    \node [block, below=of audio_cs] (data_fusion) {Data Fusion};
    \node [rect, below=of data_fusion] (output) {Output Chord Sequence};
    
    \path[line] (midi_cs) -- (data_fusion);
    \path[line] (audio_cs) -- (data_fusion);
    \path[line] (tab_cs) -- (data_fusion);
    \path[line] (data_fusion) -- (output);
\end{tikzpicture}
\caption{Diagram of \textsc{decibel}'s data fusion subsystem. The set of \textsc{midi}- audio- and tab-based audio-timed chord label sequences forms a rich harmonic representation. Our data fusion combines all chord label sequences for a song into a single output chord label sequence.}
\label{fig:diagram-df}
\end{figure}

\section{A rich harmonic representation}\label{sec:harmonic-representation}
In the previous chapters we have seen how \textsc{decibel}'s subsystems estimate chord label sequences from audio, \textsc{midi} and tab representations of a song. These three representations store musical content of the same song in fundamentally different ways: as we have seen in Section~\ref{sec:musicrepresentations}, audio files are obtained by digitizing the waveform; \textsc{midi} files are a series of note on and note off events; whereas tab files show the guitar fingering or chord labels, aligned to the lyrics. The timing information in audio is represented in seconds, based on the underlying performance; in \textsc{midi} files, timing is measured in ticks; in tabs, timing information is missing. Also, the source of information differs: audio files are recordings of a performance; \textsc{midi} files are either score-based or transcriptions of a recording; tabs can be considered as (untimed) chord label sequences, manually annotated by music enthusiasts.

However, the chord label estimation and synchronization steps performed by \textsc{decibel}'s subsystems transform these three heterogeneous representations into a homogeneous format, i.e. as a series of $\langle$start time, end time, chord label$\rangle$ 3-tuples. The combination of the set of chord label sequences from each of the representations of a song results in a rich harmonic representation. 

This rich harmonic representation is a very interesting by-product of the \textsc{decibel} system, as it allows for cross-version analysis, i.e. comparing analysis results from different representations. The chord label sequences can easily be visualized, which makes it very easy to see the consistencies and inconsistencies in chord labels between different representations. As identified earlier in research by \citet{ewert2012towards} and \citet{konz2013cross}, having such a unified view of different analysis results deepens the understanding of both the algorithm's behavior and the properties of the underlying musical material.

\begin{figure}[ht]
\centering
\includegraphics[width=\textwidth]{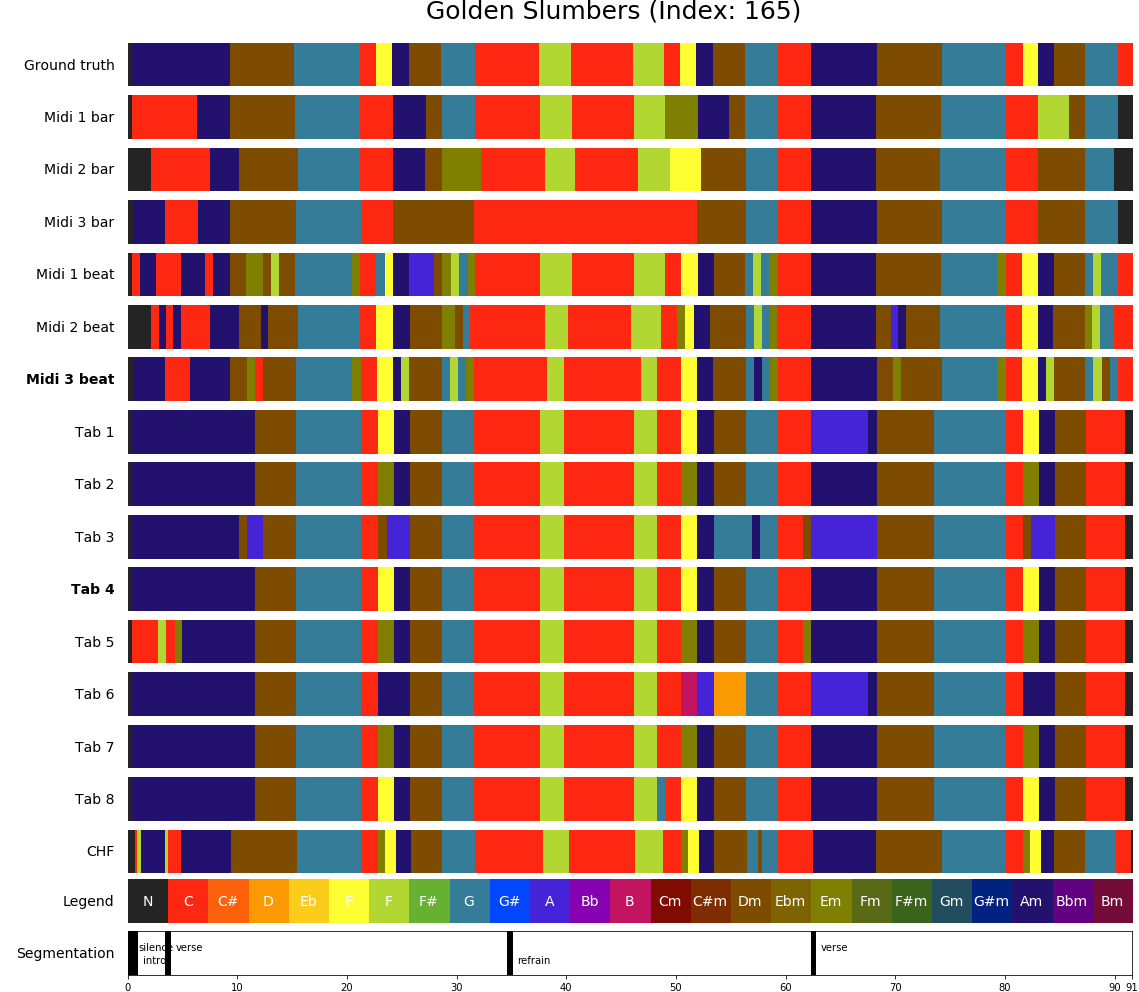}
\caption{Multiple perspectives on the Beatles song \textit{Golden Slumbers}: we can consider the start of the song as a long A minor chord or as a sequence of A minor and C major chords and there are multiple opinions on e.g. the modality of the chord starting after 62 seconds.}\label{fig:165}
\end{figure}

Consider for example the visualization of the chord label sequence for the Beatles song \textit{Golden Slumbers} in Figure~\ref{fig:165}. In our data set, we have matched three \textsc{midi} files and eight tab files to the audio version of the song. \textsc{Cassette}, our \textsc{midi ace} method, analyzes the chords both on a bar and a beat level, so we have six chord label sequences based on the three \textsc{midi} files. In combination with the eight chord label sequences based on the tab files and a single analysis by the Chordify algorithm based on the audio file, we have $6 + 8 + 1 = 15$ chord label sequences. Each of these estimated chord sequences, as well as the ground truth, is visually represented by a horizontal bar in which the color represent the chord label and time in seconds can be read from the position on the $x$-axis. In this example, we observe there are some parts of the song for which (most of) the 15 estimated chord sequences agree on the labels, e.g. the D minor chord starting after 11 seconds; the G major chord sounding immediately thereafter or the C major chord that starts at 32 seconds. However, in other musical passages there is some disagreement. If there is a single deviating chord sequence estimation, this is typically either due to some error in the \textsc{midi} or tab file or due to an error in the estimation method. For example, when estimating the chord sounding in the 38th second from our third \textsc{midi} file, using \textsc{cassette} on the bar level, we falsely find a F major chord instead of a C major chord. In some passages, we observe multiple ``clusters'' of possible chord labels, i.e. different perspectives on the harmonic content of a musical passage, each of them supported by multiple chord label estimations. For example, we can consider the start of the song either as a long A minor chord or as a sequence of A minor and C major chords. Also, there exist different perspectives on the chord in the 23rd second (is it a C major, E major, F minor or something else?) and the modality of the A (major or minor) chord starting after 62 seconds.

\textsc{Midi} and tab files implicitly incorporate musical knowledge, as they are (in the popular music genre) typically transcriptions made by music enthusiasts. Therefore, the comparison of analysis results from different music representations reveals passages in the music for which there are multiple possible perspectives. Studying these passages can give us a deeper understanding of subjectivity issues in the musical material.

In this section, we have seen that the chord label sequences estimated on different music representations of the same song together form a rich harmonic representation. Visualization of this harmonic representation can give valuable insights in different perspectives on chord labels within a specific segment of the song. This makes large-scale cross-version analysis feasible. Moreover, it bridges the gap between technical, audio-oriented, musical signal processing and non-technical, score-oriented musicology. A detailed analysis on the different harmonic perspectives on the Beatles and Queen songs from our data set is beyond the scope of this research, but would be an interesting direction of future work which would provide better insights in the subjectivity issues in harmonic analysis of popular music.

\section{Related Work on data fusion}\label{sec:df-related-work}
In the remainder of this chapter, we study data fusion methods to combine chord label sequence obtained from \textsc{decibel}'s audio, \textsc{midi} and tab subsystems into a single output sequence. The integration of heterogeneous output of multiple \textsc{ace} algorithms using data fusion is a new idea, which was recently proposed by \citet{koops2016integration}. In this study, the authors experiment with three different techniques to combine chord sequence estimates from different sources (the \textsc{mirex} 2013 \textsc{ace} submissions, applied to the Billboard data set) into one final output sequence for each song. They show that their proposed data fusion method yields the best results in terms of \textsc{wcsr}. Also, they show that the output sequence found by their data fusion method is an improvement to the best scoring team, with an increase between 3.6 percentage point and 5.4 percentage point compared to the best team.

The three techniques used for combining chord label sequences are random picking (\textsc{rnd}), majority voting (\textsc{mv}) and data fusion (\textsc{df}). In all three techniques, the system proposed in \citet{koops2016integration} first samples the chord label sequence from each source in segments of 10 milliseconds. In the \textsc{rnd} technique, the output chord label for each segment is found by picking the corresponding label from a randomly chosen source. For \textsc{mv}, the output chord label for each segment is the most frequent label of all sources. If multiple labels are most frequent, the output label is picked randomly from these most frequent labels. The \textsc{df} technique takes three things into account when integrating the output of multiple \textsc{ace} algorithms:\begin{itemize}
\item The accuracy of sources;
\item The probabilities of the values provided by the sources; and
\item The probability of dependency between sources.
\end{itemize}

The \textbf{source accuracy} $A(S_i)$ is the arithmetic mean of the probabilities of all chord labels a source $S_i$ provides. Initially, these chord label probabilities may be based on a frequency count. A source with a high source accuracy inherently has a lot of chord labels that agree with other sources, and therefore can be considered more trustworthy. Assuming that the sources are independent, the source accuracy is the probability that a source provides the appropriate chord. On the other hand, the probability that a source provides an inappropriate chord can be computed with $\frac{1 - A(S_i)}{n}$, given $n$ possible inappropriate chord labels. The \textbf{source vote count} $VS(S_i)$ combines these probabilities and is computed as follows: $VS(S_i) = \ln \frac{n \cdot A(S_i)}{1 - A(S_i)}$. Chord labels of sources with higher source vote counts are more likely to be picked than labels of sources with low source vote counts, as we will see shortly.

The goal of the data fusion method is to determine the most likely chord label for each segment, given a number of sources. To this end, the data fusion method computes a \textbf{chord label vote count} $VC(\mathcal{L})$ for each chord label $\mathcal{L}$. The chord label vote count is dependent on both the number of sources with a matching chord label and the source vote count of these sources, as expressed in the formula $VC(\mathcal{L}) = \sum_{\sigma \in S^\mathcal{L}} VS(\sigma)$, in which $S^\mathcal{L}$ is the set of sources with chord label $\mathcal{L}$. 

From the label vote counts, we can compute \textbf{chord label probabilities} for each chord label $\mathcal{L}$ by $P(\mathcal{L}) = \frac{\exp(VC(\mathcal{L}))}{\sum_{l \in D} \exp(VC(l))}$. As we divide the chord label vote count of our chord label by the chord label vote count of all possible chord labels $D$, this results in a probability value between 0 and 1.

The data fusion method proposed by \citet{koops2016integration} also takes the dependency between sources into account, following the intuition that sources that are dependent of each other share many uncommon chord labels. Based on the number of shared uncommon chords with other sources, the authors compute a \textbf{source dependency} weight $\mathcal{I}(S_i, \mathcal{L})$, which represents the probability that a source $S_i$ provides a chord label $\mathcal{L}$ independently. 

We now have seen the definitions of chord label probabilities, source accuracy and source dependency. Note that they are all defined in terms of each other. As a solution to this paradoxical situation, \citet{koops2016integration} propose to initialize the chord label probabilities with equal probabilities and iteratively compute source dependency, chord label probabilities and source accuracy until convergence is reached. Finally, their data fusion method selects for each segment the chord label with the highest chord label probability.

In their experiments, the authors compare the performance of \textsc{df} to \textsc{rnd} and \textsc{mv} for two subsets of the Billboard data set. Each of these three chord sequence combination methods is computed on all twelve \textsc{mirex} 2013 \textsc{ace} submissions for both Billboard subsets. The Billboard annotations serve as ground truth. Their results show that the \textsc{df} method significantly outperforms both the \textsc{rnd} and \textsc{mv} methods as well as each of the individual submissions. We can conclude that combining chord label sequences from multiple sources using data fusion is a promising strategy to improve \textsc{ace}.

\section{Data fusion experiments}\label{sec:df-experiments}
In the previous section, we have seen that the heterogeneous output of multiple \textsc{ace} systems can be integrated into a single output chord sequence, using data fusion. Although the results obtained in previous work are solely based on audio data, the data fusion method can also be applied to chord label sequences that are based on symbolic data. This is exactly what \textsc{decibel}'s data fusion subsystem does: the subsystem uses the chord sequences from audio, \textsc{midi} and tabs and integrates them in a final output sequence. In contrast to earlier work, in our case the number of sources differs per song. That is why we experiment with a selection strategy that elects only one source per representation. In this section, we will explain \textsc{decibel}'s data fusion implementations.



In order to test if \textsc{decibel} improves current audio \textsc{ace} systems, we need to combine these chord label sequences into one final sequence. \textsc{Decibel} achieves this using a data fusion step. In this section, we motivate the chosen data fusion method by an evaluation of various methods. These methods are based on earlier work by  \cite{koops2016integration}. In contrast to earlier work, \textsc{decibel} needs to combine a varying number of sources per song. That is why we experimented with a selection strategy that selects only one source per representation. We compare two selection strategies in combination with three different integration methods. The two selection strategies are \textsc{all} and \textsc{best}: \textsc{all} takes the chord sequences of all tabs and \textsc{midi} files as sources. \textsc{Best} only uses the sources of the expected best tab and \textsc{midi} file for each song, as described in Section~\ref{sec:midi-selection} and \ref{sec:tab-selection}. There are eight songs for which no \textsc{Midi} is selected, since all \textsc{midi}s were badly aligned. For these songs, \textsc{Best} only combines the audio and tab sequences.

The three integration methods are based on earlier work \citep{koops2016integration}. We first divide each input chord sequence in 10 millisecond samples. Then we integrate the sources, selected using \textsc{all}/\textsc{best}, with either random picking (\textsc{rnd}), majority voting (\textsc{mv}) or data fusion (\textsc{df}). 
The implementations of \textsc{rnd} and \textsc{mv} are unchanged compared to \cite{koops2016integration}: given a specific sample, \textsc{rnd} takes the chord label of an arbitrary source, while \textsc{mv} assigns the chord label used by most of the sources.

Our implementation of the \textsc{df} technique takes into account both the expected accuracy of sources and the probability of the labels provided by the sources.
Let $\mathcal{X}$ be the set of samples, $\mathcal{V}$ be the chord vocabulary, $\mathcal{S}$ be the sources selected by the integration method and let $\mathcal{L}: \mathcal{S} \times \mathcal{X} \times \mathcal{V} \mapsto \{0, 1\}$ be a labelling function such that $\mathcal{L}(s,x,v) = 1$ if source $s$ assigns chord $v$ to segment $x$ and $0$ otherwise. 

The source accuracy $A[s]$ is the probability that a source $s$ provides appropriate chords; the chord label probability $P[x, v]$ is the probability of chord label $v$ at segment $x$, given the chord labels of the sources. The chord label probability is defined based on the labelling $\mathcal{L}$ and the chord label vote count $VC[x,v]$. This value determines the influence of each source on the final label, based on its source accuracy. These values are iteratively computed as follows:

\begin{equation}\label{eq:source-accuracy}
    A[s] = \frac{\sum_{x \in \mathcal{X}} \sum_{v \in \mathcal{V}} P[x,v] \cdot \mathcal{L}(s,x,v)}{|\mathcal{X}|}
\end{equation}
\begin{equation}\label{eq:vote-count}
    \textit{VC}[x,v] = \sum_{s \in \mathcal{S}} \mathcal{L}(s,x,v) \cdot \ln{\frac{(|\mathcal{V}|-1) A[s]}{1 - A[s]}}
\end{equation}
\begin{equation}\label{eq:chord-label-probability}
    P[x,v] = \frac{\exp{\textit{VC}[x,v]}}{\sum_{v' \in \mathcal{V}} \exp{\textit{VC}[x,v']}}
\end{equation}

$A[s]$, $\textit{VC}[x,v]$ and $P[x,v]$ are defined in terms of each other. Therefore, we use an alternative equation for the initial computation of the chord label probability $P_0[x,v]$:
\begin{equation}\label{eq:chord-label-probability-initial}
    P_0[x,v] = \frac{\sum_{s \in \mathcal{S}} \mathcal{L}(s, x, v)}{|\mathcal{S}|}
\end{equation}

The final chord labelling for \textsc{df} is obtained by first computing Equation~$\ref{eq:chord-label-probability-initial}$ and then repeatedly updating Equations~\ref{eq:source-accuracy}--\ref{eq:chord-label-probability}. Whereas~\cite{koops2016integration} repeat until a fixed point is reached, we always repeat 5 times. After this, we assign the chord label $v$ with the highest probability $P[x,v]$ to each segment $x$ to obtain the final \textsc{df} chord sequence.

\section{Data fusion results}
\begin{figure}[p!]
 \centering
  \begin{subfigure}[b]{0.49\textwidth}
 	\includegraphics[width=\textwidth]{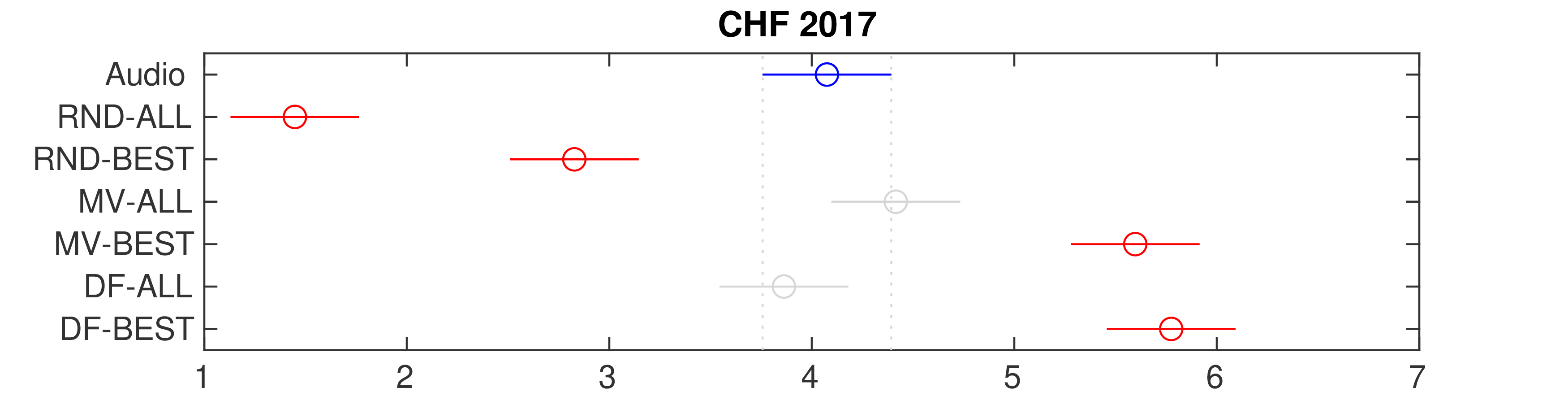}
    \caption{\textsc{chf} (2017)}
 \end{subfigure}
 \begin{subfigure}[b]{0.49\textwidth}
 	\includegraphics[width=\textwidth]{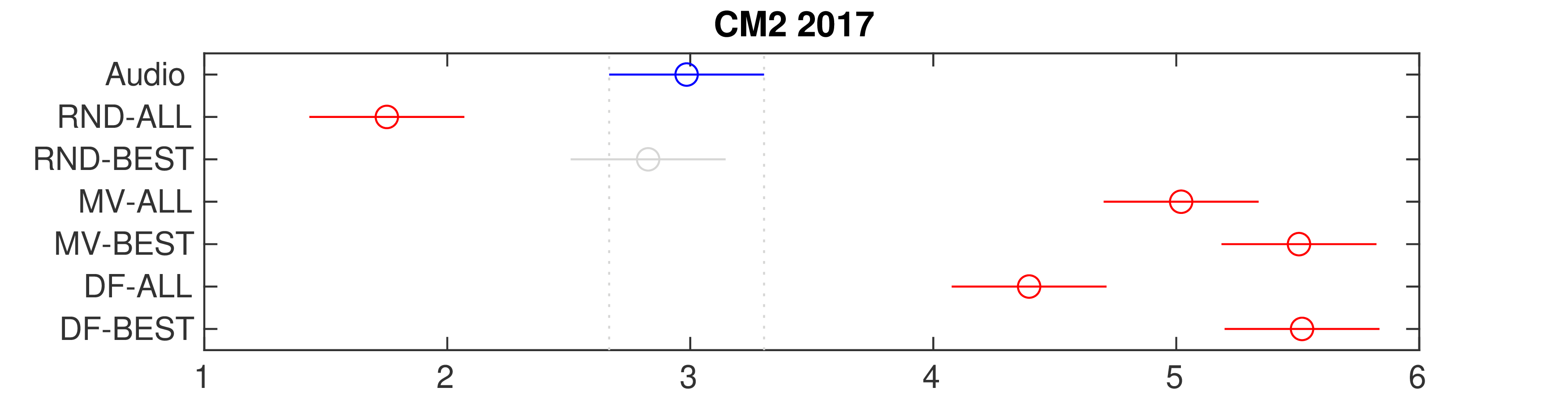}
    \caption{\textsc{cm2} (2017) / \textsc{cm1} (2018)}
 \end{subfigure}
 \begin{subfigure}[b]{0.49\textwidth}
 	\includegraphics[width=\textwidth]{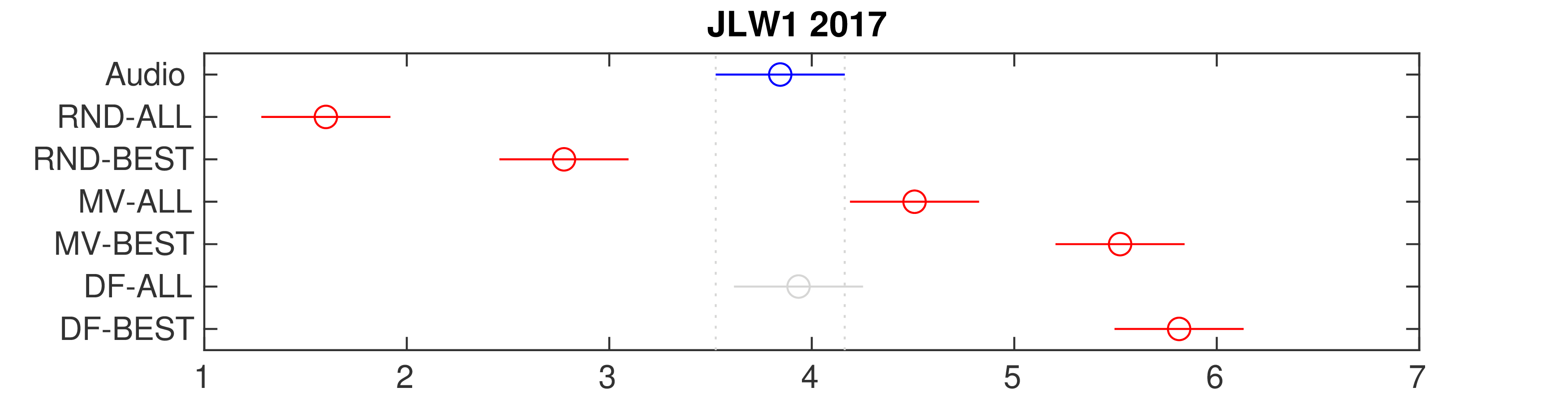}
    \caption{\textsc{jlw1} (2017)}
 \end{subfigure}
 \begin{subfigure}[b]{0.49\textwidth}
 	\includegraphics[width=\textwidth]{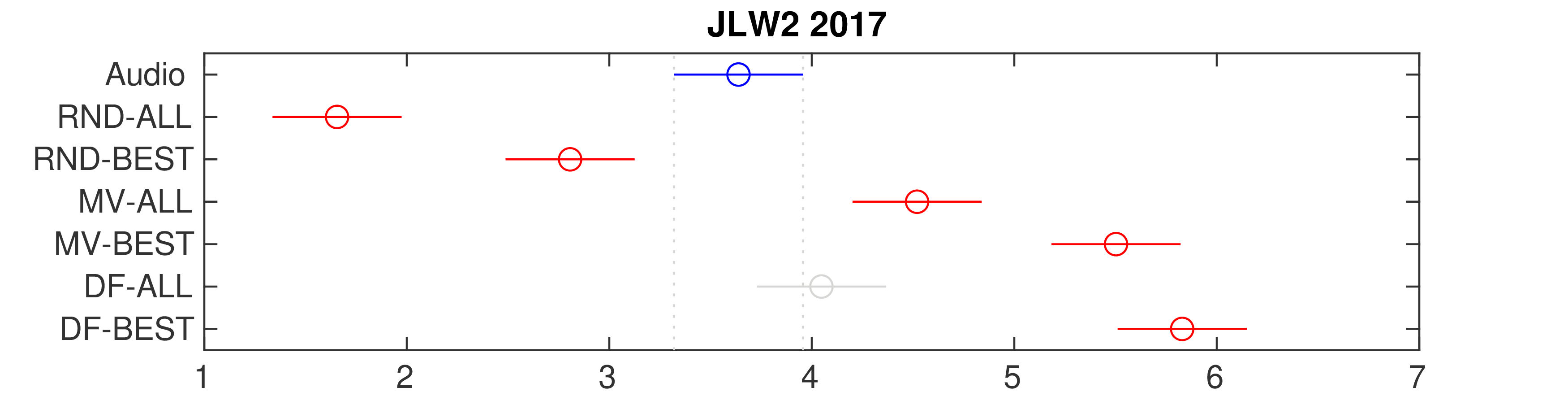}
    \caption{\textsc{jlw2} (2017)}
 \end{subfigure}
 \begin{subfigure}[b]{0.49\textwidth}
 	\includegraphics[width=\textwidth]{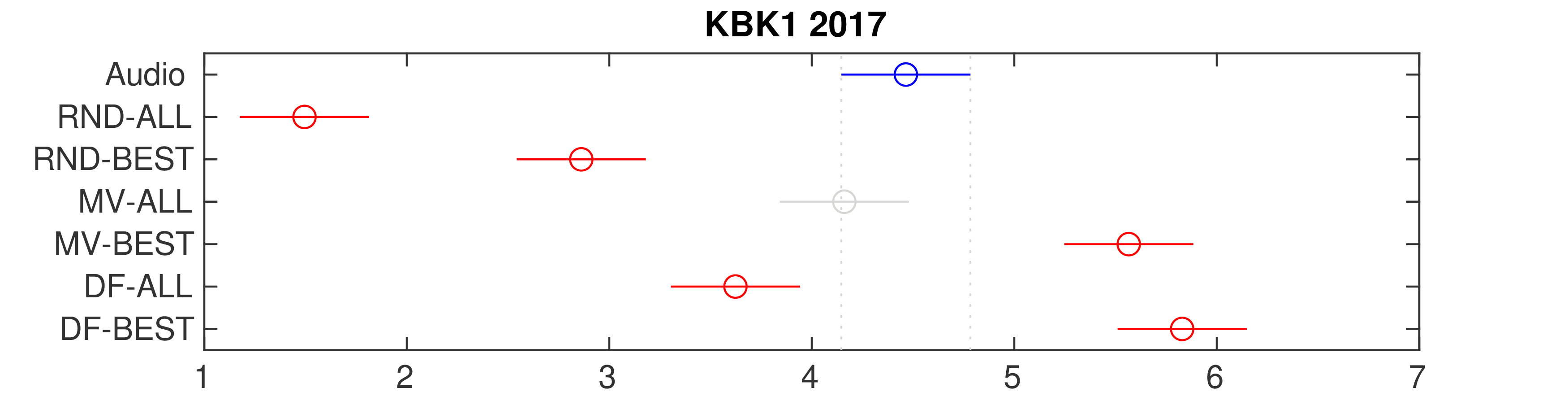}
    \caption{\textsc{kbk1} (2017)}
 \end{subfigure}
 \begin{subfigure}[b]{0.49\textwidth}
 	\includegraphics[width=\textwidth]{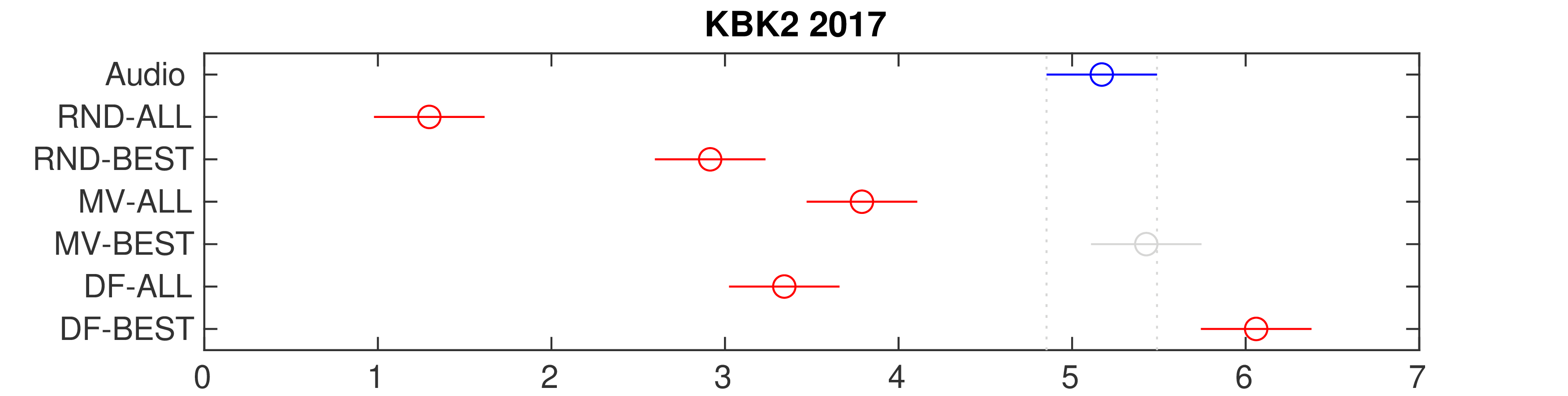}
    \caption{\textsc{kbk2} (2017) / \textsc{fk2} (2018)}
 \end{subfigure}
 \begin{subfigure}[b]{0.49\textwidth}
 	\includegraphics[width=\textwidth]{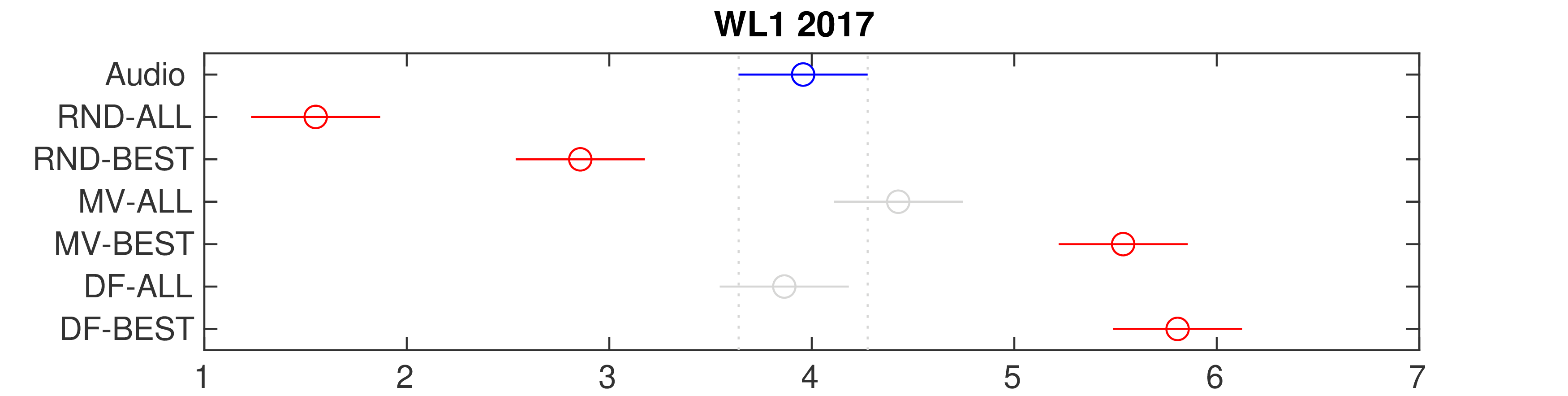}
    \caption{\textsc{wl1} (2017)}
 \end{subfigure}
  \begin{subfigure}[b]{0.49\textwidth}
 	\includegraphics[width=\textwidth]{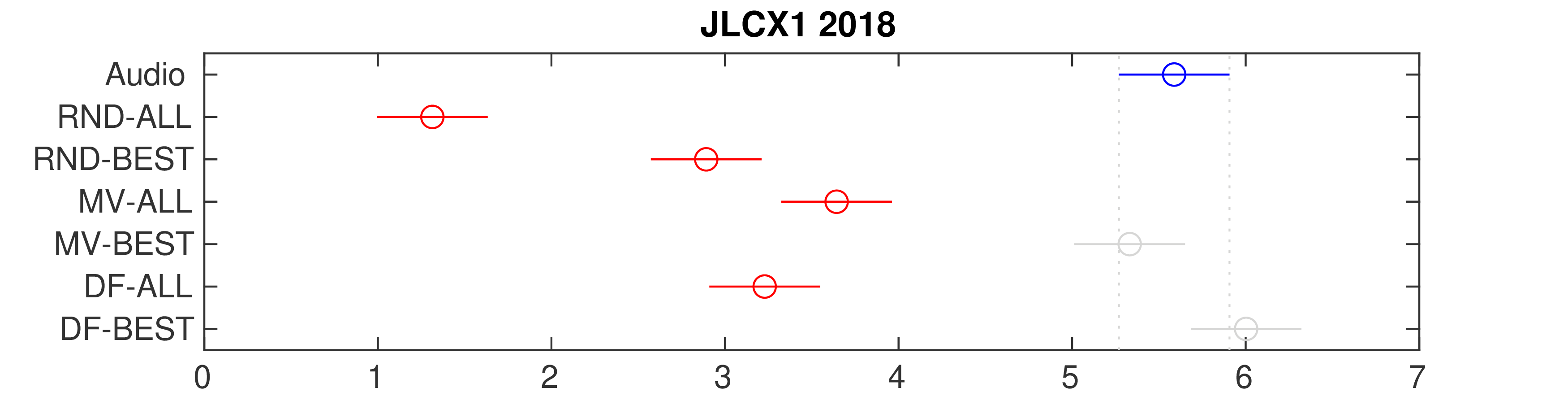}
    \caption{\textsc{jlcx1} (2018)}
 \end{subfigure}
 \begin{subfigure}[b]{0.49\textwidth}
 	\includegraphics[width=\textwidth]{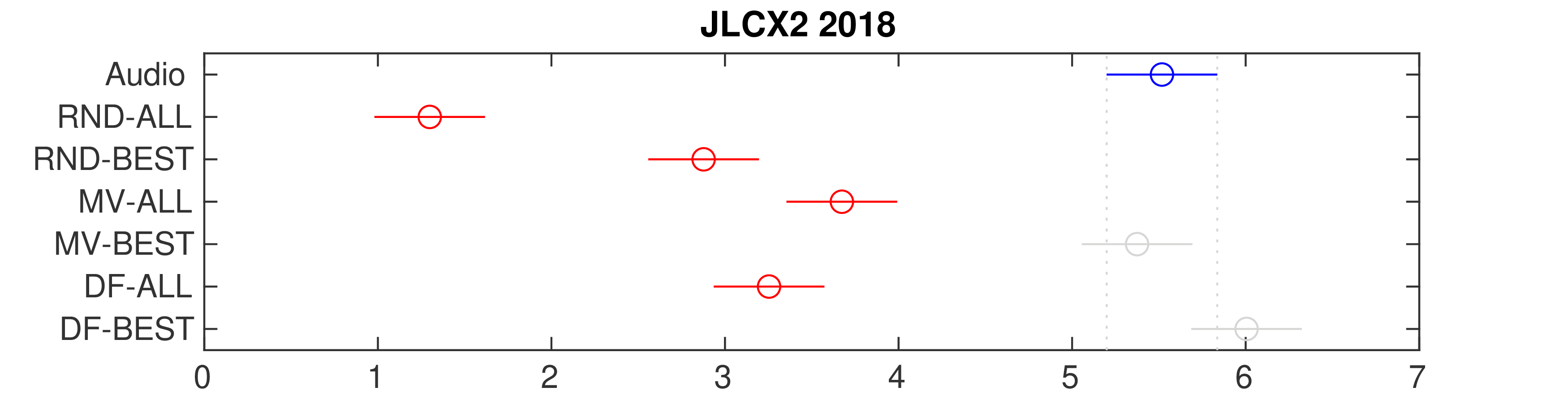}
    \caption{\textsc{jlcx2} (2018)}
 \end{subfigure}
  \begin{subfigure}[b]{0.49\textwidth}
 	\includegraphics[width=\textwidth]{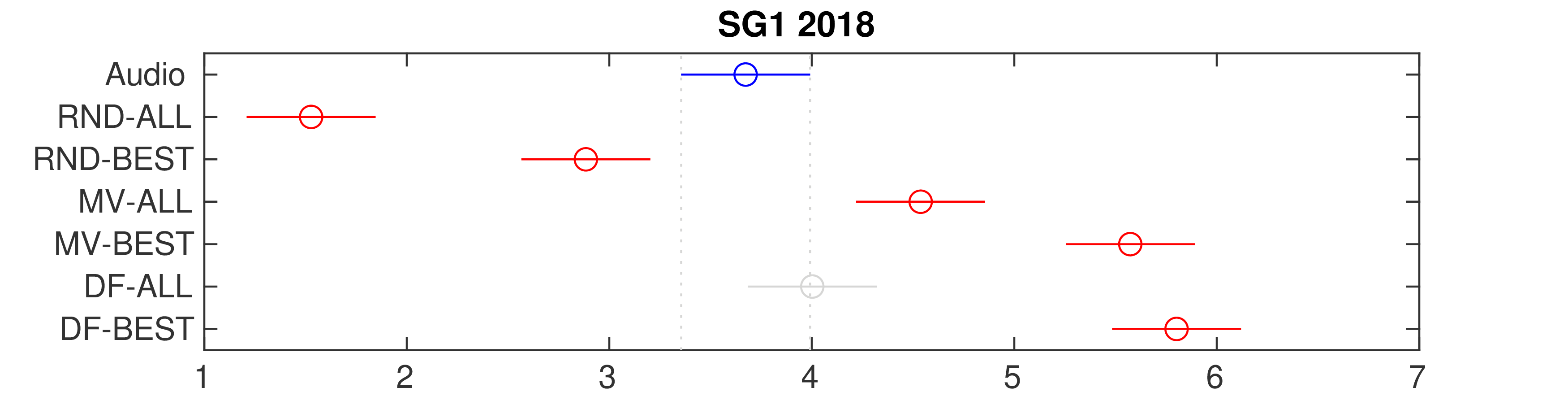}
    \caption{\textsc{sg1} (2018)}
 \end{subfigure}
 \caption{Visual representation of the differences in terms of Chord Symbol Recall between different data fusion methods, for \textsc{chf} and each of the \textsc{mirex} 2017 and/or 2018 \textsc{ace} submissions. For each of the horizontal lines that do \textbf{not} overlap in the $y$-direction, the difference in \textsc{csr} between the corresponding data fusion methods is significant.}\label{fig:csr-df-mirex2017}
\end{figure}

We now compare the performance of each of the six combinations of integration methods and selection strategies (i.e. \textsc{rnd-all}, \textsc{rnd-best}, \textsc{mv-all}, \textsc{mv-best}, \textsc{df-all} and \textsc{df-best}), applied to each of our ten audio \textsc{ace} systems. Friedman tests~\citep{friedman1937use} for each of these ten audio \textsc{ace} show that the integration methods and selection strategies give significantly different results in terms of \textsc{csr}. Consequently, we perform a Tukey's Honest Significant Difference post-hoc test~\citep{tukey1949comparing} to identify which integration-selection combinations are significantly different. The results are visually represented in Figure~\ref{fig:csr-df-mirex2017}. In these figures, we can easily observe which differences are significant because the corresponding horizontal lines do not overlap in the $y$-direction. Consider for example Figure~\ref{fig:csr-df-mirex2017}a. Here we see that both \textsc{rnd} methods are significantly worse than the original audio \textsc{ace} system (\textsc{chf}). There is no significant difference between \textsc{chf} and \textsc{df-all}. \textsc{Mv-all} is better than \textsc{chf}, but the difference is not found significant by Tukey's HSD test. \textsc{Mv-best} and \textsc{df-best} perform significantly better than the original \textsc{chf} system, and \textsc{df-best} turns out to be the winner.

When examining the test results for all ten audio \textsc{ace} systems for all ten state-of-the-art audio \textsc{ace} systems in combination with all two selection strategies and all three integration methods, we make four observations.

\textit{First}, the random picking integration method always performs worse than the original audio \textsc{ace} system, regardless of the chosen selection strategy. This is consistent with findings in earlier work \citep{koops2016integration}. 
From this observation, we conclude that agreements between sources should be taken into account.

\textit{Second}, the \textsc{best} selection method always performs better than the \textsc{all} selection method. The explanation for this is as follows: given a poor \textsc{midi} or tab file, \textsc{decibel}'s \textsc{midi} or tab subsystem will find a poor chord label sequence. In the \textsc{all} selection method, this chord label sequence will, to a greater or lesser extent, be integrated in the final output sequence, which deteriorates the output sequence's quality, whereas the \textsc{best} selection method (hopefully) ignores this poor chord label sequence. 

\textit{Third}, we observe that the difference between \textsc{best} and \textsc{all} is even larger for the \textsc{df} integration method than for \textsc{mv}. A possible explanation for this is that chord label sequences from tabs are often undersegmented, for example because chord changes in instrumental parts or very short chords are often not detected. This is typical for tabs and therefore can occur at multiple (independent) tabs of the same song. Consequently, suboptimal chord sequences from tabs could get a high source accuracy in \textsc{df-all}, and therefore substantially influence the final output sequence. \textsc{Df-best} only considers the expected best tab and \textsc{midi} file. If the expected best tab is undersegmented, it will probably get a low source accuracy, because chord sequences from \textsc{midi} and audio have less undersegmentation issues. The difference between \textsc{mv-all} and \textsc{mv-best} is smaller, as \textsc{mv} does not use source accuracies.

Our \textit{fourth} and final observation is that \textsc{df-best} performs better than \textsc{mv-best}, although the difference is not always significant. However, \textsc{df-best} is always better than the original audio \textsc{ace} system. As shown in Table~\ref{tab:wcsr}, the improvement is significant for all tested audio methods except \textsc{jlcx1} and \textsc{jlcx2}.

\begin{figure}[ht]
\centering
\includegraphics[width=\textwidth]{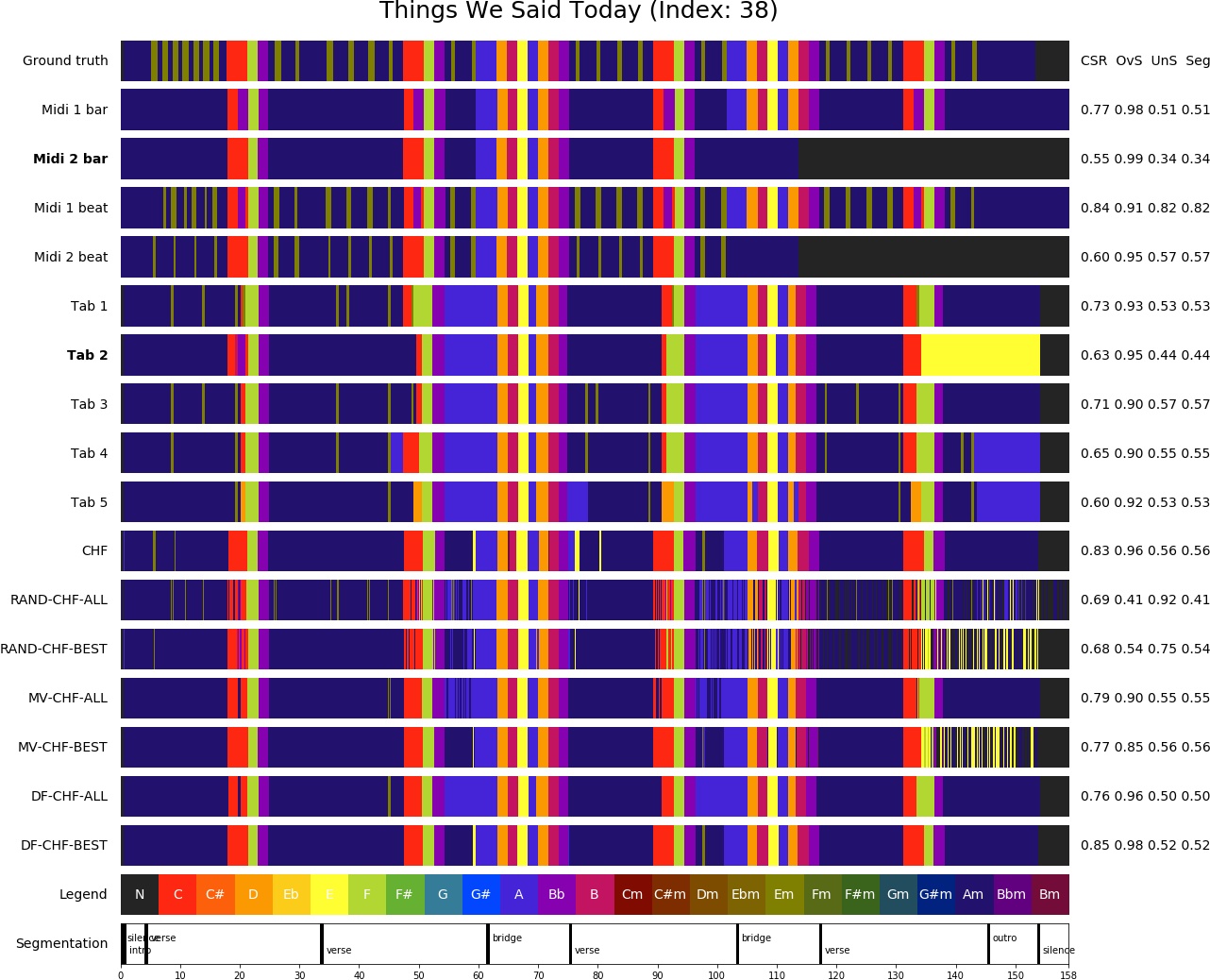}
\caption{Chord label sequences for the Beatles song \textit{Things We Said Today}. \textsc{Csr} = Chord Symbol Recall; OvS = oversegmentation; UnS = undersegmentation; and Seg = segmentation. Note that \textsc{rand} and \textsc{mv} tend to oversegment, in contrast to \textsc{df}.}\label{fig:38}
\end{figure}

When comparing the oversegmentation measure for our integration methods, we can do another observation: \textsc{rand} and \textsc{mv} tend to oversegment, in contrast to \textsc{df}. This becomes clear when looking at the chord sequences for the song \textit{Things We Said Today} in Figure~\ref{fig:38}. Note that both \textsc{rnd} chord sequences have way too many chord changes, resulting in a low oversegmentation score. Both \textsc{mv} chord sequences are oversegmented as well, albeit to a lower degree. On the other hand, we see no oversegmentation issues in the \textsc{df} chord label sequence.

From these observations, we can conclude that \textsc{df-best} is the best selection-integration combination. Table~\ref{tab:wcsr} shows a comparison between the \textsc{wcsr} of each of the ten audio \textsc{ace} systems and \textsc{df-best} applied to each system. The table shows that using \textsc{df-best} improves \textsc{ace} \textsc{wcsr} on average by 3.05\%. 

\begin{table}
\centering
\begin{tabular}{ l l r r r r }
\textbf{Audio ACE}&\textbf{\textsc{Mirex}}&\textbf{Audio \textsc{wcsr}}&\textbf{\textsc{Df-best wcsr}}&\textbf{Improvement}&\textbf{\textsc{Df-gt-best wcsr}}\\
\hline
\textsc{chf}&-&82.0\%&84.6\%&\textbf{2.6\%}&86.3\%\\
\textsc{cm2/cm1}&2017, 2018&75.7\%&81.7\%&\textbf{6.0\%}&84.1\%\\
\textsc{jlw1}&2017&79.0\%&83.2\%&\textbf{4.2\%}&84.6\%\\
\textsc{jlw2}&2017&78.5\%&83.0\%&\textbf{4.5\%}&84.3\%\\
\textsc{kbk1}&2017&82.8\%&85.5\%&\textbf{2.6\%}&86.8\%\\
\textsc{kbk2/fk2}&2017, 2018&87.3\%&88.2\%&\textbf{0.8\%}&88.9\%\\
\textsc{wl1}&2017&79.9\%&83.6\%&\textbf{3.8\%}&85.0\%\\
\textsc{jlcx1}&2018&86.3\%&87.1\%&0.9\%&88.1\%\\
\textsc{jlcx2}&2018&86.5\%&87.1\%&0.6\%&88.1\%\\
\textsc{sg1}&2018&79.5\%&84.0\%&\textbf{4.6\%}&85.8\%\\
\end{tabular}
\caption{\textsc{wcsr} of audio \textsc{ace} systems and \textsc{df-best}. Note that two of the 2017 systems were resubmitted in \textsc{mirex} 2018. Significant improvements are shown in bold. Using \textsc{df-best} improves \textsc{ace} \textsc{wcsr} on average by 3.05\%.}\label{tab:wcsr}
\end{table}

\begin{figure}[p!]
\centering
\includegraphics[width=0.9\textwidth]{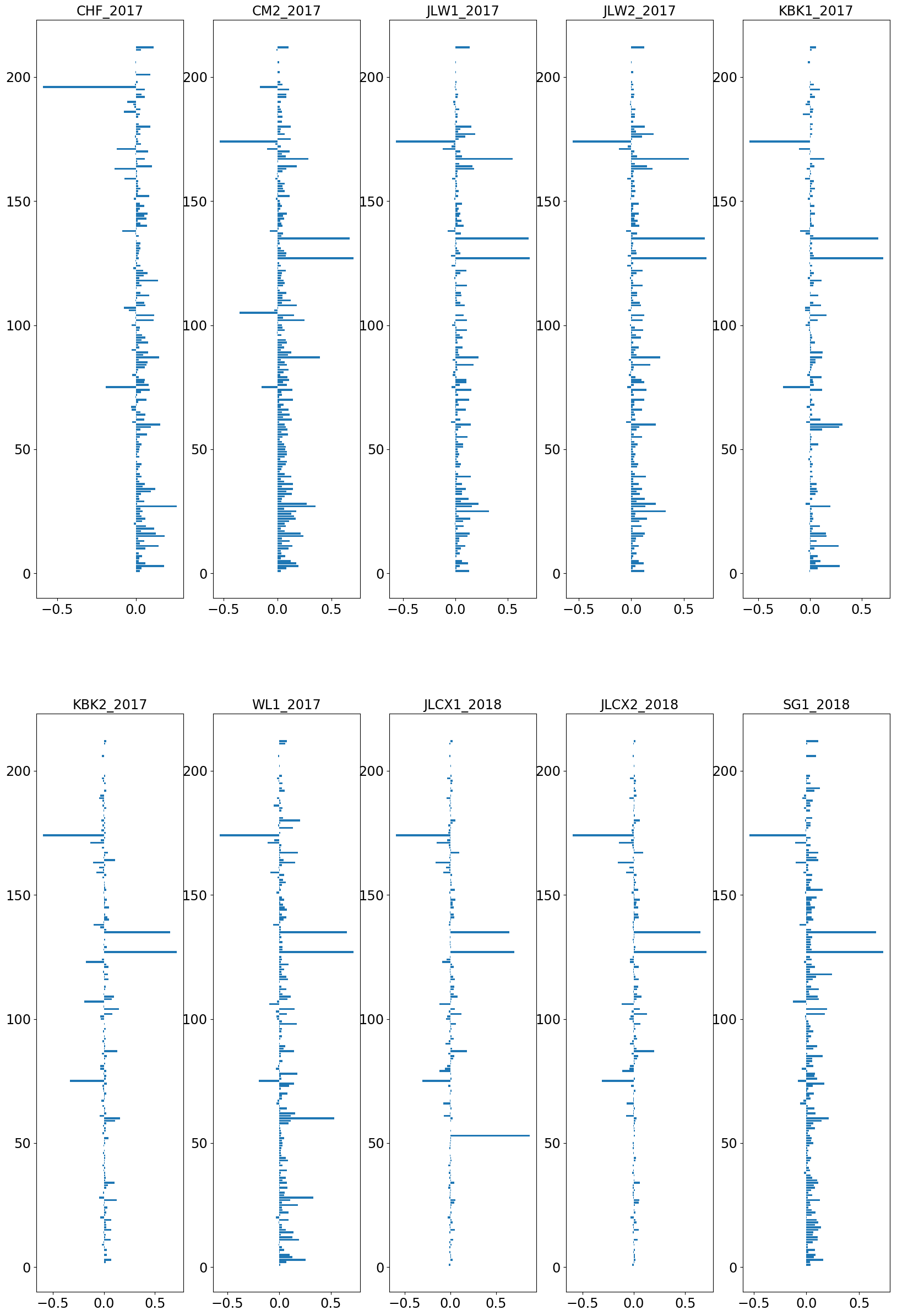}
\caption{\textsc{Csr} of \textsc{df-best} compared to the \textsc{csr} of the original audio \textsc{ace} system for each song in the dataset. The difference between the \textsc{csr} of \textsc{df-best} and the \textsc{csr} of the original audio \textsc{ace} system is given on the $x$-axis; the $y$-axis gives the index of the song in our data set. Horizontal lines on the right side of $x=0$ correspond to songs for which \textsc{df-best} improves the original system in terms of \textsc{csr}, while the horizontal lines on the left side of $x=0$ correspond to songs for which the original \textsc{ace} system produced better results in terms of \textsc{csr}.}\label{fig:song-wise-comparision}
\end{figure}

\begin{figure}[p!]
\centering
\includegraphics[width=\textwidth]{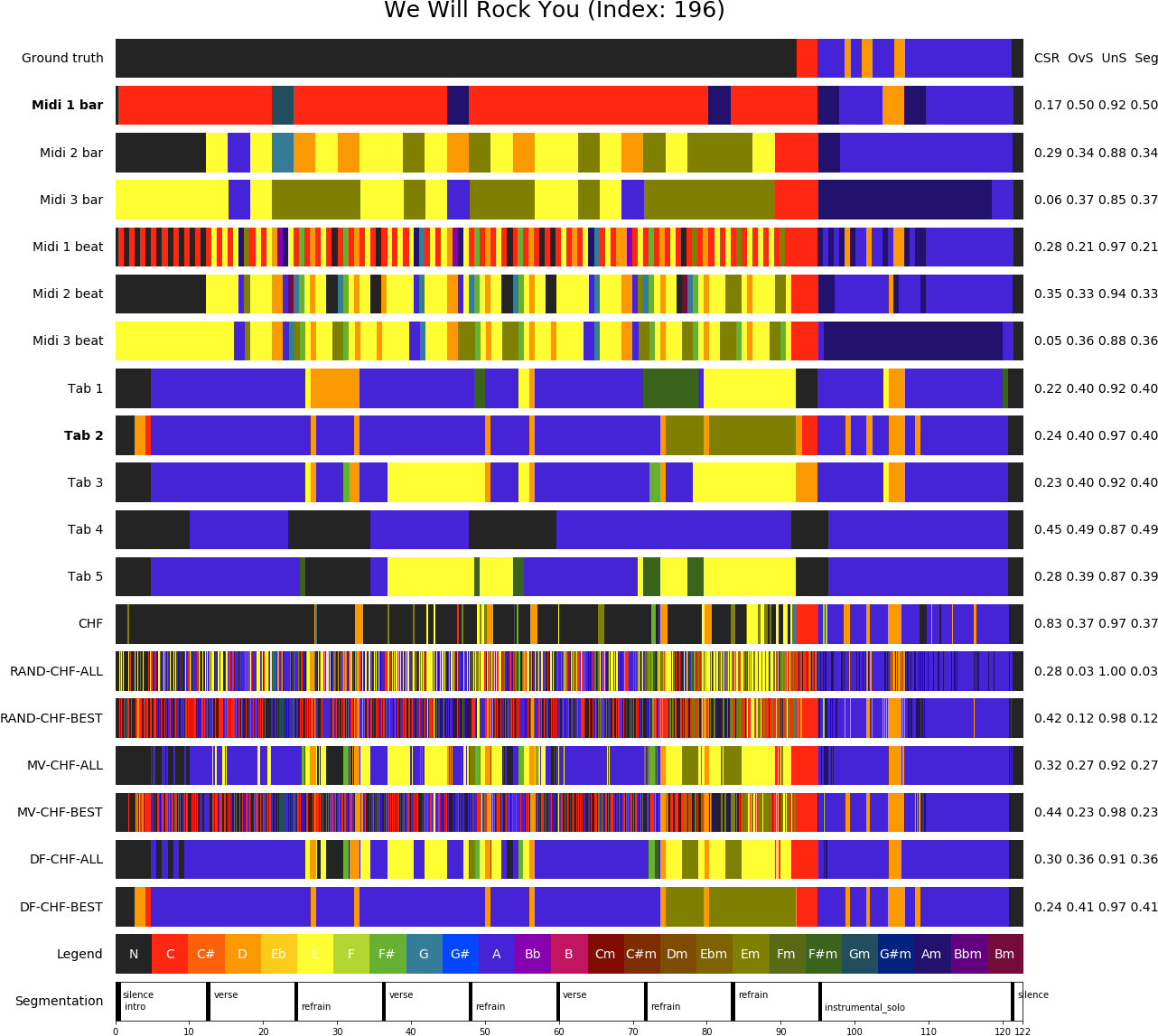}
\caption{Chord label sequences for the Queen song \textit{We will rock you}. Note: \textsc{csr} = Chord Symbol Recall; OvS = oversegmentation; UnS = undersegmentation; and Seg = segmentation.}\label{fig:wwry}
\end{figure}

Finally, let us look at the song-wise performance of \textsc{df-best} compared to the original audio file. Figure~\ref{fig:song-wise-comparision} shows the difference between the \textsc{csr} of the chord label sequence obtained by \textsc{df-best} and the \textsc{csr} of the chord label sequence found by the original audio \textsc{ace} system. The song keys are given on the $y$-axis, and the length of the horizontal bars correspond to the difference in \textsc{csr}. If a bar is located on the right side of $x = 0$, then the chord label sequence found by \textsc{df-best} is better than the sequence found by the original audio \textsc{ace} system. We see that this is the case for the vast majority of songs in our dataset. On the other hand, if a bar is located on the left side of $x = 0$, \textsc{df-best} performs worse than the original audio \textsc{ace} system. From all ten plots, we see one song for which \textsc{df-best} performs considerably worse. This is song number 174, \textit{Let it be} by The Beatles. This song has an issue similar to \textit{Don't pass me by}, as we saw in Figure~\ref{fig:ace_dont_pass_me_by}: the audio file that we used was shifted compared to the audio file that was used in the \textsc{mirex} competition. That is why the chord label sequence found by \textsc{df-best} is shifted and therefore is not consistent with the Isophonics annotations any more. We see another `peak' to the left at song 196, \textit{We will rock you} by Queen, for the \textsc{chf} algorithm. The reason for this difference is that \textsc{chf} estimates the chord label sequence for this song quite well, as shown in Figure~\ref{fig:wwry}. However, the best found \textsc{midi} and tab file find very different chord labels for the first 90 seconds of the song and therefore the result found by \textsc{df-best} is based on the tab file, which apparently had the highest source accuracy. In practice, this song starts with a lot of percussion and little harmonic content. Therefore, it is not so surprising that there are multiple views on the chord labels for this song.

In this section we have seen that \textsc{df-best} is the best combination of the selection strategy and integration method. \textsc{Df-best} performs better than the original audio algorithm in terms of \textsc{wcsr} and does not suffer from oversegmentation.

\chapter{Conclusion}\label{ch:conclusion}
The goal of this study was to test whether exploiting \textsc{midi}s and tabs improves Automatic Chord Estimation (\textsc{ace}). The \textsc{ace} task is concerned with estimating chords in audio recordings or symbolic representations of music and has applications in both music performance and in Music Information Retrieval (\textsc{mir}). Many systems for audio \textsc{ace} exist, but \textsc{ace} is not yet a solved problem: as we have identified in Section~\ref{sec:stagnation-subjectivity}, existing methods suffer from stagnation and overfitting to subjective annotations. There is need for a new strategy that overcomes existing stagnation in \textsc{ace} without further overfitting to existing (subjective) data sets. An improved system for \textsc{ace} would have positive consequences for millions of performers and \textsc{mir} systems for e.g. cover song identification and genre classification.

For this purpose, we created a new system, \textsc{decibel}, that (1) aligns \textsc{midi} and tab files to audio recordings and uses representation-specific chord estimation techniques for both symbolic and audio formats to estimate chord sequences for each file; and (2) integrates the resulting heterogeneous chord sequences into one final output sequence. In this concluding chapter, we will give an overview of our contributions and some directions for future work.

\section{Contributions}
The main contributions of this research are twofold: we designed a system that automatically extracts a rich harmonic representation and used this representation to improve state-of-the-art audio \textsc{ace} systems. In addition, we extended the existing Isophonics Reference Annotations with a large set of \textsc{midi} and tab files, which we manually matched to the annotations. As a final contribution, we made the implementation of our system available on GitHub. 

\subsection{Rich harmonic representation}
As we have seen in Chapters~\ref{ch:subsystem-audio}, \ref{ch:subsystem-midi} and \ref{ch:subsystem-tabs}, \textsc{decibel} has three subsystems that estimate chord label sequences from audio, \textsc{midi} and tab files, using representation-specific \textsc{ace} and alignment techniques. The resulting chord label sequences form a rich harmonic representation that is a contribution by itself, as it can easily be visualized and enables large-scale cross-version analysis of popular music, giving more insights on subjectivity issues. The potential use of this rich harmonic representation is described in Section~\ref{sec:harmonic-representation}.

\subsection{Improvement of state-of-the-art audio \textsc{ace} systems}
In Chapter~\ref{ch:data-fusion}, we described our experiments on data fusion. \textsc{Decibel} uses data fusion methods to combine the chord label sequences from the aforementioned rich harmonic representation into a single output sequence. The method \textsc{df-best} turned out to be the best-performing method. Using this data fusion method, \textsc{decibel} improves each of the ten original state-of-the-art audio \textsc{ace} systems. The average \textsc{wcsr} improvement is as much as 3.05\%. Supported by statistical tests, we can confirm the hypothesis that audio \textsc{ace} can be improved by integrating symbolic music formats.

Although \textsc{decibel} needs a data set of symbolic formats and has a longer calculation time than the original audio \textsc{ace} systems, it also has two major advantages beyond the \textsc{wcsr} improvement. As a first advantage, by using \textsc{midi} and tab files that are scraped from the Internet, \textsc{decibel} implicitly incorporates musical knowledge: \textsc{midi} and tab files can be considered as human-made transcriptions of musical notes and chords. As a second advantage, \textsc{decibel} requires a minimum amount additional training and therefore prohibits overfitting to subjective data sets. There is little training needed for the HMM used by Jump Alignment, but apart from that, neither the tab or \textsc{midi} subsystem needs to be trained. 

\subsection{Extended Isophonics data set}
As a third contribution, we extended the widely used Isophonics Reference Annotations with 770 \textsc{midi} files and 1668 tabs. These symbolic music files are scraped from the Internet and manually matched to the corresponding song in the Isophonics data set. In this research project, we learned that \textsc{midi} files and tabs are highly useful for the \textsc{ace} task. We expect that the extended data set can also contribute to advances in other \textsc{mir} tasks, such as structural segmentation or lyrics-to-audio alignment.

\subsection{Python implementations}
As a fourth and final contribution, we made our Python implementation of \textsc{decibel} available to the \textsc{mir} community. The complete code repository can be found on \url{https://github.com/DaphneO/DECIBEL}. As part of the complete \textsc{decibel} system, this repository contains new Python implementations for existing algorithms, such as Jump Alignment \citep{mcvicar2011using} and data fusion \citep{koops2016integration}, as well as implementations for new algorithms, such as our \textsc{midi} chord recognizer \textsc{cassette} and a tab parser.

\section{Future work}
From the results of this research project, we can conclude that the integration of multiple symbolic formats and audio for the improvement of \textsc{ace} is an interesting research direction that deserves more attention in the future. In this section, we give some directions for future work.

As a first suggestion for future work, it would be interesting to test \textsc{decibel}s performance on a larger data set, for example the Billboard data set. The Isophonics data set was convenient to use, because there were myriad \textsc{midi} files and tabs available for music by The Beatles and Queen. \textsc{Decibel}'s performance may drop when testing on less popular or more modern artists, because there may be less symbolic formats available for their songs. It would also be interesting to test \textsc{decibel} on music of another genre, for example jazz music. Very recently, a new data set for jazz, consisting of 113 tracks, was released by \citet{eremenko2018audio}. In the accompanying paper, \citet{eremenko2018audio} evaluate two audio \textsc{ace} methods on their new data set and observe that the \textsc{csr}'s for these methods were quite low. Extending the data set with \textsc{midi} and tab files for these jazz songs and running the \textsc{decibel} system on the extended data set may give better results.

As a second suggestion, we recommend using a larger chord vocabulary in future work. It would be interesting to see if the integration of symbolic formats also helps in recognizing more complex chords, such as seventh chords. Using \textsc{midi} and tab files might be very helpful in recognizing complex chords, as these representations directly encode notes and chord labels respectively.

Our third suggestion for future work is to experiment various techniques for \textsc{decibel}'s subtasks. In the current implementation, we decided to use state-of-the-art audio \textsc{ace} methods in combination with relatively simple methods for alignment and \textsc{midi} chord estimation. The rationale behind this is as follows: the audio \textsc{ace} methods should be as ``good'' as possible, as it may be trivial to improve a poorly performing audio \textsc{ace} method. On the other hand, the techniques for tab alignment, \textsc{midi} alignment and \textsc{midi} chord recognition should be simple: if we can show that simple methods already improve the audio \textsc{ace} method, we can assume that more sophisticated techniques will give even better results. Besides, simpler methods are easier to understand and implement, and typically do not require a lot of training. However, as we now know that applying simple techniques on \textsc{midi} and tab files already improves audio \textsc{ace}, it would be interesting to see how much we can improve our results by using more sophisticated techniques. For example, a \textsc{midi} chord recognizer that takes the musical context into account will probably give better results than our proposed \textsc{cassette} algorithm.

As a fourth and final suggestion for future work, we propose to test \textsc{decibel}'s performance by doing user tests. We saw that \textsc{decibel} improves each of the ten tested state-of-the-art audio \textsc{ace} systems. Based on these results, we expect that performers prefer \textsc{decibel}'s estimated chord label sequences to the sequences found by audio \textsc{ace} systems. However, recall from Section~\ref{sec:performance-evaluation} that \textsc{wcsr} is not always a good measure for the quality of a chord sequence. Therefore, it would be good to verify the improvement by doing user tests.

\bibliography{biblio}

\appendix
\chapter{Data set}\label{appendix:data-set}
\begin{longtable}{ll}
\endfirsthead
\endhead
\hline
\multicolumn{2}{r}{\textit{The Beatles - Please Please Me}} \\
\hline
\textbf{Index} & \textbf{Song title} \\
1 & I Saw Her Standing There \\
2 & Misery \\
3 & Anna (Go To Him) \\
4 & Chains \\
5 & Boys \\
6 & Ask Me Why \\
7 & Please Please Me \\
8 & Love Me Do \\
9 & P. S. I Love You \\
10 & Baby It's You \\
11 & Do You Want To Know A Secret \\
12 & A Taste Of Honey \\
13 & There's A Place \\
14 & Twist And Shout \\
\hline
\multicolumn{2}{r}{\textit{The Beatles - With the Beatles}} \\
\hline
\textbf{Index} & \textbf{Song title} \\
15 & It Won't Be Long \\
16 & All I've Got To Do \\
17 & All My Loving \\
18 & Don't Bother Me \\
19 & Little Child \\
20 & Till There Was You \\
21 & Please Mister Postman \\
22 & Roll Over Beethoven \\
23 & Hold Me Tight \\
24 & You Really Got A Hold On Me \\
25 & I Wanna Be Your Man \\
26 & Devil In Her Heart \\
27 & Not A Second Time \\
28 & Money \\
\hline
\multicolumn{2}{r}{\textit{The Beatles - A Hard Day's Night}} \\
\hline
\textbf{Index} & \textbf{Song title} \\
29 & A Hard Day's Night \\
30 & I Should Have Known Better \\
31 & If I Fell \\
32 & I'm Happy Just To Dance With You \\
33 & And I Love Her \\
34 & Tell Me Why \\
35 & Can't Buy Me Love \\
36 & Any Time At All \\
37 & I'll Cry Instead \\
38 & Things We Said Today \\
39 & When I Get Home \\
40 & You Can't Do That \\
41 & I'll Be Back \\
\hline
\multicolumn{2}{r}{\textit{The Beatles - Beatles for Sale}} \\
\hline
\textbf{Index} & \textbf{Song title} \\
42 & No Reply \\
43 & I'm a Loser \\
44 & Baby's In Black \\
45 & Rock and Roll Music \\
46 & I'll Follow the Sun \\
47 & Mr. Moonlight \\
48 & Kansas City- Hey, Hey, Hey, Hey \\
49 & Eight Days a Week \\
50 & Words of Love \\
51 & Honey Don't \\
52 & Every Little Thing \\
53 & I Don't Want to Spoil the Party \\
54 & What You're Doing \\
55 & Everybody's Trying to Be My Baby \\
\hline
\multicolumn{2}{r}{\textit{The Beatles - Help!}} \\
\hline
\textbf{Index} & \textbf{Song title} \\
56 & Help! \\
57 & The Night Before \\
58 & You've Got To Hide Your Love Away \\
59 & I Need You \\
60 & Another Girl \\
61 & You're Going To Lose That Girl \\
62 & Ticket To Ride \\
63 & Act Naturally \\
64 & It's Only Love \\
65 & You Like Me Too Much \\
66 & Tell Me What You See \\
67 & I've Just Seen a Face \\
68 & Yesterday \\
69 & Dizzy Miss Lizzy \\
\hline
\multicolumn{2}{r}{\textit{The Beatles - Rubber Soul}} \\
\hline
\textbf{Index} & \textbf{Song title} \\
70 & Drive My Car \\
71 & Norwegian Wood (This Bird Has Flown) \\
72 & You Won't See Me \\
73 & Nowhere Man \\
74 & Think For Yourself \\
75 & The Word \\
76 & Michelle \\
77 & What Goes On \\
78 & Girl \\
79 & I'm Looking Through You \\
80 & In My Life \\
81 & Wait \\
82 & If I Needed Someone \\
83 & Run For Your Life \\
\hline
\multicolumn{2}{r}{\textit{The Beatles - Revolver}} \\
\hline
\textbf{Index} & \textbf{Song title} \\
84 & Taxman \\
85 & Eleanor Rigby \\
86 & I'm Only Sleeping \\
87 & Love You To \\
88 & Here, There And Everywhere \\
89 & Yellow Submarine \\
90 & She Said She Said \\
91 & Good Day Sunshine \\
92 & And Your Bird Can Sing \\
93 & For No One \\
94 & Doctor Robert \\
95 & I Want To Tell You \\
96 & Got To Get You Into My Life \\
97 & Tomorrow Never Knows \\
\hline
\multicolumn{2}{r}{\textit{The Beatles - Sgt. Pepper's Lonely Hearts Club Band}} \\
\hline
\textbf{Index} & \textbf{Song title} \\
98 & Sgt. Pepper's Lonely Hearts Club Band \\
99 & With A Little Help From My Friends \\
100 & Lucy In The Sky With Diamonds \\
101 & Getting Better \\
102 & Fixing A Hole \\
103 & She's Leaving Home \\
104 & Being For The Benefit Of Mr. Kite! \\
105 & Within You Without You \\
106 & When I'm Sixty-Four \\
107 & Lovely Rita \\
108 & Good Morning Good Morning \\
109 & Sgt. Pepper's Lonely Hearts Club Band (Reprise) \\
110 & A Day In The Life \\
\hline
\multicolumn{2}{r}{\textit{The Beatles - Magical Mystery Tour}} \\
\hline
\textbf{Index} & \textbf{Song title} \\
111 & Magical Mystery Tour \\
112 & The Fool On The Hill \\
113 & Flying \\
114 & Blue Jay Way \\
115 & Your Mother Should Know \\
116 & I Am The Walrus \\
117 & Hello Goodbye \\
118 & Strawberry Fields Forever \\
119 & Penny Lane \\
120 & Baby You're A Rich Man \\
121 & All You Need Is Love \\
\hline
\multicolumn{2}{r}{\textit{The Beatles - The Beatles}} \\
\hline
\textbf{Index} & \textbf{Song title} \\
122 & Back in the USSR \\
123 & Dear Prudence \\
124 & Glass Onion \\
125 & Ob-La-Di, Ob-La-Da \\
126 & Wild Honey Pie \\
127 & The Continuing Story of Bungalow Bill \\
128 & While My Guitar Gently Weeps \\
129 & Happiness is a Warm Gun \\
130 & Martha My Dear \\
131 & I'm So Tired \\
132 & Black Bird \\
133 & Piggies \\
134 & Rocky Raccoon \\
135 & Don't Pass Me By \\
136 & Why Don't We Do It In The Road \\
137 & I Will \\
138 & Julia \\
139 & Birthday \\
140 & Yer Blues \\
141 & Mother Nature's Son \\
142 & Everybody's Got Something To Hide Except Me and My Monkey \\
143 & Sexy Sadie \\
144 & Helter Skelter \\
145 & Long Long Long \\
146 & Revolution \\
147 & Honey Pie \\
148 & Savoy Truffle \\
149 & Cry Baby Cry \\
150 & Revolution \\
151 & Good Night \\
\hline
\multicolumn{2}{r}{\textit{The Beatles - Abbey Road}} \\
\hline
\textbf{Index} & \textbf{Song title} \\
152 & Come Together \\
153 & Something \\
154 & Maxwell's Silver Hammer \\
155 & Oh! Darling \\
156 & Octopus's Garden \\
157 & I Want You \\
158 & Here Comes The Sun \\
159 & Because \\
160 & You Never Give Me Your Money \\
161 & Sun King \\
162 & Mean Mr Mustard \\
163 & Polythene Pam \\
164 & She Came In Through The Bathroom Window \\
165 & Golden Slumbers \\
166 & Carry That Weight \\
167 & The End \\
168 & Her Majesty \\
\hline
\multicolumn{2}{r}{\textit{The Beatles - Let It Be}} \\
\hline
\textbf{Index} & \textbf{Song title} \\
169 & Two of Us \\
170 & Dig a Pony \\
171 & Across the Universe \\
172 & I Me Mine \\
173 & Dig It \\
174 & Let It Be \\
175 & Maggie Mae \\
176 & I've Got A Feeling \\
177 & One After \\
178 & The Long and Winding Road \\
179 & For You Blue \\
180 & Get Back \\
\hline
\multicolumn{2}{r}{\textit{Queen - Greatest Hits I}} \\
\hline
\textbf{Index} & \textbf{Song title} \\
181 & Bohemian Rhapsody \\
182 & Another One Bites The Dust \\
184 & Fat Bottomed Girls \\
185 & Bicycle Race \\
186 & You're My Best Friend \\
187 & Don't Stop Me Now \\
188 & Save Me \\
189 & Crazy Little Thing Called Love \\
190 & Somebody To Love \\
192 & Good Old Fashioned Lover Boy \\
193 & Play The Game \\
195 & Seven Seas Of Rhye \\
196 & We Will Rock You \\
197 & We Are The Champions \\
\hline
\multicolumn{2}{r}{\textit{Queen - Greatest Hits II}} \\
\hline
\textbf{Index} & \textbf{Song title} \\
198 & A Kind Of Magic \\
201 & I Want It All \\
202 & I Want To Break Free \\
206 & Who Wants To Live Forever \\
211 & Hammer To Fall \\
212 & Friends Will Be Friends \\
\hline
\end{longtable}

\chapter{Alignment listening test results}
\begin{table}[ht]
\begin{tabular}{ccl}
\textbf{\textsc{midi} file} & \textbf{Alignment} & \textbf{Listening test result} \\
& \textbf{error} & \\
128-005 & 0.723 &  2 - Good alignment, no fade out in \textsc{midi}\\
014-003 & 0.762 &  2 - Good alignment, \textsc{midi} cut off too soon at the end\\
095-002 & 0.636 &  2 - Good alignment, but seems to be a bit shifted\\
047-001 & 0.934 &  1 - Very bad alignment\\
063-004 & 0.764 &  3 - Good alignment\\
098-004 & 0.850 &  3 - Good alignment\\
196-001 & 0.656 &  3 - Good alignment\\
099-002 & 0.959 &  1 - Very bad alignment\\
195-001 & 0.723 &  2 - Good alignment with minor issues at the end\\
036-002 & 0.903 &  1 - First verse and chorus missing in \textsc{midi}, issues at the end\\
184-001 & 0.720 &  3 - Good alignment\\
078-003 & 0.654 &  2 - Good alignment, no fade out in \textsc{midi}\\
042-001 & 0.815 &  2 - Good alignment, but sometimes seems to be a bit shifted\\
152-002 & 0.705 &  2 - Good alignment, no fade out in \textsc{midi}\\
181-005 & 0.964 &  1 - Very bad alignment (transcription with only piano)\\
002-001 & 0.644 &  2 - Good alignment, no fade out in \textsc{midi}\\
061-002 & 0.762 &  2 - Good alignment, minor issues at the end\\
155-001 & 0.880 &  1 - Mediocre alignment, last 50s of the audio missing in \textsc{midi} file\\
073-004 & 0.760 &  2 - Good alignment, but seems to be a bit shifted\\
055-002 & 0.712 &  3 - Good alignment\\
003-001 & 0.771 &  2 - Good alignment, but seems to be a bit shifted\\
010-001 & 0.709 &  2 - Good alignment, no fade out in \textsc{midi}\\
125-002 & 0.784 &  3 - Good alignment\\
202-005 & 0.641 &  3 - Good alignment\\
212-001 & 0.532 &  3 - Good alignment\\
\end{tabular}
\caption{Results of the alignment listening test}\label{tab:alignment-listening-test}
\end{table}

\chapter{Tabs in parsing evaluation set}\label{appendix:tabs-evaluation-set}
The annotated subset for the tab parsing evaluation consists of these 25 songs:
\begin{enumerate}
\item Let It Be (ver 6) Tabs
\item A Hard Days Night (ver 10) Chords
\item You're My Best Friend solo Tabs
\item Savoy Truffle (ver 2) Chords
\item Strawberry Fields Forever Acoustic Chords
\item A Taste Of Honey (ver 2) Chords
\item What You're Doing Chords
\item A Hard Days Night (ver 3) Tabs
\item Let It Be (ver 4) Chords
\item Love You To Tabs
\item Cry Baby Cry (ver 4) Chords
\item For You Blue Chords
\item Getting Better (ver 5) Chords
\item Help (ver 8) Chords
\item No Reply (ver 4) Chords
\item Happiness Is A Warm Gun Tabs
\item I'll Be Back (ver 2) Tabs
\item Tell Me Why Tabs
\item Get Back (ver 5) Tabs
\item Anna (ver 2) Chords
\item A Day In The Life Tabs
\item You Never Give Me Your Money Tabs
\item Eleanor Rigby (ver 3) Tabs
\item I'll Be Back (ver 7) Chords
\item Yesterday Chords
\end{enumerate}

\chapter{Tabs in forward and backward training set}\label{appendix:tabs-pf-pb-training-set}
The following 22 songs were used for training forward and backward probabilities:
\begin{enumerate}
\item Misery Chords
\item Please Mister Postman Chords
\item You Really Got A Hold On Me (ver 2) Chords
\item And I Love Her Tabs
\item I'll Cry Instead Tabs
\item I'll Be Back Tabs
\item Mr Moonlight Tabs
\item Help (ver 6) Chords
\item I Need You Tabs
\item You're Going To Lose That Girl Chords
\item Act Naturally (ver 3) Tabs
\item Norwegian Wood (ver 4) Chords
\item Think For Yourself Chords
\item Girl Acoustic Chords
\item Good Day Sunshine Chords
\item Sgt Peppers Lonely Hearts Club Band Reprise (ver 2) Chords
\item Penny Lane (ver 6) Chords
\item Come Together Chords
\item Mean Mr Mustard (ver 2) Chords
\item Golden Slumbers Chords
\item A Kind Of Magic Chords
\item I Want To Break Free Chords
\end{enumerate}

{
}

\end{document}